\documentclass[preprint]{ptephy_v1}%%%%%% to generate preprint number with ptep logo

\preprintnumber{XXXX-XXXX} %%% %%% Insert preprint number here

\usepackage{txfonts}

\usepackage[utf8]{inputenc}
\usepackage{hyperref}

\usepackage[T1]{fontenc} % if needed
\usepackage{amsmath}
\usepackage{graphicx,epsfig,wrapfig}
\usepackage{dutchcal}
\usepackage{makeidx}
\usepackage{amsfonts}
\usepackage{amssymb}
\usepackage{mathtools}
\usepackage{amsbsy}
\usepackage[capitalize]{cleveref}
\usepackage{mathrsfs}
\usepackage{slashed}
\usepackage{bm}
\usepackage{multirow}
\usepackage{braket}

\newcommand{\bfp}{{\bf p}_{\perp}}

\usepackage{xcolor}
\usepackage{rotating}
\begin{document}

\title{TMD Relations: Insights from a Light-Front Quark-Diquark Model}

%%%% To generate auto affiliation numbers please use \author{}\affil{} command

\author{Shubham Sharma}
\affil{Computational High Energy Physics Lab, Department of Physics, Dr. B.R. Ambedkar National Institute of Technology, Jalandhar, 144008, India \email{s.sharma.hep@gmail.com}}

\author{Satyajit Puhan}
\affil{Computational High Energy Physics Lab, Department of Physics, Dr. B.R. Ambedkar National Institute of Technology, Jalandhar, 144008, India \email{puhansatyajit@gmail.com}}

\author{Narinder Kumar}
\affil{Computational High Energy Physics Lab, Department of Physics, Dr. B.R. Ambedkar National Institute of Technology, Jalandhar, 144008, India \&
Computational Theoretical High Energy Physics Lab, Department of Physics, Doaba College, Jalandhar 144004, India\email{narinderhep@gmail.com}}

\author{Harleen Dahiya\thanks{These authors contributed equally to this work}}
\affil{Computational High Energy Physics Lab, Department of Physics, Dr. B.R. Ambedkar National Institute of Technology, Jalandhar, 144008, India \email{dahiyah@nitj.ac.in}}

% \author{Insert third author name here}
% \author[3]{Insert fourth author name here} %%% Use optional bracket [3] to change the respective address
% \affil{Insert third author address here}

% \author{Insert last author name here\thanks{These authors contributed equally to this work}}
% \affil{Insert last author address here}

%%% To include the collaborator name... Please use the command "\collaborator"
%%% For example: \collaborator{ATLAS Collaboration}

\begin{abstract}%
In this work, we have established the relations between the T-even proton transverse momentum-dependent parton distributions (TMDs) at all twist levels up to twist-4 using the light-front quark diquark model (LFQDM). From the parameterization equations of TMDs, we have found that there are multiple ways by which a particular TMD can be expressed in terms of the helicities of the proton in the initial and final states. For the first time, we have presented a parameterization table that can be applied to the derivation and recognition of proton TMDs based on their helicity. We have constructed the linear and quadratic relationships of TMDs at the intra-twist and inter-twist levels within the same model. We have also looked at the inequality relations that TMDs follow. Additionally, to provide an easy access to the calculations, amplitude matrices have been expressed in the form of TMDs over all the possible helicities of the diquark.
\end{abstract}

\subjectindex{xxxx, xxx}

\maketitle

% \section{Insert A head here}
% This demo file is intended to serve as a ``starter file''
% for ptephy journal papers produced under \LaTeX\ using
% \verb+ptephy_v1.cls+ v0.1

	\section{Introduction}
	One of the challenging tasks of contemporary physicists is to reveal the complex internal structure of the proton. Proton is supposed to be the bound state of three strongly interacting constituent quarks, locked with gluons in a pool of sea quarks \cite{Gross:2022hyw}. Perturbative quantum chromodynamics (pQCD) provides an understanding of its structure at a high $Q^2$ regime, where gluon contribution is dominant over the quarks. While in the non-perturbative regime, i.e., at low $Q^2$, the valence quarks dominate \cite{Donnachie:1993it}. Therefore, multi-dimensional distribution functions (DFs) are used in non-perturbative QCD to present some ideas regarding the internal structure of the proton \cite{Hoodbhoy:2003uu,Zhu:2011ym,Baranov:2014ewa,Lorce:2011dv,Kanazawa:2014nha}.
	 The information about the physical properties of the proton is also encoded in these DFs. 

	The deep inelastic scattering (DIS) process \cite{Collins:2004nx,Ji:2004wu,Polchinski:2002jw} is a crucial process for revealing the hadronic structure because it enables resolution down to the level of individual quarks and gluons, which are called partons. The DFs extracted from DIS experiments are parton distribution functions (PDFs) \cite{Collins:1981uw, Martin:1998sq,Gluck:1994uf,Gluck:1998xa}, which encode the information of parton's longitudinal momentum fraction ($x$) acquired from it's parent proton. Being a function of longitudinal momentum fraction $x$, the PDFs give only one-dimensional structural information of the proton and lack information about the proton's spatial and transverse structure. The transverse momentum-dependent parton distributions (TMDs) \cite{Barone:2001sp, Diehl:2015uka,Puhan:2023ekt,Puhan:2023hio} and generalized parton distributions (GPDs) \cite{Diehl:2003ny,Garcon:2002jb,Belitsky:2005qn,Sharma:2023ibp} are essential for understanding these three-dimensional structures. Along with the longitudinal momentum fraction $x$ of the quark inside the proton, the TMDs store transverse momentum information $\bfp$ of quarks. Through TMD evolution and factorization, one can access the cross-section of the  Drell-Yan process \cite{Tangerman:1994eh,Zhou:2009jm,Mulders:1995dh,Collins:2002kn,Bacchetta:2017gcc}, the semi-inclusive deep inelastic-scattering (SIDIS) \cite{Bacchetta:2006tn,Brodsky:2002cx,Ji:2004wu,Bacchetta:2017gcc} and $Z^0/W^\pm$ production processes \cite{Bacchetta:2017gcc,Catani:2015vma}. More details regarding proton TMDs will be obtained by the forthcoming electron-ion collider (EIC) \cite{Accardi:2012qut,AbdulKhalek:2021gbh,Amoroso:2022eow}. Further, the spatial structure, spin densities, charge, magnetization, form factors (FFs), radius, mechanical properties, etc., can be easily extracted by using the three-dimensional GPDs of protons \cite{Sharma:2023ibp}. The GPDs are functions of longitudinal momentum fraction ($\textit{x}$), skewness ($\zeta$), and momentum transfer between initial and final hadron ($\Delta$). The theoretical description of hard exclusive reactions such as deeply virtual meson production (DVMP) \cite{Favart:2015umi} or deeply virtual Compton scattering (DVCS) \cite{Ji:1996nm} can be obtained through the GPDs. Both the TMDs and GPDs enable us to create three-dimensional images of the proton and give us crucial details on the momentum distribution and the orbital motion of partons inside the proton. Furthermore, the six-dimensional generalized transverse-momentum dependent parton distributions (GTMDs) and five-dimensional  Wigner distribution are there to understand the dynamics of double Drell-Yan process (DDY) \cite{Lorce:2011dv,Kanazawa:2014nha,Chakrabarti:2017teq,Kumar:2017xcm,Kaur:2019kpi,Sharma:2023tre,Sharma:2023qgb}.

	In this work, we target the relations among the proton TMDs up to twist-$4$ arising from the valence quark distributions. There have been developments in theoretical prediction as well as experimental data for leading twist TMDs for proton \cite{Pasquini:2008ax,Avakian:2010br,Kanazawa:2014nha}, but higher twist TMDs have been less accessed yet. There are a total of $20$ T-even TMDs for proton, out of which twist-2, twist-3, and twist-4 TMDs are $6$, $8$, and $6$ in number respectively \cite{Meissner:2009ww}.
	The holographic model \cite{Maji:2017wwd,Lyubovitskij:2020otz}, Nambu–Jona–Lasinio (NJL) \cite{Takyi:2019ahv}, Valon \cite{AlizadehYazdi:2014dug}, light-front constituent quark (LFCQM) \cite{Pasquini:2008ax,Pasquini:2010af,Boffi:2009sh,Pasquini:2011tk}, quark-diquark model \cite{Jakob:1997wg, Gamberg:2007wm,Cloet:2007em,Bacchetta:2008af,Maji:2015vsa}, covariant parton model (CPM) \cite{Bastami:2020rxn}, chiral quark soliton model \cite{Diakonov:1997sj,Schweitzer:2001sr,Schweitzer:2013iva}, and bag model \cite{Avakian:2008dz,Courtoy:2008dn} have produced excellent results for the twist-2 TMDs. For a leading twist, certain model-independent lattice QCD calculations have also been performed \cite{Hagler:2009mb, Musch:2010ka, Musch:2011er}.
 % 
%	 Excellent findings have been obtained for twist-2 TMDs using the holographic model \cite{Maji:2017wwd,Lyubovitskij:2020otz}, Nambu–Jona–Lasinio (NJL) \cite{Takyi:2019ahv}, Valon \cite{AlizadehYazdi:2014dug}, light-front constituent quark (LFCQM) \cite{Pasquini:2008ax,Pasquini:2010af,Boffi:2009sh,Pasquini:2011tk}, quark-diquark \cite{Jakob:1997wg, Gamberg:2007wm,Cloet:2007em,Bacchetta:2008af,Maji:2015vsa}, covariant parton model (CPM) \cite{Bastami:2020rxn}, chiral quark soliton \cite{Diakonov:1997sj,Schweitzer:2001sr,Schweitzer:2013iva}, and bag models \cite{Avakian:2008dz,Courtoy:2008dn}. Some model-independent lattice QCD calculation has also been done for leading twist \cite{Hagler:2009mb, Musch:2010ka, Musch:2011er}. 
	While looking into higher twist TMD calculations, CPM \cite{Bastami:2020rxn} for sub-leading twist, quark models \cite{Lorce:2014hxa} up to twist-$4$, light-front models \cite{Pasquini:2018oyz} for twist-$3$ $e^q$ TMDs have been studied theoretically. The valence quark contribution is more significant in twist-2 TMDs as compared to the higher twist TMDs due to the factor of $1/Q$ and $1/Q^2$ in the correlator for twist-3 and twist-4 cases, respectively \cite{Bastami:2020rxn}. 

	In this work, we have calculated all the T-even TMDs up to twist-4 without taking account the gluon contributions and presented them in the overlap form of light-front wave functions (LFWFs). We have taken the Wilson line to the lowest contribution value which is unity. We have used the light-front quark diquark model (LFQDM) \cite{Maji:2016yqo,Maji:2017bcz,Kumar:2017dbf} to compute our results. LFQDM, a model influenced by AdS/QCD, links the quantum field theory in LF formalism for strongly interacting particles to the 5-dimensional Anti-de Sitter (AdS) space-time of gravitational theory. The confinement of quarks and gluons within hadrons is reflected in the behavior of string-like particles in the higher dimensional AdS space that describes hadrons \cite{Brodsky:2005en}. According to the LFQDM, the proton is characterized as a mixture of an active quark and a diquark spectator of a given mass. The Fock-state of the proton in LFQDM consists of constituent quarks only.
%	A combination of an active quark and a diquark spectator of specified mass is how the proton is defined in the LFQDM. 
LFQDM is consistent with the quark counting rule and the Drell-Yan-West relation. The scalar $(S = 0)$ and axial vector $(S = 1)$ diquarks contribute to the LFWFs, which are generated from the AdS/QCD predictions. They have a $SU(4)$ spin-flavor structure. LFQDM has been prominent in describing the Leading twist TMDs, the transverse shape of the proton, the relationship between quark densities and the first moments,  transversity $h_1 (x)$ and helicity $g_1 (x)$ PDFs for leading twist, PDFs evolution up to the scale $\mu^2 = 10^4$ GeV$^2$, SIDIS asymmetries with HERMES and COMPASS findings, leading twist GPDs and their relations with TMDs, gravitational form factors $A (Q^2)$ and B(Q$^2$), mechanical properties from D-term etc.

 The paper is arranged as follows. In Section \ref{secmodel}, we have discussed the LFQDM and the input parameters for the calculation of results. In Sections \ref{sec_corpar}, \ref{sec_ovf} and \ref{sec_exf}, we have discussed the TMDs correlator, overlap expressions for all the T-even TMDs and their explicit expressions respectively. Futher in Section \ref{sec_rel}
, we have introduced the relation among all the higher twist T-even TMDs and TMD matrix for the active (spectator) quark. We have finally concluded in Section \ref{sec_conclusion}.

	%%%%%%%%%%%%%%%%%%%%%%%%%%%%%%%%
	\section{Light-Front Quark-Diquark Model (LFQDM) \label{secmodel}}
	Proton has been defined as a combination of an active quark and a diquark witness with a specific mass in the LFQDM \cite{Maji:2016yqo,Chakrabarti:2019wjx}. Proton possess a spin-flavor $SU(4)$ structure which is characterized by an amalgam of isoscalar-scalar diquark singlet $|u~ S^0\rangle$, isoscalar-vector diquark $|u~ A^0\rangle$, and isovector-vector diquark $|d~ A^1\rangle$ states as \cite{Jakob:1997wg,Bacchetta:2008af}
	\begin{equation}
		|P; \pm\rangle = C_S|u~ S^0\rangle^\pm + C_V|u~ A^0\rangle^\pm + C_{VV}|d~ A^1\rangle^\pm. \label{PS_state}
	\end{equation}
	In this case, scalar and vector diquarks are denoted by $S$ and $A=V,VV$ respectively. 
	Their corresponding isospins are denoted by the superscripts $0$ and $1$. The coefficients $C_{i}$ of scalar and vector diquark states have been found in Ref. \cite{Maji:2016yqo} and are provided in Table \ref{tab_par}. The active quark's proportion of longitudinal momentum from the parent proton is $x=p^+/P^+$, where the momentum of quark ($p$) and diquark ($P_X$) are respectively given as
	\begin{eqnarray}
		p &&\equiv \bigg(xP^+, p^-,\bfp \bigg)\,,\label{qu} \\
		P_X &&\equiv \bigg((1-x)P^+,P^-_X,-\bfp\bigg). \label{diq}
	\end{eqnarray}
	For the scalar $|\nu~ S\rangle^\pm $ and vector diquark $|\nu~ A \rangle^\pm$ case, the expansion of the Fock-state in two particles for $J^z =\pm1/2$ can be expressed as \cite{Ellis:2008in}
	\begin{eqnarray}
		|\nu~ S\rangle^\pm &=&\sum_{\lambda^q}  \int \frac{dx~ d^2\bfp}{2(2\pi)^3\sqrt{x(1-x)}}  \psi^{\pm(\nu)}_{\lambda^q}(x,\bfp)\bigg|\lambda^{q},\lambda^{s}; xP^+,\bfp\bigg\rangle, \label{fockSD}\\
		|\nu~ A \rangle^\pm &=&\sum_{\lambda^q} \sum_{\lambda^D} \int \frac{dx~ d^2\bfp}{2(2\pi)^3\sqrt{x(1-x)}} \psi^{\pm(\nu)}_{\lambda^q \lambda^D }(x,\bfp)\bigg|\lambda^{q},\lambda^{D}; xP^+,\bfp\bigg\rangle. \label{fockVD}
	\end{eqnarray}
For the scalar case, the flavor index is $\nu ~=u$, whereas for the vector case, it is $\nu ~=u,d$. The two particle state is represented as $|\lambda^q,~\lambda^{Sp};  xP^+,\bfp\rangle$, where the spectator diquark helicity is $\lambda^{Sp}$ and the quark helicity is $\lambda^q=\pm\frac{1}{2}$. The scalar diquark's spectator helicity is $\lambda^{Sp}=\lambda^{S}=0$ (singlet), whereas the vector diquark's spectator helicity is $\lambda^{Sp}=\lambda^{D}=\pm 1,0$ (triplet). In Table \ref{tab_LFWF}, the LFWFs have been provided when $J^z=\pm1/2$, with the diquarks conceivably being a scalar or a vector  \cite{Maji:2017bcz}.
	\begin{table}[h]
		\centering % used for centering table
		\begin{tabular}{ |p{0.6cm}|p{1.1cm}|p{0.8cm}|p{0.4cm} p{4.4cm}|p{0.4cm} p{4.4cm}|}
			\hline
			&$~\lambda^q$&$~\lambda^{Sp}$&\multicolumn{2}{c|}{LFWFs for $J^z=+1/2$} & \multicolumn{2}{c|}{LFWFs for $J^z=-1/2$}\\
			\hline
			$\rm{~S}$&$+1/2$&$~0$&$\psi^{+(\nu)}_{+}$&$~=~N_S~ \varphi^{(\nu)}_{1}$&$\psi^{-(\nu)}_{+}$&$~=~N_S \bigg(\frac{p^1-ip^2}{xM}\bigg)~ \varphi^{(\nu)}_{2}$  \\
			&$-1/2$&$~0$&$\psi^{+(\nu)}_{-}$&$~=~-N_S\bigg(\frac{p^1+ip^2}{xM} \bigg)~ \varphi^{(\nu)}_{2}$&$\psi^{-(\nu)}_{-}$&$~=~N_S~ \varphi^{(\nu)}_{1}$~   \\
			%		~S No.~&~$\lambda^q$~&~$\lambda^D$~&\multicolumn{2}{c|}{LFWFs for $J^z=+1/2$} & \multicolumn{2}{c|}{LFWFs for $J^z=-1/2$}\\
			\hline
			&$+1/2$&$+1$&$\psi^{+(\nu)}_{+~+}$&$~=N^{(\nu)}_1 \sqrt{\frac{2}{3}} \bigg(\frac{p^1-ip^2}{xM}\bigg)~  \varphi^{(\nu)}_{2}$&$\psi^{-(\nu)}_{+~+}$&$~=~0$~  \\
			&$-1/2$&$+1$&$\psi^{+(\nu)}_{-~+}$&$~=~N^{(\nu)}_1 \sqrt{\frac{2}{3}}~ \varphi^{(\nu)}_{1}$&$\psi^{-(\nu)}_{-~+}$&$~=~0$~ \\
			$\rm{~A}$&$+1/2$&$~0$&$\psi^{+(\nu)}_{+~0}$&$~=~-N^{(\nu)}_0 \sqrt{\frac{1}{3}}~  \varphi^{(\nu)}_{1}$&$\psi^{-(\nu)}_{+~0}$&$~=~N^{(\nu)}_0 \sqrt{\frac{1}{3}} \bigg( \frac{p^1-ip^2}{xM} \bigg)~  \varphi^{(\nu)}_{2}$~   \\
			&$-1/2$&$~0$&$\psi^{+(\nu)}_{-~0}$&$~=N^{(\nu)}_0 \sqrt{\frac{1}{3}} \bigg(\frac{p^1+ip^2}{xM} \bigg)~ \varphi^{(\nu)}_{2}$&$\psi^{-(\nu)}_{-~0}$&$~=~N^{(\nu)}_0\sqrt{\frac{1}{3}}~  \varphi^{(\nu)}_{1}$~   \\
			&$+1/2$&$-1$&$\psi^{+(\nu)}_{+~-}$&$~=0$&$\psi^{-(\nu)}_{+~-}$&$~=~- N^{(\nu)}_1 \sqrt{\frac{2}{3}}~  \varphi^{(\nu)}_{1}$~   \\
			&$-1/2$&$-1$&$\psi^{+(\nu)}_{-~-}$&$~=0$&$\psi^{-(\nu)}_{-~-}$&~$=~N^{(\nu)}_1 \sqrt{\frac{2}{3}} \bigg(\frac{p^1+ip^2}{xM}\bigg)~  \varphi^{(\nu)}_{2}$~   \\
			\hline
		\end{tabular}
		\caption{The LFWFs for the active quark $\lambda^q$ and the spectator diquark $\lambda^{Sp}$ variations with their helicities for both the diquark possibilities  $J^z=\pm1/2$. The normalization constants are $N_S$, $N^{(\nu)}_0$ and $N^{(\nu)}_1$.
			}
		\label{tab_LFWF} % is used to refer this table in the text
	\end{table}
	Based on soft-wall AdS/QCD prediction, the generic form of LFWFs $\varphi^{(\nu)}_{i}=\varphi^{(\nu)}_{i}(x,\bfp)$ in Table \ref{tab_LFWF} have been constructed. The parameters $a^\nu_i,~b^\nu_i$ and $\delta^\nu$ were established in accordance with Ref. \cite{Maji:2017bcz}. We have
	\begin{eqnarray}
		\varphi_i^{(\nu)}(x,\bfp)=\frac{4\pi}{\kappa}\sqrt{\frac{\log(1/x)}{1-x}}x^{a_i^\nu}(1-x)^{b_i^\nu}\exp\Bigg[-\delta^\nu\frac{\bfp^2}{2\kappa^2}\frac{\log(1/x)}{(1-x)^2}\bigg].
		\label{LFWF_phi}
	\end{eqnarray}
	% For convenience, we define
	% \begin{eqnarray}
	% 	\mathbcal{T}_{ij}^{(\nu)}(x,\bfp)&=&\varphi_i^{(\nu) \dagger}(x,\bfp) \varphi_j^{(\nu)}(x,\bfp)
	% 	\label{Tij1},
	% \end{eqnarray}
	% where, $i,j=1,2$. As a direct consequence of Eq. \eqref{LFWF_phi} and Eq. \eqref{Tij1}, we get
	% \begin{eqnarray}
	% 	\mathbcal{T}_{ij}^{(\nu)}(x,\bfp)&=&\mathbcal{T}_{ji}^{(\nu)}(x,\bfp)\label{Tij2},\\
	% 	\varphi_i^{(\nu)\dagger}(x,\bfp)&=&\varphi_i^{(\nu)}(x,\bfp)\label{Tij3}.
	% \end{eqnarray}
	In the scenario of $a_i^\nu=b_i^\nu=0$  and $\delta^\nu=1.0$, the wave functions $\varphi_i^\nu ~(i=1,2)$ contract to AdS/QCD. In order to fit the parameters $a_i^{\nu}$ and $b_i^{\nu}$ at the model scale $\mu_0=0.313{\ \rm GeV}$, the Dirac and Pauli form factor data has been used \cite{Maji:2016yqo,Efremov:2009ze,Burkardt:2007rv}. For both the quark flavors, the value of parameter $\delta^{\nu}$ has been assumed to be one that has been taken from AdS/QCD \cite{Maji:2016yqo,deTeramond:2011aml}. Along with this, the normalization constants $N_{i}^{2}$ in Table {\ref{tab_LFWF}} have been obtained from Ref. \cite{Maji:2016yqo}. The model parameters for both the striking quark possibilities are presented in Table \ref{tab_par} for completeness. The value $0.4~\mathrm{GeV}$ \cite{Chakrabarti:2013dda,Chakrabarti:2013gra} has been assigned to the AdS/QCD scale parameter $\kappa$ appearing in  Eq. (\ref{LFWF_phi}). We have taken the proton mass ($M$) and the constituent quark mass ($m$) respectively as $0.938~\mathrm{GeV}$ and $0.055~\mathrm{GeV}$, in accordance with Ref. \cite{Chakrabarti:2019wjx}.
	\begin{table}[h]
		\centering
		\begin{tabular}{|c|c|c|}
			\hline
			% \hline
			$\nu$       & $u$                 & $d$                          \\ \hline 
   % \hline
			$C_{S}^{2}$ & $1.3872$            & $0$                     \\ \hline
			$C_{V}^{2}$ & $0.6128$            & $0$                     \\ \hline
			$C_{VV}^{2}$ & $0$            & $1$                     \\ \hline
			$N_{S}$     & $2.0191$            & $0$                          \\ \hline
			$N_0^{\nu}$ & $3.2050$            & $5.9423$                     \\ \hline
			$N_1^{\nu}$ & $0.9895$            & $1.1616$                     \\ \hline
			$a_1^{\nu}$ & $0.280\pm 0.001$    & $0.5850 \pm 0.0003$          \\ \hline
			$b_1^{\nu}$ & $0.1716 \pm 0.0051$ & $0.7000 \pm 0.0002$          \\ \hline
			$a_2^{\nu}$ & $0.84 \pm 0.02$     & $0.9434^{+0.0017}_{-0.0013}$ \\ \hline
			$b_2^{\nu}$ & $0.2284 \pm 0.0035$ & $0.64^{+0.0082}_{-0.0022}$   \\ \hline 
   % \hline
		\end{tabular}
		\caption{Values of coefficients, normalization constants, and model parameters \cite{Maji:2016yqo}. }
		\label{tab_par} % is used to refer this table in the text
	\end{table}
\section{TMD Correlator and Parameterization}\label{sec_corpar}
In the light-front formalism for SIDIS, the un-integrated quark-quark correlator at equal light-front time $z^+=0$ can be defined as \cite{Goeke:2005hb}
\begin{eqnarray}
\Phi_{[\Lambda^{N_i}\Lambda^{N_f}]}^{\nu [\Gamma]}(x,\textbf{p}_\perp)&=&\frac{1}{2}\int \frac{dz^- d^2z_T}{2(2\pi)^3} e^{ip.z} \langle P; \Lambda^{N_f}|\bar{\psi}^\nu (0)\Gamma \mathcal{W}_{[0,z]} \psi^\nu (z) |P;\Lambda^{N_i}\rangle\Bigg|_{z^+=0}. \label{TMDcor}
\end{eqnarray}
We choose the light-cone gauge $A^+=0$ and the frame is taken where the transverse momentum of the proton is $ P\equiv (P^+,\frac{M^2}{P^+},\textbf{0}_\perp)$. The momentum of the virtual photon is $q\equiv (x_B P^+, \frac{Q^2}{x_BP^+},\textbf{0})$, where $x_B= \frac{Q^2}{2P.q}$ is the Bjorken variable and $Q^2 = -q^2$. The helicity of proton in the initial and final state is denoted by $\Lambda^{N_i}$ and $\Lambda^{N_f}$ respectively. In our work, the value of Wilson line $(\mathcal{W}_{[0,z]})$ is taken to be $1$. By substituting the scalar and vector diquark Fock-states from Eq.~(\ref{fockSD}) and Eq.~(\ref{fockVD}) in the TMD correlator given in Eq.~(\ref{TMDcor}) via the use of proton state Eq.~(\ref{PS_state}), one can express the correlator for the scalar and vector diquark parts in the form of overlaps of the LFWFs provided in Table {\ref{tab_LFWF}} as
\begin{eqnarray} 
	\Phi_{[\Lambda^{N_i}\Lambda^{N_f}]}^{ [\Gamma](S)}(x,\textbf{p}_\perp)&=&\frac{C_{S}^{2}}{16\pi^3} \sum_{\lambda^{q_i}} \sum_{\lambda^{q_f}} \psi^{\Lambda^{N_f}\dagger}_{\lambda^{q_f}}(x,\bfp)\psi^{\Lambda^{N_i}}_{\lambda^{q_i}}(x,\bfp) \nonumber\\  &&\frac{u^{\dagger}_{\lambda^{q_f}}(x P^{+},\bfp)\gamma^{0} \Gamma u_{\lambda^{q_i}}(x P^{+},\bfp)}{2 x P^{+}}\,, \label{cors} \\
	% 
%  \end{eqnarray}
% \begin{eqnarray}
 % 
	\Phi_{[\Lambda^{N_i}\Lambda^{N_f}]}^{ [\Gamma](A)}(x,\textbf{p}_\perp)&=&\frac{C_{A}^{2}}{16\pi^3} \sum_{\lambda^{q_i}} \sum_{\lambda^{q_f}} \sum_{\lambda^{D}} \psi^{\Lambda^{N_f}\dagger}_{\lambda^{q_f} \lambda^D}(x,\bfp)\psi^{\Lambda^{N_i}}_{\lambda^{q_i}\lambda^D}(x,\bfp) \nonumber\\  &&\frac{u^{\dagger}_{\lambda^{q_f}}(x P^{+},\bfp)\gamma^{0} \Gamma u_{\lambda^{q_i}}(x P^{+},\bfp)}{2 x P^{+}}\,, \label{corv} 
\end{eqnarray} 
where, $C_A=C_V, C_{VV}$ for $u$ and $d$ quarks sequentially. $u^{\dagger}_{\lambda^{q_f}}(x P^{+},\bfp)\gamma^{0} \Gamma u_{\lambda^{q_i}}(x P^{+},\bfp)$ represents the spinor product corresponding to the Dirac matrices. Explicit Dirac spinors are given in Ref.
\cite{Harindranath:1996hq,Brodsky:1997de}. Here, $\lambda^{q_i}$ and $\lambda^{q_f}$ symbolize the quark helicity in the initial and the final state consequently. For the vector diquark, additional summation over diquark helicity $\lambda^{D}$ exists.
\par
There are $32$ TMDs in total considering twist-$2$, twist-$3$, and twist-$4$ cases, with their respective contributions of $8$, $16$, and $8$, respectively. Of these, $20$ TMDs are T-even, and the remaining $12$ are T-odd. The TMDs projection in the form of correlator Eq. (\ref{TMDcor}) in accordance with Ref. \cite{Meissner:2009ww} are represented as follows
\begin{eqnarray}
	\Phi_{[\Lambda^{N_i}\Lambda^{N_f}]}^{[\gamma^+]}
	&=& \frac{1}{2M} \, \bar{u}(P, \Lambda^{N_F}) \, \bigg[
	{\color{blue}f_1^{\nu}\left(x, \textbf{p}_{\perp}^2\right)}
	- \frac{i\sigma^{i+} \textbf{p}_{\perp}^i}{P^+} \, {\color{red}f_{1T}^{\perp \nu}\left(x, \textbf{p}_{\perp}^2\right)}
	\bigg] \, u(P, \Lambda^{N_i})
	\,, \label{par1} \\
	\Phi_{[\Lambda^{N_i}\Lambda^{N_f}]}^{[\gamma^+\gamma_5]}
	&=& \frac{1}{2M} \, \bar{u}(P, \Lambda^{N_F}) \, \bigg[
	\frac{i\sigma^{i+}\gamma_5 \textbf{p}_{\perp}^i}{P^+} \,{\color{blue}g_{1T}^{\nu}\left(x, \textbf{p}_{\perp}^2\right)}
	+ i\sigma^{+-}\gamma_5 \, {\color{blue}g_{1L}^{\nu}\left(x, \textbf{p}_{\perp}^2\right)}
	\bigg] \, u(P, \Lambda^{N_i})
	\,, \label{par2}
%	\nonumber\\
	\\
	\Phi_{[\Lambda^{N_i}\Lambda^{N_f}]}^{[i\sigma^{j+}\gamma_5]}
	&=& \frac{1}{2M} \, \bar{u}(P, \Lambda^{N_F}) \, \bigg[
	\frac{i\varepsilon_T^{ij} \textbf{p}_{\perp}^i}{M} \, {\color{red}h_1^{\perp \nu}\left(x, \textbf{p}_{\perp}^2\right)}
	+ \frac{M \, i\sigma^{j+}\gamma_5}{P^+} \, {\color{blue}h_{1T}^{\nu}\left(x, \textbf{p}_{\perp}^2\right)} \nonumber\\
	&&+ \frac{\textbf{p}_{\perp}^j \, i\sigma^{\rho+}\gamma_5 \textbf{p}_{\perp}^{\rho}}{M \, P^+} \, {\color{blue}h_{1T}^{\perp \nu}\left(x, \textbf{p}_{\perp}^2\right)} 	
	+ \frac{\textbf{p}_{\perp}^j \, i\sigma^{+-}\gamma_5}{M} \, {\color{blue}h_{1L}^{\perp \nu}\left(x, \textbf{p}_{\perp}^2\right)}
	\bigg] \, u(P, \Lambda^{N_i})
	\,, \label{par3}\\
	\Phi_{[\Lambda^{N_i}\Lambda^{N_f}]}^{[1]}
	&=& \frac{1}{2P^+} \, \bar{u}(P, \Lambda^{N_F}) \, \bigg[
	{\color{blue}e^{\nu}\left(x, \textbf{p}_{\perp}^2\right)}
	- \frac{i\sigma^{i+} \textbf{p}_{\perp}^i}{P^+} {\color{red}e_{T}^{\perp\nu}\left(x, \textbf{p}_{\perp}^2\right)}
	\bigg] \,{u}(P, \Lambda^{N_i})
	\,,
	%		\nonumber
	%		\\*
	\label{e:gtmd_4}\\
	\Phi_{[\Lambda^{N_i}\Lambda^{N_f}]}^{[\gamma_5]}
	&=& \frac{1}{2P^+} \, \bar{u}(P, \Lambda^{N_F}) \, \bigg[
	- \frac{i\sigma^{i+}\gamma_5 \textbf{p}_{\perp}^i}{P^+} {\color{red}e_{T}^{\nu}\left(x, \textbf{p}_{\perp}^2\right)}
	- i\sigma^{+-}\gamma_5 {\color{red}e_{L}^{\nu}\left(x, \textbf{p}_{\perp}^2\right)}
	\bigg] \, {u}(P, \Lambda^{N_i})
	\,,
%	\nonumber\\*
	\label{e:gtmd_5}\\	
	\Phi_{[\Lambda^{N_i}\Lambda^{N_f}]}^{[\gamma^j]}
	&=& \frac{1}{2P^+} \, \bar{u}(P, \Lambda^{N_F}) \, \bigg[
	\frac{\textbf{p}_{\perp}^j}{M} {\color{blue}f^{\perp \nu}\left(x, \textbf{p}_{\perp}^2\right)}
	+ \frac{M \, i\sigma^{j+}}{P^+} {\color{red}f_{T}^{' \nu}\left(x, \textbf{p}_{\perp}^2\right)} \nonumber\\*
	&&+ \frac{\textbf{p}_{\perp}^j \, i\sigma^{k+} \textbf{p}_{\perp}^k}{M \, P^+}  {\color{red}f_{T}^{\perp \nu}\left(x, \textbf{p}_{\perp}^2\right)} + \frac{i\sigma^{ij} \textbf{p}_{\perp}^i}{M} {\color{red}f_{L}^{\perp\nu}\left(x, \textbf{p}_{\perp}^2\right)}
	\bigg] \,{u}(P, \Lambda^{N_i})
	\,,
			\nonumber
	\\
%  \end{eqnarray}
% \begin{eqnarray}
	\Phi_{[\Lambda^{N_i}\Lambda^{N_f}]}^{[\gamma^j\gamma_5]}
	&=& \frac{1}{2P^+} \, \bar{u}(P, \Lambda^{N_F}) \, \bigg[
	\frac{i\varepsilon_T^{ij} \textbf{p}_{\perp}^i}{M} {\color{red}g^{\perp\nu}\left(x, \textbf{p}_{\perp}^2\right)}
	+ \frac{M \, i\sigma^{j+}\gamma_5}{P^+} {\color{blue}g_{T}^{\prime\nu}\left(x, \textbf{p}_{\perp}^2\right)} \nonumber\\*
	&&+ \frac{\textbf{p}_{\perp}^j \, i\sigma^{k+}\gamma_5 \textbf{p}_{\perp}^k}{M \, P^+} {\color{blue}g_{T}^{\perp\nu}\left(x, \textbf{p}_{\perp}^2\right)} + \frac{\textbf{p}_{\perp}^j \, i\sigma^{+-}\gamma_5}{M} {\color{blue}g_{L}^{\perp\nu}\left(x, \textbf{p}_{\perp}^2\right)}
	\bigg] \,{u}(P, \Lambda^{N_i})
	\,, 
	%		\\*
	\label{e:gtmd_7}\\
	\Phi_{[\Lambda^{N_i}\Lambda^{N_f}]}^{[i\sigma^{ij}\gamma_5]}
	&=& - \frac{i\varepsilon_T^{ij}}{2P^+} \, \bar{u}(P, \Lambda^{N_F}) \, \bigg[-
	{\color{red}h^{\nu}\left(x, \textbf{p}_{\perp}^2\right)}
	+ \frac{i\sigma^{k+} \textbf{p}_{\perp}^k}{P^+} {\color{blue}h_{T}^{\perp\nu}\left(x, \textbf{p}_{\perp}^2\right)}
	\bigg] \,{u}(P, \Lambda^{N_i})
	\,, 
	\label{e:gtmd_8}\\
	\Phi_{[\Lambda^{N_i}\Lambda^{N_f}]}^{[i\sigma^{+-}\gamma_5]}
	&=& \frac{1}{2P^+} \, \bar{u}(P, \Lambda^{N_F}) \, \bigg[
	\frac{i\sigma^{i+}\gamma_5 \textbf{p}_{\perp}^i}{P^+} {\color{blue}h_{T}^{\nu}\left(x, \textbf{p}_{\perp}^2\right)} 
	%		\nonumber\\*& & 
	+ i\sigma^{+-}\gamma_5 {\color{blue}h_{L}^{\nu}\left(x, \textbf{p}_{\perp}^2\right)}
	\bigg] \,{u}(P, \Lambda^{N_i})
	\,,
	\label{e:gtmd_9}\\
	%	\end{eqnarray}
%%
% \begin{equation}
	\Phi_{[\Lambda^{N_i}\Lambda^{N_f}]}^{[\gamma^-]}
	&=& \frac{M}{2(P^+)^2} \, \bar{u}(P, \Lambda^{N_F}) \, \bigg[
	{\color{blue}f_3^{\nu}\left(x, \textbf{p}_{\perp}^2\right)}
	- \frac{i\sigma^{i+} \textbf{p}_{\perp}^i}{P^+} \, {\color{red}f_{3T}^{\perp \nu}\left(x, \textbf{p}_{\perp}^2\right)}
	\bigg] \, u(P, \Lambda^{N_i})
	\,, \label{par10} 
 \\
	\Phi_{[\Lambda^{N_i}\Lambda^{N_f}]}^{[\gamma^-\gamma_5]}
	&=& \frac{M}{2(P^+)^2} \, \bar{u}(P, \Lambda^{N_F}) \, \bigg[
	\frac{i\sigma^{i+}\gamma_5 \textbf{p}_{\perp}^i}{P^+} \,{\color{blue}g_{3T}^{\nu}\left(x, \textbf{p}_{\perp}^2\right)}
	+ i\sigma^{+-}\gamma_5 \, {\color{blue}g_{3L}^{\nu}\left(x, \textbf{p}_{\perp}^2\right)}
	\bigg] \, u(P, \Lambda^{N_i})
	\,, \label{par11}\nonumber\\
	% \\
	%	
  \end{eqnarray}
\begin{eqnarray}	\Phi_{[\Lambda^{N_i}\Lambda^{N_f}]}^{[i\sigma^{j-}\gamma_5]}
	&=& \frac{M}{2(P^+)^2} \, \bar{u}(P, \Lambda^{N_F}) \, \bigg[
	\frac{i\varepsilon_T^{ij} \textbf{p}_{\perp}^i}{M} \, {\color{red}h_3^{\perp \nu}\left(x, \textbf{p}_{\perp}^2\right)}
	+ \frac{M \, i\sigma^{j+}\gamma_5}{P^+} \, {\color{blue}h_{3T}^{\nu}\left(x, \textbf{p}_{\perp}^2\right)} \nonumber\\
	&&+ \frac{\textbf{p}_{\perp}^j \, i\sigma^{k+}\gamma_5 \textbf{p}_{\perp}^k}{M \, P^+} \, {\color{blue}h_{3T}^{\perp \nu}\left(x, \textbf{p}_{\perp}^2\right)} 	
	+ \frac{\textbf{p}_{\perp}^j \, i\sigma^{+-}\gamma_5}{M} \, {\color{blue}h_{3L}^{\perp \nu}\left(x, \textbf{p}_{\perp}^2\right)}
	\bigg] \, u(P, \Lambda^{N_i})
	\,. \label{par12}
\end{eqnarray}
	\par
	The color code for T-even and T-odd TMDs is {\textcolor{blue}{blue}} and {\textcolor{red}{red}} respectively. In above equations, $\bfp^2=|\vec{p}_\perp|^2$ and the transverse directions are indicated by the indices $i$, $j$, and $\rho$. The conventional notation is adopted for $\sigma^{i j}=\frac{i}{2}\left[\gamma^{i}, \gamma^{j}\right]$. We have used the definition $\varepsilon_{\perp}^{i j}=\varepsilon^{-+i j}$, where $\varepsilon_\perp^{12}=-\varepsilon_\perp^{21}=1$ and is zero when $i$ and $j$ are identical. Again, we would like to emphasize that the study presented here is concerned only with T-even TMDs at Wandzura-Wilczek approximation \cite{Wandzura:1977qf}.
	\par
 On solving the parameterization from Eqs. {\eqref{par1}}-{\eqref{par12}} for the different possibilities of the proton helicities in the initial and final states, we have obtained the parameterization Table \ref{tab_param}. In this Table, the first column refers to the different possibilities of the Dirac matrix structures appearing at various levels of twist. Columns $2$ to $5$ refer to the different proton helicity combinations of the quark-quark TMD correlator. The first combination refers to the unpolarized TMDs, the second refers to longitudinally polarized TMDs, whereas the third and fourth combination denotes the transversely polarized TMDs.
\begin{table}[h]
	\centering % used for centering table
	\begin{tabular}{|c|c|c|c|c|}
		\hline
%		\hline
		$\Gamma$~~&~~$  \Phi_{++}^{[\Gamma]}+\Phi_{--}^{[\Gamma]}  $~~&~~$  \Phi_{++}^{[\Gamma]}-\Phi_{--}^{[\Gamma]} $~~&~~$  \Phi_{-+}^{[\Gamma]}+\Phi_{+-}^{[\Gamma]} $~~&~~$ \Phi_{-+}^{[\Gamma]}-\Phi_{+-}^{[\Gamma]}  $\\
		\hline
%		\hline
		$\gamma^+$~~&~~$ 2  \color{blue}{f_1}   $~~&~~$ 0 $~~&~~$ \frac{-2i \textbf{p}_{y}}{M} \color{red}{f_{1T}^\perp} $~~&~~$ \frac{2 \textbf{p}_{x}}{M} \color{red}{f_{1T}^\perp} $\\
		\hline
		$\gamma^+\gamma_5$~~&~~$ 0   $~~&~~$ 4  \color{blue}{g_{1L}} $~~&~~$ \frac{2 \textbf{p}_{x}}{M} \color{blue}{g_{1T}} $~~&~~$ \frac{-2i \textbf{p}_{y}}{M} \color{blue}{g_{1T}} $\\
		\hline
		$i\sigma^{1+}\gamma_5$~~&~~$  \frac{-2i \textbf{p}_{y}}{M} \color{red}{h_{1}^\perp}  $~~&~~$  \frac{4 \textbf{p}_{x}}{M} \color{blue}{h_{1L}^\perp} $~~&~~$ 2 \big[  {\color{blue}{h_{1T}}} + \frac{\textbf{p}_{x}^2}{M^2}  {\color{blue}{h_{1T}^{\perp}}} \big] $~~&~~$ \frac{-2i \textbf{p}_{x} \textbf{p}_{y}}{M^2} \color{blue}{h_{1T}^\perp} $\\
		\hline
		$i\sigma^{2+}\gamma_5$~~&~~$  \frac{2i \textbf{p}_{x}}{M} \color{red}{h_{1}^\perp}  $~~&~~$ \frac{4 \textbf{p}_{y}}{M} \color{blue}{h_{1L}^\perp} $~~&~~$ \frac{2 \textbf{p}_{x} \textbf{p}_{y}}{M^2} \color{blue}{h_{1T}^\perp} $~~&~~$ -2i \big[  {\color{blue}{h_{1T}}} + \frac{\textbf{p}_{y}^2}{M^2}  {\color{blue}{h_{1T}^{\perp}}} \big] $\\
		\hline
		$1$~~&~~$ \frac{2 M}{P^+}  \color{blue}{e}   $~~&~~$ 0 $~~&~~$ \frac{-2i \textbf{p}_{y}}{P^+} \color{red}{e_{T}^\perp} $~~&~~$ \frac{2 \textbf{p}_{x} }{P^+} \color{red}{e_{T}^\perp} $\\
		\hline
		$\gamma_5$~~&~~$ 0   $~~&~~$ \frac{-4 M}{P^+}  \color{red}{e_{L}} $~~&~~$ \frac{-2 \textbf{p}_{x}}{P^+} \color{red}{e_{T}} $~~&~~$ \frac{2i \textbf{p}_{y} }{P^+} \color{red}{e_{T}} $\\
		\hline
		$\gamma^1$~~&~~$   \frac{2 \textbf{p}_{x} }{P^+} \color{blue}{f^\perp}  $~~&~~$  \frac{2i \textbf{p}_{y}}{P^+} \color{red}{f_{L}^\perp} $~~&~~$ \frac{2i\textbf{p}_{x} \textbf{p}_{y}}{M P^+}  \color{red}{f_{T}^{\perp}} $~~&~~$ \frac{-2M}{P^+}\big[  {\color{red}{f_T'}} + \frac{\textbf{p}_{x}^2}{M^2}  {\color{red}{f_{T}^{\perp}}} \big] $\\
		\hline
		$\gamma^2$~~&~~$  \frac{2 \textbf{p}_{y} }{P^+} \color{blue}{f^\perp}   $~~&~~$\frac{2i \textbf{p}_{x}}{P^+} \color{red}{f_{L}^\perp} $~~&~~$ \frac{2iM}{P^+}\big[  {\color{red}{f_T'}} + \frac{\textbf{p}_{y}^2}{M^2}  {\color{red}{f_{T}^{\perp}}} \big]$~~&~~$ \frac{-2i\textbf{p}_{x} \textbf{p}_{y}}{M P^+}  \color{red}{f_{T}^{\perp}} $\\
		\hline
		$\gamma^1\gamma_5$~~&~~$ \frac{-2i \textbf{p}_{y} }{P^+} \color{red}{g^\perp}   $~~&~~$\frac{4 \textbf{p}_{x}}{P^+} \color{blue}{g_{L}^\perp} $~~&~~$ \frac{2M}{P^+}\big[  {\color{blue}{g_T'}} + \frac{\textbf{p}_{x}^2}{M^2}  {\color{blue}{g_{T}^{\perp}}} \big]$~~&~~$ \frac{-2i\textbf{p}_{x} \textbf{p}_{y}}{M P^+}  \color{blue}{g_{T}^{\perp}} $\\
		\hline
		$\gamma^2\gamma_5$~~&~~$  \frac{2i \textbf{p}_{x} }{P^+} \color{red}{g^\perp}  $~~&~~$  \frac{4 \textbf{p}_{y}}{P^+} \color{blue}{g_{L}^\perp} $~~&~~$ \frac{2\textbf{p}_{x} \textbf{p}_{y}}{M P^+}  \color{blue}{g_{T}^{\perp}} $~~&~~$ \frac{-2iM}{P^+}\big[  {\color{blue}{g_T'}} + \frac{\textbf{p}_{y}^2}{M^2}  {\color{blue}{g_{T}^{\perp}}} \big] $\\
		\hline
		$i\sigma^{12}\gamma_5$~~&~~$  \frac{2iM}{P^+} \color{red}{h}  $~~&~~$ 0 $~~&~~$ \frac{2 \textbf{p}_{y} }{P^+} \color{blue}{h_{T}^\perp} $~~&~~$ \frac{2i \textbf{p}_{x}}{P^+}  \color{blue}{h_{T}^{\perp}} $\\
		\hline
		$i\sigma^{21}\gamma_5$~~&~~$  \frac{-2iM}{P^+} \color{red}{h}  $~~&~~$ 0 $~~&~~$ \frac{-2 \textbf{p}_{y} }{P^+} \color{blue}{h_{T}^\perp} $~~&~~$ \frac{-2i \textbf{p}_{x}}{P^+}  \color{blue}{h_{T}^{\perp}} $\\
		\hline
		$i\sigma^{+-}\gamma_5$~~&~~$  0  $~~&~~$ \frac{4M}{(P)^2} \color{blue}{h_{L}} $~~&~~$ \frac{2 \textbf{p}_{x}}{P^+} \color{blue}{h_{T}} $~~&~~$ \frac{-2i \textbf{p}_{y}}{P^+} \color{blue}{h_{T}} $\\
		\hline
		$\gamma^-$~~&~~$ \frac{2 M^2}{(P^+)^2}  \color{blue}{f_3}   $~~&~~$ 0 $~~&~~$ \frac{-2i \textbf{p}_{y} M}{(P^+)^2} \color{red}{f_{3T}^\perp} $~~&~~$ \frac{2 \textbf{p}_{x} M}{(P^+)^2} \color{red}{f_{3T}^\perp} $\\
		\hline
		$\gamma^-\gamma_5$~~&~~$ 0   $~~&~~$ \frac{4 M^2}{(P^+)^2}  \color{blue}{g_{3L}} $~~&~~$ \frac{2 \textbf{p}_{x} M}{(P^+)^2} \color{blue}{g_{3T}} $~~&~~$ \frac{-2i \textbf{p}_{y} M}{(P^+)^2} \color{blue}{g_{3T}} $\\
		\hline
		$i\sigma^{1-}\gamma_5$~~&~~$  \frac{-2i \textbf{p}_{y} M}{(P^+)^2} \color{red}{h_{3}}^\perp  $~~&~~$  \frac{4 \textbf{p}_{x} M}{(P^+)^2} \color{blue}{h_{3L}^\perp} $~~&~~$ \frac{2M^2}{(P^+)^2} \big[  {\color{blue}{h_{3T}}} + \frac{\textbf{p}_{x}^2}{M^2}  {\color{blue}{h_{3T}^{\perp}}} \big] $~~&~~$ \frac{-2i \textbf{p}_{x} \textbf{p}_{y}}{(P^+)^2} \color{blue}{h_{3T}^\perp} $\\
		\hline
		$i\sigma^{2-}\gamma_5$~~&~~$  \frac{2i \textbf{p}_{x} M}{(P^+)^2} \color{red}{h_{3}^\perp}  $~~&~~$ \frac{4 \textbf{p}_{y} M}{(P^+)^2} \color{blue}{h_{3L}^\perp} $~~&~~$ \frac{2 \textbf{p}_{x} \textbf{p}_{y}}{(P^+)^2} \color{blue}{h_{3T}^\perp} $~~&~~$ \frac{-2iM^2}{(P^+)^2} \big[  {\color{blue}{h_{3T}}} + \frac{\textbf{p}_{y}^2}{M^2}  {\color{blue}{h_{3T}^{\perp}}} \big] $\\
%		\hline
		\hline
	\end{tabular}
	\caption{Solution of parameterization Eqs. {\eqref{par1}}-{\eqref{par12}} for the unpolarized, longitudinally polarized, and transversely polarized proton helicities and for quark helicities which are specified by various Dirac matrices $\Gamma$.}
	\label{tab_param} % is used to refer to this table in the text
\end{table}
From Table \ref{tab_param}, one can find that few transversely polarized T-even TMDs of the proton are clubbed naturally by the structure of their parameterization, which is represented by the following TMDs as \cite{Sharma:2023llg,Liu:2021ype,Sharma:2022caa}
\begin{eqnarray}
	h_{1}(x, {\bf p_\perp^2}) &=&  {h_{1T}(x, {\bf p_\perp^2})} + \frac{\bf p_\perp^2}{2 M^2}  {h_{1T}^{\perp}(x, {\bf p_\perp^2})} ,\\
	g_T(x, {\bf p_\perp^2}) &=& {g_T'(x, {\bf p_\perp^2})} + \frac{\bf p_\perp^2}{2M^2}  {g_{T}^{\perp}(x, {\bf p_\perp^2})} ,\\
	h_{3}(x, {\bf p_\perp^2}) &=&  {h_{3T}(x, {\bf p_\perp^2})} + \frac{\bf p_\perp^2}{2 M^2}  {h_{3T}^{\perp}(x, {\bf p_\perp^2})} .
\end{eqnarray}
However, the total number of independent TMDs remains the same. Therefore, any $2$ TMDs can be chosen from any of the equations mentioned above depending on one's convenience \cite{Sharma:2023llg,Liu:2021ype,Sharma:2022caa}. Before moving further, it should be clarified that we do not intend to claim the already published results of twist-$2$ TMDs \cite{Maji:2017bcz}.
Also, recently we have worked on twist-$3$ \cite{Sharma:2023wha,Sharma:2023fbb,Sharma:2022caa} and twist-$4$ \cite{Sharma:2023llg,Sharma:2023fbb,Sharma:2022ylk} TMDs.
This work's main idea is to express the other available ways of correlator combination and parameterization equations by which the TMD results can be obtained. We focus on deriving the TMD expressions and exploring the linear and quadratic relations among these TMDs and the TMD amplitude matrix. This objective is possible only when we establish the results of TMDs at each twist in every possible way.

\section{Overlap Expressions of TMDs}
	\label{sec_ovf}
The TMDs are usually represented as wave functions or overlap forms, representing the initial and final spin states of protons and quarks. We must substitute the scalar diquark Fock-state Eq. \eqref{fockSD} in the TMD correlator Eq. \eqref{TMDcor} via proton state Eq. \eqref{PS_state} 
with the proper polarization to obtain the overlap form of TMDs for the scalar diquark. Once a specific correlation has been chosen (for $\Gamma=
\gamma^{+}, \gamma^{+} \gamma_{5}, i \sigma^{j+} \gamma_{5}, 1, \gamma_{5}, \gamma^{j}, \gamma^{j} \gamma_{5}, i \sigma^{i j} \gamma_{5}, i \sigma^{+-} \gamma_{5}, \gamma^{-}, \gamma^{-} \gamma_{5},$ and $ i \sigma^{j-} \gamma_{5}$) using parameterization from Table \ref{tab_param}, we can compute a specific TMD by choosing the appropriate combination of proton polarization.
For example, TMD $e^{\nu}(x, {\bf p_\perp^2}), 
g_{L}^{\perp\nu}(x, {\bf p_\perp^2})$, and 
$h_{3T}^{\perp\nu}(x, {\bf p_\perp^2})$ can be written from Table \ref{tab_param} as 
\begin{eqnarray}
	e^{\nu}(x, {\bf p_\perp^2}) &=& \frac{P^+}{2 M} \Bigg(\Phi_{++}^{[1]}(x, {\bf p_\perp})+\Phi_{--}^{[1]}(x, {\bf p_\perp})\Bigg)\, \label{g1.1} ,\\ 
	g_{L}^{\perp\nu}(x, {\bf p_\perp^2}) &=& \frac{P^+}{4~ {\textbf{p}_{x}}}\Bigg(\Phi_{++}^{[\gamma_1\gamma_5]}(x, {\bf p_\perp})-\Phi_{--}^{[\gamma_1\gamma_5]}(x, {\bf p_\perp})\Bigg)\, \label{g1.2} ,\\  
	h_{3T}^{\perp\nu}(x, {\bf p_\perp^2}) &=& \frac{(P^+)^2}{2~ {\textbf{p}_{x}}{\textbf{p}_{y}}}\Bigg(\Phi_{-+}^{[i\sigma^{2-}\gamma_5]}(x, {\bf p_\perp})+\Phi_{+-}^{[i\sigma^{2-}\gamma_5]}(x, {\bf p_\perp})\Bigg)\, \,. \label{g1.3}
\end{eqnarray}
As one can quickly point out from Table \ref{tab_param}, that there is at least one way a TMD can be expressed. Also, in a few cases of Dirac matrix structure, there is a connection between correlators having different proton helicities. For example,
\begin{eqnarray}
	\Phi_{++}^{[\gamma^+]}(x, {\bf p_\perp})&=&\Phi_{--}^{[\gamma^+]}(x, {\bf p_\perp}),\\
	\Phi_{++}^{[\gamma^+\gamma_5]}(x, {\bf p_\perp})&=& -\Phi_{--}^{[\gamma^+\gamma_5]}(x, {\bf p_\perp}).
\end{eqnarray}
Taking into account all these aspects, the overlap form for scalar diquark can be written as
\begin{eqnarray}
	f_{1}^{\nu(S)}(x, {\bf p_\perp^2}) &=&  \frac{C_{S}^{2}}{16 \pi^3} \Bigg[|\psi ^{+\nu}_+(x,\textbf{p}_{\perp})|^2+|\psi ^{ + \nu}_-(x,\textbf{p}_{\perp})|^2\Bigg], \label{oef1s1}\\
	% \\
	 &=&  \frac{C_{S}^{2}}{16 \pi^3} \Bigg[|\psi ^{-\nu}_+(x,\textbf{p}_{\perp})|^2+|\psi ^{- \nu}_-(x,\textbf{p}_{\perp})|^2\Bigg], \label{oef1s2}\\
	%	
	%g1ls1
	%
	g_{1L}^{\nu(S)}(x, {\bf p_\perp^2}) &=& \frac{C_{S}^{2}}{32 \pi^3} \Bigg[|\psi ^{+\nu}_{+}(x,\textbf{p}_{\perp})|^2-|\psi ^{ + \nu}_{-}(x,\textbf{p}_{\perp})|^2 \Bigg], \label{oeg1lps1} \\
	&=& \frac{C_{S}^{2}}{32 \pi^3} \Bigg[|\psi ^{-\nu}_{+}(x,\textbf{p}_{\perp})|^2-|\psi ^{ - \nu}_{-}(x,\textbf{p}_{\perp})|^2 \Bigg], \label{oeg1lps2} \\
	%
	%
	%g1tv1
	%
	g_{1T}^{\nu(S)}(x, {\bf p_\perp^2}) &=&  \frac{C_{S}^{2} M}{32 \pi^3  {\textbf{p}_{x}}} \Bigg[
	\psi ^{+\nu \dagger}_{+}(x,\textbf{p}_{\perp})\psi^{- \nu}_{+}(x,\textbf{p}_{\perp}) - \psi ^{+\nu \dagger}_{-}(x,\textbf{p}_{\perp}) \psi ^{-\nu}_{-}(x,\textbf{p}_{\perp}) \nonumber\\
	&&+ \psi ^{-\nu \dagger}_{+}(x,\textbf{p}_{\perp}) \psi ^{+\nu}_{+}(x,\textbf{p}_{\perp})-\psi^{- \nu \dagger}_{-}(x,\textbf{p}_{\perp}) \psi ^{+\nu}_{-}(x,\textbf{p}_{\perp})\Bigg], \label{oeg1ts1} \\
	%
	%
	%g1tv2
	%
	% g_{1T}^{\nu(S)}(x, {\bf p_\perp^2}) 
 &=&  \frac{\iota C_{S}^{2} M}{32 \pi^3 {\textbf{p}_{y}} } \Bigg[
	\psi ^{+\nu \dagger}_{+}(x,\textbf{p}_{\perp})\psi^{- \nu}_{+}(x,\textbf{p}_{\perp}) - \psi ^{+\nu \dagger}_{-}(x,\textbf{p}_{\perp}) \psi ^{-\nu}_{-}(x,\textbf{p}_{\perp}) \nonumber\\
	&&- \psi ^{-\nu \dagger}_{+}(x,\textbf{p}_{\perp}) \psi ^{+\nu}_{+}(x,\textbf{p}_{\perp})+\psi^{- \nu \dagger}_{-}(x,\textbf{p}_{\perp}) \psi ^{+\nu}_{-}(x,\textbf{p}_{\perp})\Bigg], \label{oeg1ts2} \\
	%h1lps1
	h_{1L}^{\perp\nu(S)}(x, {\bf p_\perp^2}) &=&  \frac{C_{S}^{2} M}{32 \pi^3 {\textbf{p}_{x}}} \Bigg[\psi ^{+\nu \dagger}_{+}(x,\textbf{p}_{\perp})\psi^{+ \nu}_{-}(x,\textbf{p}_{\perp}) +  \psi ^{+\nu \dagger}_{-}(x,\textbf{p}_{\perp})\psi^{+ \nu}_{+}(x,\textbf{p}_{\perp})\Bigg],
\nonumber\\
	\label{oeh1lps1}
 % \\
	%\\
	%
   \end{eqnarray}
\begin{eqnarray}
	%h1lps2
	%
	% h_{1L}^{\perp\nu(S)}(x, {\bf p_\perp^2}) 
 &=& \frac{-C_{S}^{2} M}{32 \pi^3 {\textbf{p}_{x}}} \Bigg[\psi ^{-\nu \dagger}_{+}(x,\textbf{p}_{\perp})\psi^{- \nu}_{-}(x,\textbf{p}_{\perp}) 
	%\nonumber \\
	%&&
	+ \psi ^{-\nu \dagger}_{-}(x,\textbf{p}_{\perp})\psi^{- \nu}_{+}(x,\textbf{p}_{\perp})\Bigg],
	\nonumber\\ 
	\label{oeh1lps2}\\
	%\\
	%
	%
	%h1lps3
	%
	% h_{1L}^{\perp\nu(S)}(x, {\bf p_\perp^2}) 
 &=&  \frac{-\iota C_{S}^{2} M}{32 \pi^3 {\textbf{p}_{y}} } \Bigg[
	\psi ^{+\nu \dagger}_{+}(x,\textbf{p}_{\perp})\psi^{+ \nu}_{-}(x,\textbf{p}_{\perp})
	-  \psi ^{+\nu \dagger}_{-}(x,\textbf{p}_{\perp})\psi^{+ \nu}_{+}(x,\textbf{p}_{\perp})\Bigg],
	\nonumber\\ 
\label{oeh1lps3}\\
	%h1lpv4
	%
	% h_{1L}^{\perp\nu(S)}(x, {\bf p_\perp^2}) 
 &=&  \frac{ \iota C_{S}^{2} M}{32 \pi^3 {\textbf{p}_{y}}} \Bigg[
	\psi ^{-\nu \dagger}_{+}(x,\textbf{p}_{\perp})\psi^{- \nu}_{-}(x,\textbf{p}_{\perp}) - \psi ^{-\nu \dagger}_{-}(x,\textbf{p}_{\perp})\psi^{- \nu}_{+}(x,\textbf{p}_{\perp})\Bigg],
	\nonumber\\
	\label{oeh1lps4}
%	\\
 % \end{eqnarray}
 % \begin{eqnarray}
	%h1ts1
	%
	h_{1T}^{\nu(S)}(x, {\bf p_\perp^2})+\frac{{\textbf{p}_{x}}^2}{M^2}h_{1T}^{\perp\nu(S)}(x, {\bf p_\perp^2})
	&=&  \frac{C_{S}^{2}}{32 \pi^3} \Bigg[ \psi ^{+\nu \dagger}_{+}(x,\textbf{p}_{\perp})\psi^{- \nu}_{-}(x,\textbf{p}_{\perp}) + \psi ^{-\nu \dagger}_{+}(x,\textbf{p}_{\perp})\psi^{+ \nu}_{-}(x,\textbf{p}_{\perp}) \nonumber \\
	&&+  \psi ^{+\nu \dagger}_{-}(x,\textbf{p}_{\perp})\psi^{- \nu}_{+}(x,\textbf{p}_{\perp}) + \psi ^{-\nu \dagger}_{-}(x,\textbf{p}_{\perp})\psi^{+ \nu}_{+}(x,\textbf{p}_{\perp})\Bigg],
 % \nonumber\\
	\label{oeh1ts1}\\
 % \end{eqnarray}
 % \begin{eqnarray}
	%h1ts2
	%
	h_{1T}^{\nu(S)}(x, {\bf p_\perp^2})+\frac{{\textbf{p}_{y}}^2}{M^2}h_{1T}^{\perp\nu(S)}(x, {\bf p_\perp^2})
	&=&  \frac{C_{S}^{2}}{32 \pi^3} \Bigg[\psi ^{+\nu \dagger}_{+}(x,\textbf{p}_{\perp})\psi^{- \nu}_{-}(x,\textbf{p}_{\perp}) - \psi ^{-\nu \dagger}_{+}(x,\textbf{p}_{\perp})\psi^{+ \nu}_{-}(x,\textbf{p}_{\perp}) \nonumber \\
	&&-\psi ^{+\nu \dagger}_{-}(x,\textbf{p}_{\perp})\psi^{- \nu}_{+}(x,\textbf{p}_{\perp}) + \psi ^{-\nu \dagger}_{-}(x,\textbf{p}_{\perp})\psi^{+ \nu}_{+}(x,\textbf{p}_{\perp})\Bigg], 
 % \nonumber\\
	\label{oeh1ts2}	\\
% \end{eqnarray}
% \begin{eqnarray}
	%h1tps1
	h_{1T}^{\perp\nu(S)}(x, {\bf p_\perp^2}) &=&  \frac{\iota C_{S}^{2} M^2}{32 \pi^3 {\textbf{p}_{x}}{\textbf{p}_{y}}} \Bigg[ \psi ^{+\nu \dagger}_{+}(x,\textbf{p}_{\perp})\psi^{- \nu}_{-}(x,\textbf{p}_{\perp}) - \psi ^{-\nu \dagger}_{+}(x,\textbf{p}_{\perp})\psi^{+ \nu}_{-}(x,\textbf{p}_{\perp}) \nonumber \\
	&&+  \psi ^{+\nu \dagger}_{-}(x,\textbf{p}_{\perp})\psi^{- \nu}_{+}(x,\textbf{p}_{\perp}) - \psi ^{-\nu \dagger}_{-}(x,\textbf{p}_{\perp})\psi^{+ \nu}_{+}(x,\textbf{p}_{\perp})\Bigg],
	%\nonumber\\
	\label{oeh1tps1}\\
	%
	%
	%h1tps2
	%
	% h_{1T}^{\perp\nu(S)}(x, {\bf p_\perp^2}) 
 &=&  \frac{-\iota C_{S}^{2} M^2}{32 \pi^3 {\textbf{p}_{x}}{\textbf{p}_{y}}} \Bigg[ \psi ^{+\nu \dagger}_{+}(x,\textbf{p}_{\perp})\psi^{- \nu}_{-}(x,\textbf{p}_{\perp}) + \psi ^{-\nu \dagger}_{+}(x,\textbf{p}_{\perp})\psi^{+ \nu}_{-}(x,\textbf{p}_{\perp}) \nonumber \\
	&&-  \psi ^{+\nu \dagger}_{-}(x,\textbf{p}_{\perp})\psi^{- \nu}_{+}(x,\textbf{p}_{\perp}) - \psi ^{-\nu \dagger}_{-}(x,\textbf{p}_{\perp})\psi^{+ \nu}_{+}(x,\textbf{p}_{\perp})\Bigg],
	%\nonumber\\
	\label{oeh1tps2}\\
	%
% \end{eqnarray}
% \begin{eqnarray}
	%h1v
	%
	%h1tpv1
	%
	%
	%
	%h1v
	%
	h_{1}^{\nu(S)}(x, {\bf p_\perp^2}) &=&  \frac{C_{S}^{2}}{32 \pi^3} \Bigg[ \psi ^{+\nu \dagger}_{+}(x,\textbf{p}_{\perp})\psi^{- \nu}_{-}(x,\textbf{p}_{\perp})+ \psi ^{-\nu \dagger}_{-}(x,\textbf{p}_{\perp})\psi^{+ \nu}_{+}(x,\textbf{p}_{\perp})\Bigg],
	\nonumber\\
	\label{oeh1v}
	%\\
	%
% \end{eqnarray}	
% Twist-3 Scalar Overlap Form
% \begin{eqnarray}
	x \ e^{\nu(S)}(x,\textbf{p}_{\perp}^2)&=&\frac{C_{S}^{2}}{16\pi^3}\frac{m}{M}\bigg[|\psi ^{+\nu}_+(x,\textbf{p}_{\perp})|^2+|\psi ^{ + \nu}_-(x,\textbf{p}_{\perp})|^2\bigg],\label{es1}  \\
	%		
	%
	% x \ e^{\nu(S)}(x,\textbf{p}_{\perp}^2)
 &=&\frac{C_{S}^{2}}{16\pi^3}\frac{m}{M}\bigg[|\psi ^{-\nu}_+(x,\textbf{p}_{\perp})|^2+|\psi ^{ - \nu}_-(x,\textbf{p}_{\perp})|^2\bigg],\label{es2}  
 \\
	%
%    \end{eqnarray}
% \begin{eqnarray}
	% f perp
	{x}~f^{\perp\nu(S)}(x,\textbf{p}_{\perp}^2)&=&\frac{C_{S}^{2}}{16\pi^3}\bigg[|\psi ^{+\nu}_+(x,\textbf{p}_{\perp})|^2+|\psi ^{ + \nu}_-(x,\textbf{p}_{\perp})|^2\bigg], \\
	%		
	%
	% f perp
	% {x}~f^{\perp\nu(S)}(x,\textbf{p}_{\perp}^2)
 &=&\frac{C_{S}^{2}}{16\pi^3}\bigg[|\psi ^{-\nu}_+(x,\textbf{p}_{\perp})|^2+|\psi ^{ - \nu}_-(x,\textbf{p}_{\perp})|^2\bigg], 
 % \\
	%
   \end{eqnarray}
\begin{eqnarray}
	%	% gL perp 1
	{x}~ g_{L}^{\perp\nu(S)}(x, {\bf p_\perp^2}) &=&  \frac{C_{S}^{2}}{32\pi^3 {\textbf{p}_{x}}}\bigg({\textbf{p}_{x}}\bigg[|\psi ^{+\nu}_{+ }(x,\textbf{p}_{\perp})|^2-|\psi ^{ + \nu}_{- }(x,\textbf{p}_{\perp})|^2\bigg] \nonumber\\
	&& + m\bigg[\psi ^{+\nu \dagger}_{+ }(x,\textbf{p}_{\perp})\psi^{+ \nu}_{- }(x,\textbf{p}_{\perp}) + \psi ^{+\nu \dagger}_{- }(x,\textbf{p}_{\perp})\psi^{+ \nu}_{+ }(x,\textbf{p}_{\perp})\bigg]\bigg), \nonumber\\
	\label{glperp1v11} \\
	% {x}~ g_{L}^{\perp\nu(S)}(x, {\bf p_\perp^2}) 
 &=&  \frac{-C_{S}^{2}}{32\pi^3 {\textbf{p}_{x}}}\bigg({\textbf{p}_{x}}\bigg[|\psi ^{-\nu}_{+ }(x,\textbf{p}_{\perp})|^2-|\psi ^{ - \nu}_{- }(x,\textbf{p}_{\perp})|^2\bigg] \nonumber\\
	&& + m\bigg[\psi ^{-\nu \dagger}_{+ }(x,\textbf{p}_{\perp})\psi^{- \nu}_{- }(x,\textbf{p}_{\perp}) + \psi ^{-\nu \dagger}_{- }(x,\textbf{p}_{\perp})\psi^{- \nu}_{+ }(x,\textbf{p}_{\perp})\bigg]\bigg), \nonumber\\
	\label{glperp1s2} \\
	%
	%
	% gL perp 3
	% {x}~ g_{L}^{\perp\nu(S)}(x, {\bf p_\perp^2}) 
 &=&  \frac{C_{S}^{2}}{32\pi^3 {\textbf{p}_{y}}}\bigg({\textbf{p}_{y}}\bigg[|\psi ^{+\nu}_{+ }(x,\textbf{p}_{\perp})|^2-|\psi ^{ + \nu}_{- }(x,\textbf{p}_{\perp})|^2\bigg] \nonumber\\
	&& - \iota m\bigg[\psi ^{+\nu \dagger}_{+ }(x,\textbf{p}_{\perp})\psi^{+ \nu}_{- }(x,\textbf{p}_{\perp}) - \psi ^{+\nu \dagger}_{- }(x,\textbf{p}_{\perp})\psi^{+ \nu}_{+ }(x,\textbf{p}_{\perp})\bigg]\bigg), \nonumber\\
	\label{glperp1s3}
 \\
	%	
%   \end{eqnarray}
% \begin{eqnarray}
	%
	% {x}~ g_{L}^{\perp\nu(S)}(x, {\bf p_\perp^2}) 
 &=&  \frac{-C_{S}^{2}}{32\pi^3 {\textbf{p}_{y}}}\bigg({\textbf{p}_{y}}\bigg[|\psi ^{-\nu}_{+ }(x,\textbf{p}_{\perp})|^2-|\psi ^{ - \nu}_{- }(x,\textbf{p}_{\perp})|^2\bigg] \nonumber\\
	&& - \iota m\bigg[\psi ^{-\nu \dagger}_{+ }(x,\textbf{p}_{\perp})\psi^{- \nu}_{- }(x,\textbf{p}_{\perp}) - \psi ^{-\nu \dagger}_{- }(x,\textbf{p}_{\perp})\psi^{- \nu}_{+ }(x,\textbf{p}_{\perp})\bigg]\bigg), \nonumber\\
	\label{glperp1s4} \\
% \end{eqnarray}
% \begin{eqnarray}
	% gT 1
	{x}\bigg(g_{T}^{'\nu(S)}(x, {\bf p_\perp^2})+\frac{\textbf{p}_{x}^2}{M^2}g_{T}^{\perp (S)}(x, {\bf p_\perp^2})\bigg) &=&\frac{C_{S}^{2}}{32\pi^3 M}\bigg(\textbf{p}_{x}\bigg[\psi ^{+\nu \dagger}_{+ }(x,\textbf{p}_{\perp})\psi^{- \nu}_{+ }(x,\textbf{p}_{\perp}) \nonumber \\
	&&- \psi ^{+\nu \dagger}_{- }(x,\textbf{p}_{\perp})\psi^{- \nu}_{- }(x,\textbf{p}_{\perp})+ \psi ^{-\nu \dagger}_{+ }(x,\textbf{p}_{\perp})\psi^{+ \nu}_{+ }(x,\textbf{p}_{\perp})\nonumber \\
	&&- \psi ^{-\nu \dagger}_{- }(x,\textbf{p}_{\perp})\psi^{+ \nu}_{- }(x,\textbf{p}_{\perp})\bigg] 
	+m\bigg[\psi ^{+\nu \dagger}_{+ }(x,\textbf{p}_{\perp})\psi^{- \nu}_{- }(x,\textbf{p}_{\perp}) \nonumber \\
	&&+ \psi ^{+\nu \dagger}_{- }(x,\textbf{p}_{\perp})\psi^{- \nu}_{+ }(x,\textbf{p}_{\perp})+\psi ^{-\nu \dagger}_{+ }(x,\textbf{p}_{\perp})\psi^{+ \nu}_{- }(x,\textbf{p}_{\perp}) \nonumber \\
	&&+ \psi ^{-\nu \dagger}_{- }(x,\textbf{p}_{\perp})\psi^{+ \nu}_{+ }(x,\textbf{p}_{\perp}) \bigg] \bigg), \label{gt1s1}  \\
	%		
	%
	% gT 2
	{x}\bigg(g_{T}^{'\nu(S)}(x, {\bf p_\perp^2})+\frac{\textbf{p}_{y}^2}{M^2}g_{T}^{\perp (S)}(x, {\bf p_\perp^2})\bigg) &=&\frac{\iota C_{S}^{2}}{32\pi^3 M}\bigg(\textbf{p}_{y}\bigg[\psi ^{+\nu \dagger}_{+ }(x,\textbf{p}_{\perp})\psi^{- \nu}_{+ }(x,\textbf{p}_{\perp}) \nonumber \\
	&&- \psi ^{+\nu \dagger}_{- }(x,\textbf{p}_{\perp})\psi^{- \nu}_{- }(x,\textbf{p}_{\perp})- \psi ^{-\nu \dagger}_{+ }(x,\textbf{p}_{\perp})\psi^{+ \nu}_{+ }(x,\textbf{p}_{\perp})\nonumber \\
	&&+ \psi ^{-\nu \dagger}_{- }(x,\textbf{p}_{\perp})\psi^{+ \nu}_{- }(x,\textbf{p}_{\perp})\bigg] -\iota m\bigg[\psi ^{+\nu \dagger}_{+ }(x,\textbf{p}_{\perp})\psi^{- \nu}_{- }(x,\textbf{p}_{\perp}) \nonumber \\
	&&- \psi ^{+\nu \dagger}_{- }(x,\textbf{p}_{\perp})\psi^{- \nu}_{+ }(x,\textbf{p}_{\perp})-\psi ^{-\nu \dagger}_{+ }(x,\textbf{p}_{\perp})\psi^{+ \nu}_{- }(x,\textbf{p}_{\perp}) \nonumber \\
	&&+ \psi ^{-\nu \dagger}_{- }(x,\textbf{p}_{\perp})\psi^{+ \nu}_{+ }(x,\textbf{p}_{\perp}) \bigg] \bigg), \label{gt1s2} 
 % \\
	%
	% 
\end{eqnarray}
\begin{eqnarray}
	% gT perp 1
	%		
	{x}~g_{T}^{\perp\nu (S)}(x, {\bf p_\perp^2}) &=& \frac{C_{S}^{2} M}{32\pi^3 {\bf p_x} \ {\bf p_y}}\bigg[\textbf{p}_{y}\bigg[\psi ^{+\nu \dagger}_{+}(x,\textbf{p}_{\perp})\psi^{- \nu}_{+}(x,\textbf{p}_{\perp}) - \psi ^{+\nu \dagger}_{-}(x,\textbf{p}_{\perp})\psi^{- \nu}_{-}(x,\textbf{p}_{\perp})\nonumber \\
	&&+ \psi^{-\nu \dagger}_{+}(x,\textbf{p}_{\perp})\psi^{+ \nu}_{+}(x,\textbf{p}_{\perp})- \psi^{-\nu \dagger}_{-}(x,\textbf{p}_{\perp})\psi^{+ \nu}_{-}(x,\textbf{p}_{\perp})\bigg]  \nonumber \\
	&& -\iota m\bigg[\psi ^{+\nu \dagger}_{+}(x,\textbf{p}_{\perp})\psi^{- \nu}_{-}(x,\textbf{p}_{\perp}) - \psi ^{+\nu \dagger}_{-}(x,\textbf{p}_{\perp})\psi^{- \nu}_{+}(x,\textbf{p}_{\perp}) \nonumber \\
	&&+\psi ^{-\nu \dagger}_{+}(x,\textbf{p}_{\perp})\psi^{+ \nu}_{-}(x,\textbf{p}_{\perp}) -\psi ^{-\nu \dagger}_{-}(x,\textbf{p}_{\perp})\psi^{+ \nu}_{+}(x,\textbf{p}_{\perp}) \bigg] \bigg],  \label{gtperp1s}
	\\
	%		
	%
	% gT perp2
	%				
	% {x}~g_{T}^{\perp\nu (S)}(x, {\bf p_\perp^2}) 
 &=& \frac{\iota C_{S}^{2} M}{32\pi^3 {\bf p_x} \ {\bf p_y}}\bigg[\textbf{p}_{x}\bigg[\psi ^{+\nu \dagger}_{+}(x,\textbf{p}_{\perp})\psi^{- \nu}_{+}(x,\textbf{p}_{\perp}) - \psi ^{+\nu \dagger}_{-}(x,\textbf{p}_{\perp})\psi^{- \nu}_{-}(x,\textbf{p}_{\perp})\nonumber \\
	&&- \psi^{-\nu \dagger}_{+}(x,\textbf{p}_{\perp})\psi^{+ \nu}_{+}(x,\textbf{p}_{\perp})+ \psi^{-\nu \dagger}_{-}(x,\textbf{p}_{\perp})\psi^{+ \nu}_{-}(x,\textbf{p}_{\perp})\bigg]  \nonumber \\
	&&+ m\bigg[\psi ^{+\nu \dagger}_{+}(x,\textbf{p}_{\perp})\psi^{- \nu}_{-}(x,\textbf{p}_{\perp}) + \psi ^{+\nu \dagger}_{-}(x,\textbf{p}_{\perp})\psi^{- \nu}_{+}(x,\textbf{p}_{\perp}) \nonumber \\
	&&-\psi ^{-\nu \dagger}_{+}(x,\textbf{p}_{\perp})\psi^{+ \nu}_{-}(x,\textbf{p}_{\perp})-\psi ^{-\nu \dagger}_{-}(x,\textbf{p}_{\perp})\psi^{+ \nu}_{+}(x,\textbf{p}_{\perp}) \bigg] \bigg],  \label{gtperp2s}	\\
	{x}~g_{T}^{\nu(S)}(x, {\bf p_\perp^2})&=&\frac{C_{S}^{2}}{64\pi^3 M}\bigg(\textbf{p}_{x}\bigg[\psi ^{+\nu \dagger}_{+}(x,\textbf{p}_{\perp})\psi^{- \nu}_{+}(x,\textbf{p}_{\perp}) - \psi ^{+\nu \dagger}_{-}(x,\textbf{p}_{\perp})\psi^{- \nu}_{-}(x,\textbf{p}_{\perp})\nonumber \\
	&&+ \psi ^{-\nu \dagger}_{+}(x,\textbf{p}_{\perp})\psi^{+ \nu}_{+}(x,\textbf{p}_{\perp})- \psi ^{-\nu \dagger}_{-}(x,\textbf{p}_{\perp})\psi^{+ \nu}_{-}(x,\textbf{p}_{\perp})\bigg]  \nonumber \\
	&&+ \iota\textbf{p}_{y}\bigg[\psi ^{+\nu \dagger}_{+}(x,\textbf{p}_{\perp})\psi^{- \nu}_{+}(x,\textbf{p}_{\perp}) - \psi ^{+\nu \dagger}_{-}(x,\textbf{p}_{\perp})\psi^{- \nu}_{-}(x,\textbf{p}_{\perp})\nonumber \\
	&&- \psi ^{-\nu \dagger}_{+}(x,\textbf{p}_{\perp})\psi^{+ \nu}_{+}(x,\textbf{p}_{\perp})+ \psi ^{-\nu \dagger}_{-}(x,\textbf{p}_{\perp})\psi^{+ \nu}_{-}(x,\textbf{p}_{\perp})\bigg]  \nonumber \\
	&& +2m\bigg[\psi ^{+\nu \dagger}_{+}(x,\textbf{p}_{\perp})\psi^{- \nu}_{-}(x,\textbf{p}_{\perp}) + \psi ^{-\nu \dagger}_{-}(x,\textbf{p}_{\perp})\psi^{+ \nu}_{+}(x,\textbf{p}_{\perp}) \bigg] \bigg), \label{gt1v31}  
  \\
	%		
%  \end{eqnarray}
% \begin{eqnarray}
 % 
	{x}~h_{T}^{\perp\nu(S)}(x, {\bf p_\perp^2}) &=&\frac{\iota C_{S}^{2}}{32\pi^3 {\textbf{p}_{y}}} \bigg(({\textbf{p}_{x}}-{\iota}{\textbf{p}_{y}})\bigg[\psi ^{+\nu \dagger}_{+}(x,\textbf{p}_{\perp})\psi^{- \nu}_{-}(x,\textbf{p}_{\perp}) + \psi ^{-\nu \dagger}_{+}(x,\textbf{p}_{\perp})\psi^{+ \nu}_{-}(x,\textbf{p}_{\perp})\bigg] \nonumber\\
	&&  -({\textbf{p}_{x}}+{\iota}{\textbf{p}_{y}})\bigg[\psi ^{+\nu \dagger}_{-}(x,\textbf{p}_{\perp})\psi^{- \nu}_{+}(x,\textbf{p}_{\perp}) + \psi ^{-\nu \dagger}_{-}(x,\textbf{p}_{\perp})\psi^{+ \nu}_{+}(x,\textbf{p}_{\perp})\bigg]\bigg),
	% \nonumber\\
	\label{htperp1s1}\\
	%	
	%		
	% hT perp 2
	% {x}~h_{T}^{\perp\nu(S)}(x, {\bf p_\perp^2}) 
 &=&\frac{C_{S}^{2}}{32\pi^3 {\textbf{p}_{x}}} \bigg(({\textbf{p}_{x}}-{\iota} {\textbf{p}_{y}})\bigg[\psi ^{+\nu \dagger}_{+}(x,\textbf{p}_{\perp})\psi^{- \nu}_{-}(x,\textbf{p}_{\perp}) - \psi ^{-\nu \dagger}_{+}(x,\textbf{p}_{\perp})\psi^{+ \nu}_{-}(x,\textbf{p}_{\perp})\bigg] \nonumber\\
	&&  -({\textbf{p}_{x}}+{\iota}{\textbf{p}_{y}})\bigg[\psi ^{+\nu \dagger}_{-}(x,\textbf{p}_{\perp})\psi^{- \nu}_{+}(x,\textbf{p}_{\perp}) - \psi ^{-\nu \dagger}_{-}(x,\textbf{p}_{\perp})\psi^{+ \nu}_{+}(x,\textbf{p}_{\perp})\bigg]\bigg),
	% \nonumber\\
	\label{htperp1s2}
 % \\
 \end{eqnarray}
 \begin{eqnarray}
	{x}~h_{L}^{\nu(S)}(x, {\bf p_\perp^2}) &=&\frac{C_{S}^{2}}{16\pi^3  M}\bigg(m\bigg[|\psi ^{+\nu}_{+}(x,\textbf{p}_{\perp})|^2-|\psi ^{ + \nu}_{-}(x,\textbf{p}_{\perp})|^2\bigg] \nonumber \\
	&& - ({\textbf{p}_{x}}-\iota {\textbf{p}_{y}})\bigg[\psi ^{+\nu \dagger}_{+}(x,\textbf{p}_{\perp})\psi^{+ \nu}_{-}(x,\textbf{p}_{\perp})\bigg]- ({\textbf{p}_{x}}+\iota {\textbf{p}_{y}})\nonumber\\
	&& \bigg[\psi ^{+\nu \dagger}_{-}(x,\textbf{p}_{\perp})\psi^{+ \nu}_{+}(x,\textbf{p}_{\perp})\bigg]\bigg), \label{hls1}
  \\
	%
	%
%  \end{eqnarray}
% \begin{eqnarray}
	%
	% {x}~h_{L}^{\nu(S)}(x, {\bf p_\perp^2})
 &=&\frac{-C_{S}^{2}}{16\pi^3  M}\bigg(m\bigg[|\psi ^{-\nu}_{+}(x,\textbf{p}_{\perp})|^2-|\psi ^{ -\nu}_{-}(x,\textbf{p}_{\perp})|^2\bigg] \nonumber \\
	&& - ({\textbf{p}_{x}}-\iota {\textbf{p}_{y}})\bigg[\psi ^{-\nu \dagger}_{+}(x,\textbf{p}_{\perp})\psi^{- \nu}_{-}(x,\textbf{p}_{\perp})\bigg]- ({\textbf{p}_{x}}+\iota {\textbf{p}_{y}})\nonumber\\
	&& \bigg[\psi ^{-\nu \dagger}_{-}(x,\textbf{p}_{\perp})\psi^{- \nu}_{+}(x,\textbf{p}_{\perp})\bigg]\bigg), \label{hls2}\\
	{x}~h_{T}^{\nu(S)}(x, {\bf p_\perp^2}) 
 &=&\frac{C_{S}^{2}}{16\pi^3 {\textbf{p}_{x}}}\bigg(m\bigg[\psi ^{+\nu \dagger}_{+}(x,\textbf{p}_{\perp})\psi^{- \nu}_{+}(x,\textbf{p}_{\perp}) - \psi ^{+\nu \dagger}_{-}(x,\textbf{p}_{\perp})\psi^{- \nu}_{-}(x,\textbf{p}_{\perp})\nonumber \\
	&& +\psi ^{-\nu \dagger}_{+}(x,\textbf{p}_{\perp})\psi^{+ \nu}_{+}(x,\textbf{p}_{\perp}) -\psi ^{-\nu \dagger}_{-}(x,\textbf{p}_{\perp})\psi^{+ \nu}_{-}(x,\textbf{p}_{\perp}) \bigg] \nonumber\\
	&& -({\textbf{p}_{x}}-\iota {\textbf{p}_{y}})\bigg[\psi ^{+\nu \dagger}_{+}(x,\textbf{p}_{\perp})\psi^{- \nu}_{-}(x,\textbf{p}_{\perp}) + \psi ^{-\nu \dagger}_{+}(x,\textbf{p}_{\perp})\psi^{+ \nu}_{-}(x,\textbf{p}_{\perp})\bigg] \nonumber \\
	&& -({\textbf{p}_{x}}+\iota {\textbf{p}_{y}})\bigg[\psi ^{+\nu \dagger}_{-}(x,\textbf{p}_{\perp})\psi^{- \nu}_{+}(x,\textbf{p}_{\perp}) + \psi ^{-\nu \dagger}_{-}(x,\textbf{p}_{\perp})\psi^{+ \nu}_{+}(x,\textbf{p}_{\perp})\bigg]\bigg),
	% \nonumber\\
	\label{ht1s1} \\
	%		\\
	%
	% hT 2
	% {x}~h_{T}^{\nu(S)}(x, {\bf p_\perp^2}) 
 &=&\frac{\iota C_{S}^{2}}{16\pi^3 {\textbf{p}_{y}}}\bigg(m\bigg[\psi ^{+\nu \dagger}_{+}(x,\textbf{p}_{\perp})\psi^{- \nu}_{+}(x,\textbf{p}_{\perp}) - \psi ^{+\nu \dagger}_{-}(x,\textbf{p}_{\perp})\psi^{- \nu}_{-}(x,\textbf{p}_{\perp})\nonumber \\
	&& -\psi ^{-\nu \dagger}_{+}(x,\textbf{p}_{\perp})\psi^{+ \nu}_{+}(x,\textbf{p}_{\perp}) +\psi ^{-\nu \dagger}_{-}(x,\textbf{p}_{\perp})\psi^{+ \nu}_{-}(x,\textbf{p}_{\perp}) \bigg] \nonumber\\
	&& -({\textbf{p}_{x}}-\iota {\textbf{p}_{y}})\bigg[\psi ^{+\nu \dagger}_{+}(x,\textbf{p}_{\perp})\psi^{- \nu}_{-}(x,\textbf{p}_{\perp}) - \psi ^{-\nu \dagger}_{+}(x,\textbf{p}_{\perp})\psi^{+ \nu}_{-}(x,\textbf{p}_{\perp})\bigg] \nonumber \\
	&& -({\textbf{p}_{x}}+\iota {\textbf{p}_{y}})\bigg[\psi ^{+\nu \dagger}_{-}(x,\textbf{p}_{\perp})\psi^{- \nu}_{+}(x,\textbf{p}_{\perp}) - \psi ^{-\nu \dagger}_{-}(x,\textbf{p}_{\perp})\psi^{+ \nu}_{+}(x,\textbf{p}_{\perp})\bigg]\bigg),
	% \nonumber\\
	\label{ht1s2} \\
	x^2 f_{3}^{\nu(S)}(x, {\bf p_\perp^2}) &=&  \frac{C_{S}^{2}}{16 \pi^3} \bigg(\frac{{\bf p_\perp^2}+m^2}{M^2}\bigg)\Bigg[|\psi ^{+\nu}_+(x,\textbf{p}_{\perp})|^2+|\psi ^{ + \nu}_-(x,\textbf{p}_{\perp})|^2\Bigg], \label{oef3s1}\\
% 
	%
	% x^2 f_{3}^{\nu(S)}(x, {\bf p_\perp^2}) 
 &=&  \frac{C_{S}^{2}}{16 \pi^3} \bigg(\frac{{\bf p_\perp^2}+m^2}{M^2}\bigg)\Bigg[|\psi ^{-\nu}_+(x,\textbf{p}_{\perp})|^2+|\psi ^{- \nu}_-(x,\textbf{p}_{\perp})|^2\Bigg], \label{oef3s2}
 \\
	%g3ls
%  \end{eqnarray}
% \begin{eqnarray} 
	%\ee
	%\be
	%
	%
	%g3lv
	%
	x^2 g_{3L}^{\nu(S)}(x, {\bf p_\perp^2}) &=& \frac{C_{S}^{2}}{32 \pi^3 M^2} \Bigg[\big({{\bf p_\perp^2}-m^2}\big)\bigg[|\psi ^{+\nu}_{+}(x,\textbf{p}_{\perp})|^2-|\psi ^{ + \nu}_{-}(x,\textbf{p}_{\perp})|^2\bigg] \nonumber\\
	&& + 2m ({\textbf{p}_{x}}-\iota {\textbf{p}_{y}})\bigg[\psi ^{+\nu \dagger}_{+}(x,\textbf{p}_{\perp})\psi^{+ \nu}_{-}(x,\textbf{p}_{\perp})\bigg] \nonumber \\
	&&+ 2m ({\textbf{p}_{x}}+\iota {\textbf{p}_{y}}) \bigg[\psi ^{+\nu \dagger}_{-}(x,\textbf{p}_{\perp})\psi^{+ \nu}_{+}(x,\textbf{p}_{\perp})\bigg]\Bigg], \label{oeg3lps1}
  % \\
	%
    \end{eqnarray}
 \begin{eqnarray}
	%
	% x^2 g_{3L}^{\nu(S)}(x, {\bf p_\perp^2}) 
 &=& \frac{-C_{S}^{2}}{32 \pi^3 M^2} \Bigg[\big({{\bf p_\perp^2}-m^2}\big)\bigg[|\psi ^{-\nu}_{+}(x,\textbf{p}_{\perp})|^2-|\psi ^{- \nu}_{-}(x,\textbf{p}_{\perp})|^2\bigg] \nonumber\\
	&& + 2m ({\textbf{p}_{x}}-\iota {\textbf{p}_{y}})\bigg[\psi ^{-\nu \dagger}_{+}(x,\textbf{p}_{\perp})\psi^{- \nu}_{-}(x,\textbf{p}_{\perp})\bigg] \nonumber \\
	&&+ 2m ({\textbf{p}_{x}}+\iota {\textbf{p}_{y}}) \bigg[\psi ^{-\nu \dagger}_{-}(x,\textbf{p}_{\perp})\psi^{- \nu}_{+}(x,\textbf{p}_{\perp})\bigg]\Bigg], \label{oeg3lps2} \\
	%
	%g3ts1
	%
	x^2 g_{3T}^{\nu(S)}(x, {\bf p_\perp^2}) &=&  \frac{C_{S}^{2}}{32 \pi^3 M  {\textbf{p}_{x}}} \Bigg[\big({{\bf p_\perp^2}-m^2}\big)\bigg[
	\psi ^{+\nu \dagger}_{+}(x,\textbf{p}_{\perp})\psi^{- \nu}_{+}(x,\textbf{p}_{\perp}) - \psi ^{+\nu \dagger}_{-}(x,\textbf{p}_{\perp}) \psi ^{-\nu}_{-}(x,\textbf{p}_{\perp}) \nonumber\\
	&&+ \psi ^{-\nu \dagger}_{+}(x,\textbf{p}_{\perp}) \psi ^{+\nu}_{+}(x,\textbf{p}_{\perp})-\psi^{- \nu \dagger}_{-}(x,\textbf{p}_{\perp}) \psi ^{+\nu}_{-}(x,\textbf{p}_{\perp})\bigg] \nonumber\\
	&& + 2m ({\textbf{p}_{x}}-\iota {\textbf{p}_{y}})\bigg[\psi ^{+\nu \dagger}_{+}(x,\textbf{p}_{\perp})\psi^{- \nu}_{-}(x,\textbf{p}_{\perp}) + \psi ^{-\nu \dagger}_{+}(x,\textbf{p}_{\perp})\psi^{+ \nu}_{-}(x,\textbf{p}_{\perp})\bigg] \nonumber \\
	&&+ 2m ({\textbf{p}_{x}}+\iota {\textbf{p}_{y}}) \bigg[\psi ^{+\nu \dagger}_{-}(x,\textbf{p}_{\perp})\psi^{- \nu}_{+}(x,\textbf{p}_{\perp}) + \psi ^{-\nu \dagger}_{-}(x,\textbf{p}_{\perp})\psi^{+ \nu}_{+}(x,\textbf{p}_{\perp})\bigg]\Bigg], \label{oeg3ts1} \nonumber\\
	\\
	%
	%
	%g3ts2
	%
	% x^2 g_{3T}^{\nu(S)}(x, {\bf p_\perp^2}) 
 &=&  \frac{\iota C_{S}^{2}}{32 \pi^3 M {\textbf{p}_{y}} } \Bigg[\big({{\bf p_\perp^2}-m^2}\big)\bigg[
	\psi ^{+\nu \dagger}_{+}(x,\textbf{p}_{\perp})\psi^{- \nu}_{+}(x,\textbf{p}_{\perp}) - \psi ^{+\nu \dagger}_{-}(x,\textbf{p}_{\perp}) \psi ^{-\nu}_{-}(x,\textbf{p}_{\perp}) \nonumber\\
	&&- \psi ^{-\nu \dagger}_{+}(x,\textbf{p}_{\perp}) \psi ^{+\nu}_{+}(x,\textbf{p}_{\perp})+\psi^{- \nu \dagger}_{-}(x,\textbf{p}_{\perp}) \psi ^{+\nu}_{-}(x,\textbf{p}_{\perp})\bigg] \nonumber\\
	&& + 2m ({\textbf{p}_{x}}-\iota {\textbf{p}_{y}})\bigg[\psi ^{+\nu \dagger}_{+}(x,\textbf{p}_{\perp})\psi^{- \nu}_{-}(x,\textbf{p}_{\perp}) - \psi ^{-\nu \dagger}_{+}(x,\textbf{p}_{\perp})\psi^{+ \nu}_{-}(x,\textbf{p}_{\perp})\bigg] \nonumber \\
	&&+ 2m ({\textbf{p}_{x}}+\iota {\textbf{p}_{y}}) \bigg[\psi ^{+\nu \dagger}_{-}(x,\textbf{p}_{\perp})\psi^{- \nu}_{+}(x,\textbf{p}_{\perp}) - \psi ^{-\nu \dagger}_{-}(x,\textbf{p}_{\perp})\psi^{+ \nu}_{+}(x,\textbf{p}_{\perp})\bigg]\Bigg], \label{oeg3ts2}\nonumber \\
	%\\
	%
	%h3lps1
	%
	x^2~h_{3L}^{\perp\nu(S)}(x, {\bf p_\perp^2}) &=&  \frac{C_{S}^{2}}{32 \pi^3 {\textbf{p}_{x}} M} \Bigg[2m{\textbf{p}_{x}} \bigg[|\psi ^{+\nu}_{+}(x,\textbf{p}_{\perp})|^2-|\psi ^{ + \nu}_{-}(x,\textbf{p}_{\perp})|^2\bigg] \nonumber\\
	&& + \bigg( m^2 -({\textbf{p}_{x}}-\iota {\textbf{p}_{y}})^2 \bigg)
	\bigg[\psi ^{+\nu \dagger}_{+}(x,\textbf{p}_{\perp})\psi^{+ \nu}_{-}(x,\textbf{p}_{\perp})\bigg] \nonumber \\
	&&+ \bigg( m^2 -({\textbf{p}_{x}}+\iota {\textbf{p}_{y}})^2 \bigg) \bigg[\psi ^{+\nu \dagger}_{-}(x,\textbf{p}_{\perp})\psi^{+ \nu}_{+}(x,\textbf{p}_{\perp})\bigg]\Bigg], \label{oeh3lps1}\\
%  \\
%     \end{eqnarray}
% \begin{eqnarray}
	%
	%h3lps2
	%
	% x^2~h_{3L}^{\perp\nu(S)}(x, {\bf p_\perp^2}) 
 &=& \frac{-C_{S}^{2}}{32 \pi^3 {\textbf{p}_{x}} M} \Bigg[2m{\textbf{p}_{x}} \bigg[|\psi ^{-\nu}_{+}(x,\textbf{p}_{\perp})|^2-|\psi ^{ - \nu}_{-}(x,\textbf{p}_{\perp})|^2\bigg] \nonumber\\
	&& + \bigg( m^2 -({\textbf{p}_{x}}-\iota {\textbf{p}_{y}})^2 \bigg)
	\bigg[\psi ^{-\nu \dagger}_{+}(x,\textbf{p}_{\perp})\psi^{- \nu}_{-}(x,\textbf{p}_{\perp})\bigg] \nonumber \\
	&&+ \bigg( m^2 -({\textbf{p}_{x}}+\iota {\textbf{p}_{y}})^2 \bigg) \bigg[\psi ^{-\nu \dagger}_{-}(x,\textbf{p}_{\perp})\psi^{- \nu}_{+}(x,\textbf{p}_{\perp})\bigg]\Bigg], \label{oeh3lps2}
 % \\
	%\\
	%
  \end{eqnarray}
\begin{eqnarray}
	%
	%h3lps3
	%
	% x^2~h_{3L}^{\perp\nu(S)}(x, {\bf p_\perp^2}) 
 &=&  \frac{C_{S}^{2}}{32 \pi^3 {\textbf{p}_{y}} M} \Bigg[2m{\textbf{p}_{y}} \bigg[|\psi ^{+\nu}_{+}(x,\textbf{p}_{\perp})|^2-|\psi ^{ + \nu}_{-}(x,\textbf{p}_{\perp})|^2\bigg] \nonumber\\
	&&-\iota \bigg( m^2 +({\textbf{p}_{x}}-\iota {\textbf{p}_{y}})^2 \bigg)
	\bigg[\psi ^{+\nu \dagger}_{+}(x,\textbf{p}_{\perp})\psi^{+ \nu}_{-}(x,\textbf{p}_{\perp})\bigg] \nonumber \\
	&&+\iota \bigg( m^2 +({\textbf{p}_{x}}+\iota {\textbf{p}_{y}})^2 \bigg) \bigg[\psi ^{+\nu \dagger}_{-}(x,\textbf{p}_{\perp})\psi^{+ \nu}_{+}(x,\textbf{p}_{\perp})\bigg]\Bigg],  \label{oeh3lps3}
 \nonumber \\
 \\
	%\\
	%
	%h3lps4
	%
	% x^2~h_{3L}^{\perp\nu(S)}(x, {\bf p_\perp^2}) 
 &=&  \frac{-C_{S}^{2}}{32 \pi^3 {\textbf{p}_{y}} M} \Bigg[2m{\textbf{p}_{y}} \bigg[|\psi ^{-\nu}_{+}(x,\textbf{p}_{\perp})|^2-|\psi ^{ - \nu}_{-}(x,\textbf{p}_{\perp})|^2\bigg] \nonumber\\
	&& -\iota \bigg( m^2 +({\textbf{p}_{x}}-\iota {\textbf{p}_{y}})^2 \bigg)
	\bigg[\psi ^{-\nu \dagger}_{+}(x,\textbf{p}_{\perp})\psi^{- \nu}_{-}(x,\textbf{p}_{\perp})\bigg] \nonumber \\
	&&+\iota \bigg( m^2 +({\textbf{p}_{x}}+\iota {\textbf{p}_{y}})^2 \bigg) \bigg[\psi ^{-\nu \dagger}_{-}(x,\textbf{p}_{\perp})\psi^{- \nu}_{+}(x,\textbf{p}_{\perp})\bigg]\Bigg], \label{oeh3lps4} \nonumber \\
	\\
	%\\
	%
% \end{eqnarray}
% \begin{eqnarray}
	%h3ts1
	%
	x^2 \bigg[h_{3T}^{\nu(S)}(x, {\bf p_\perp^2})+\frac{{\textbf{p}_{x}}^2}{M^2}h_{3T}^{\perp\nu(S)}(x, {\bf p_\perp^2})\bigg]
	&=&  \frac{C_{S}^{2}}{32 \pi^3 M^2} \Bigg[2m{\textbf{p}_{x}} \bigg[\psi ^{+\nu \dagger}_{+}(x,\textbf{p}_{\perp})\psi^{- \nu}_{+}(x,\textbf{p}_{\perp}) \nonumber\\
	&&- \psi ^{+\nu \dagger}_{-}(x,\textbf{p}_{\perp}) \psi ^{-\nu}_{-}(x,\textbf{p}_{\perp}) + \psi ^{-\nu \dagger}_{+}(x,\textbf{p}_{\perp}) \psi ^{+\nu}_{+}(x,\textbf{p}_{\perp})\nonumber\\
	&&-\psi^{- \nu \dagger}_{-}(x,\textbf{p}_{\perp}) \psi ^{+\nu}_{-}(x,\textbf{p}_{\perp})\bigg]  + \bigg( m^2 -({\textbf{p}_{x}}-\iota {\textbf{p}_{y}})^2 \bigg)\nonumber\\
	&&	\bigg[\psi ^{+\nu \dagger}_{+}(x,\textbf{p}_{\perp})\psi^{- \nu}_{-}(x,\textbf{p}_{\perp}) + \psi ^{-\nu \dagger}_{+}(x,\textbf{p}_{\perp})\psi^{+ \nu}_{-}(x,\textbf{p}_{\perp})\bigg] \nonumber \\
	&&+ \bigg( m^2 -({\textbf{p}_{x}}+\iota {\textbf{p}_{y}})^2 \bigg) \bigg[\psi ^{+\nu \dagger}_{-}(x,\textbf{p}_{\perp})\psi^{- \nu}_{+}(x,\textbf{p}_{\perp}) \nonumber\\
	&&+ \psi ^{-\nu \dagger}_{-}(x,\textbf{p}_{\perp})\psi^{+ \nu}_{+}(x,\textbf{p}_{\perp})\bigg]\Bigg], 
	%	\nonumber\\
	\label{oeh3tv1}\\
	%
	%
	%
	%h3ts2
	%
	x^2 \bigg[h_{3T}^{\nu(S)}(x, {\bf p_\perp^2})+\frac{{\textbf{p}_{y}}^2}{M^2}h_{3T}^{\perp\nu(S)}(x, {\bf p_\perp^2})\bigg]
	&=&  \frac{\iota C_{S}^{2}}{32 \pi^3 M^2} \Bigg[2m{\textbf{p}_{y}} \bigg[\psi ^{+\nu \dagger}_{+}(x,\textbf{p}_{\perp})\psi^{- \nu}_{+}(x,\textbf{p}_{\perp}) \nonumber\\
	&&- \psi ^{+\nu \dagger}_{-}(x,\textbf{p}_{\perp}) \psi ^{-\nu}_{-}(x,\textbf{p}_{\perp})- \psi ^{-\nu \dagger}_{+}(x,\textbf{p}_{\perp}) \psi ^{+\nu}_{+}(x,\textbf{p}_{\perp})\nonumber\\
	&&+\psi^{- \nu \dagger}_{-}(x,\textbf{p}_{\perp}) \psi ^{+\nu}_{-}(x,\textbf{p}_{\perp})\bigg]  -\iota \bigg( m^2 +({\textbf{p}_{x}}-\iota {\textbf{p}_{y}})^2 \bigg)\nonumber\\
	&&
	\bigg[\psi ^{+\nu \dagger}_{+}(x,\textbf{p}_{\perp})\psi^{- \nu}_{-}(x,\textbf{p}_{\perp}) - \psi ^{-\nu \dagger}_{+}(x,\textbf{p}_{\perp})\psi^{+ \nu}_{-}(x,\textbf{p}_{\perp})\bigg] \nonumber \\
	&&+\iota \bigg( m^2 +({\textbf{p}_{x}}+\iota {\textbf{p}_{y}})^2 \bigg) \bigg[\psi ^{+\nu \dagger}_{-}(x,\textbf{p}_{\perp})\psi^{- \nu}_{+}(x,\textbf{p}_{\perp}) \nonumber\\
	&&- \psi ^{-\nu \dagger}_{-}(x,\textbf{p}_{\perp})\psi^{+ \nu}_{+}(x,\textbf{p}_{\perp})\bigg]\Bigg],
	%	 \nonumber\\
	\label{oeh3ts2}
	% \\
	%
 \end{eqnarray}
 \begin{eqnarray}
	%h3tps1
	%
	x^2~h_{3T}^{\perp\nu(S)}(x, {\bf p_\perp^2}) &=&  \frac{C_{S}^{2}}{32 \pi^3 {\textbf{p}_{x}}{\textbf{p}_{y}}} \Bigg[2m{\textbf{p}_{y}} \bigg[\psi ^{+\nu \dagger}_{+}(x,\textbf{p}_{\perp})\psi^{- \nu}_{+}(x,\textbf{p}_{\perp}) - \psi ^{+\nu \dagger}_{-}(x,\textbf{p}_{\perp}) \psi ^{-\nu}_{-}(x,\textbf{p}_{\perp}) \nonumber\\
	&&+ \psi ^{-\nu \dagger}_{+}(x,\textbf{p}_{\perp}) \psi ^{+\nu}_{+}(x,\textbf{p}_{\perp})-\psi^{-\nu \dagger}_{-}(x,\textbf{p}_{\perp}) \psi ^{+\nu}_{-}(x,\textbf{p}_{\perp})\bigg]  -\iota \bigg( m^2 \nonumber\\
	&&+({\textbf{p}_{x}}-\iota {\textbf{p}_{y}})^2 \bigg)
	\bigg[\psi ^{+\nu \dagger}_{+}(x,\textbf{p}_{\perp})\psi^{- \nu}_{-}(x,\textbf{p}_{\perp}) + \psi ^{-\nu \dagger}_{+}(x,\textbf{p}_{\perp})\psi^{+ \nu}_{-}(x,\textbf{p}_{\perp})\bigg] +\iota \bigg( m^2 \nonumber\\
	&&+({\textbf{p}_{x}}+\iota {\textbf{p}_{y}})^2 \bigg) \bigg[\psi ^{+\nu \dagger}_{-}(x,\textbf{p}_{\perp})\psi^{- \nu}_{+}(x,\textbf{p}_{\perp}) + \psi ^{-\nu \dagger}_{-}(x,\textbf{p}_{\perp})\psi^{+ \nu}_{+}(x,\textbf{p}_{\perp})\bigg]\Bigg],
 % \nonumber\\
	\label{oeh3tps1} \\
% 
%  \end{eqnarray}
% \begin{eqnarray}
	%
	%
	%h3tps2
	%
	% x^2~h_{3T}^{\perp\nu(S)}(x, {\bf p_\perp^2}) 
 &=&  \frac{\iota C_{S}^{2}}{32 \pi^3 {\textbf{p}_{x}}{\textbf{p}_{y}}} \Bigg[2m{\textbf{p}_{x}} \bigg[\psi ^{+\nu \dagger}_{+}(x,\textbf{p}_{\perp})\psi^{- \nu}_{+}(x,\textbf{p}_{\perp}) - \psi ^{+\nu \dagger}_{-}(x,\textbf{p}_{\perp}) \psi ^{-\nu}_{-}(x,\textbf{p}_{\perp}) \nonumber\\
	&&- \psi ^{-\nu \dagger}_{+}(x,\textbf{p}_{\perp}) \psi ^{+\nu}_{+}(x,\textbf{p}_{\perp})+\psi^{- \nu \dagger}_{-}(x,\textbf{p}_{\perp}) \psi ^{+\nu}_{-}(x,\textbf{p}_{\perp})\bigg] + \bigg( m^2 \nonumber\\
	&&-({\textbf{p}_{x}}-\iota {\textbf{p}_{y}})^2 \bigg)
	\bigg[\psi ^{+\nu \dagger}_{+}(x,\textbf{p}_{\perp})\psi^{- \nu}_{-}(x,\textbf{p}_{\perp}) - \psi ^{-\nu \dagger}_{+}(x,\textbf{p}_{\perp})\psi^{+ \nu}_{-}(x,\textbf{p}_{\perp})\bigg] + \bigg( m^2 \nonumber\\
	&&-({\textbf{p}_{x}}+\iota {\textbf{p}_{y}})^2 \bigg) \bigg[\psi ^{+\nu \dagger}_{-}(x,\textbf{p}_{\perp})\psi^{- \nu}_{+}(x,\textbf{p}_{\perp}) - \psi ^{-\nu \dagger}_{-}(x,\textbf{p}_{\perp})\psi^{+ \nu}_{+}(x,\textbf{p}_{\perp})\bigg]\Bigg],
 % \nonumber\\
	\label{oeh3tps2}\\
	%
	%
	%h3v
	%
	x^2 h_{3}^{\nu(S)}(x, {\bf p_\perp^2})
	&=&  \frac{C_{S}^{2}}{64 \pi^3 M^2} \Bigg[2m\bigg[{\textbf{p}_{x}} \Big[\psi ^{+\nu \dagger}_{+}(x,\textbf{p}_{\perp})\psi^{- \nu}_{+}(x,\textbf{p}_{\perp}) - \psi ^{+\nu \dagger}_{-}(x,\textbf{p}_{\perp}) \psi ^{-\nu}_{-}(x,\textbf{p}_{\perp}) \nonumber\\
	&&+ \psi ^{-\nu \dagger}_{+}(x,\textbf{p}_{\perp}) \psi ^{+\nu}_{+}(x,\textbf{p}_{\perp})-\psi^{- \nu \dagger}_{-}(x,\textbf{p}_{\perp}) \psi ^{+\nu}_{-}(x,\textbf{p}_{\perp})\Big] \nonumber\\
	&&+\iota{\textbf{p}_{y}} \Big[\psi ^{+\nu \dagger}_{+}(x,\textbf{p}_{\perp})\psi^{- \nu}_{+}(x,\textbf{p}_{\perp}) - \psi ^{+\nu \dagger}_{-}(x,\textbf{p}_{\perp}) \psi ^{-\nu}_{-}(x,\textbf{p}_{\perp}) \nonumber\\
	&&- \psi ^{-\nu \dagger}_{+}(x,\textbf{p}_{\perp}) \psi ^{+\nu}_{+}(x,\textbf{p}_{\perp})+\psi^{- \nu \dagger}_{-}(x,\textbf{p}_{\perp}) \psi ^{+\nu}_{-}(x,\textbf{p}_{\perp})\Big]\bigg] \nonumber\\
	&& + \bigg( m^2 -({\textbf{p}_{x}}-\iota {\textbf{p}_{y}})^2 \bigg)
	\bigg[\psi ^{+\nu \dagger}_{+}(x,\textbf{p}_{\perp})\psi^{- \nu}_{-}(x,\textbf{p}_{\perp}) + \psi ^{-\nu \dagger}_{+}(x,\textbf{p}_{\perp})\psi^{+ \nu}_{-}(x,\textbf{p}_{\perp})\bigg] \nonumber \\
	&&+ \bigg( m^2 -({\textbf{p}_{x}}+\iota {\textbf{p}_{y}})^2 \bigg) \bigg[\psi ^{+\nu \dagger}_{-}(x,\textbf{p}_{\perp})\psi^{- \nu}_{+}(x,\textbf{p}_{\perp}) + \psi ^{-\nu \dagger}_{-}(x,\textbf{p}_{\perp})\psi^{+ \nu}_{+}(x,\textbf{p}_{\perp})\bigg], \nonumber\\
	&& + \bigg( m^2 +({\textbf{p}_{x}}-\iota {\textbf{p}_{y}})^2 \bigg)
	\bigg[\psi ^{+\nu \dagger}_{+}(x,\textbf{p}_{\perp})\psi^{- \nu}_{-}(x,\textbf{p}_{\perp}) - \psi ^{-\nu \dagger}_{+}(x,\textbf{p}_{\perp})\psi^{+ \nu}_{-}(x,\textbf{p}_{\perp})\bigg] \nonumber \\
	&&- \bigg( m^2 +({\textbf{p}_{x}}+\iota {\textbf{p}_{y}})^2 \bigg) \bigg[\psi ^{+\nu \dagger}_{-}(x,\textbf{p}_{\perp})\psi^{- \nu}_{+}(x,\textbf{p}_{\perp}) - \psi ^{-\nu \dagger}_{-}(x,\textbf{p}_{\perp})\psi^{+ \nu}_{+}(x,\textbf{p}_{\perp})\bigg]\Bigg].
	\nonumber\\
	\label{oeh3v}
	%\\
	%
\end{eqnarray}
Now, to get the overlap form of TMDs for the vector diquark, we have to substitute the vector diquark Fock-state Eq. \eqref{fockVD} with suitable polarization in TMD correlator Eq. \eqref{TMDcor}. TMDs in terms of LFWFs for the vector diquark can be written as
\begin{eqnarray}
	f_{1}^{\nu(A)}(x, {\bf p_\perp^2}) &=& \sum_{\lambda^D} \frac{C_{A}^{2}}{16 \pi^3} \Bigg[|\psi ^{+\nu}_{+\lambda^D}(x,\textbf{p}_{\perp})|^2+|\psi ^{ + \nu}_{-\lambda^D}(x,\textbf{p}_{\perp})|^2\Bigg], \label{oef1v1}
 % \\
	%\\
	%
    \end{eqnarray}
\begin{eqnarray}
	% f_{1}^{\nu(A)}(x, {\bf p_\perp^2}) 
 &=& \sum_{\lambda^D} \frac{C_{A}^{2}}{16 \pi^3} \Bigg[|\psi ^{-\nu}_{+\lambda^D}(x,\textbf{p}_{\perp})|^2+|\psi ^{- \nu}_{-\lambda^D}(x,\textbf{p}_{\perp})|^2\Bigg], \label{oef1v2}\\
	%	
	%g1ls1
	%
	g_{1L}^{\nu(A)}(x, {\bf p_\perp^2}) &=& \sum_{\lambda^D}\frac{C_{A}^{2}}{32 \pi^3} \Bigg[|\psi ^{+\nu}_{+\lambda^D}(x,\textbf{p}_{\perp})|^2-|\psi ^{ + \nu}_{-\lambda^D}(x,\textbf{p}_{\perp})|^2 \Bigg], \label{oeg1lpv1} \\
	%
	%g1ls2
	%
	% g_{1L}^{\nu(A)}(x, {\bf p_\perp^2})
 &=&\sum_{\lambda^D} \frac{C_{A}^{2}}{32 \pi^3} \Bigg[|\psi ^{-\nu}_{+\lambda^D}(x,\textbf{p}_{\perp})|^2-|\psi ^{ - \nu}_{-\lambda^D}(x,\textbf{p}_{\perp})|^2 \Bigg], \label{oeg1lpv2} \\
	%
	%
	%g1tv1
	%
	g_{1T}^{\nu(A)}(x, {\bf p_\perp^2}) &=&  \frac{C_{A}^{2} M}{32 \pi^3  {\textbf{p}_{x}}} \Bigg[
	\psi ^{+\nu \dagger}_{+0}(x,\textbf{p}_{\perp})\psi^{- \nu}_{+0}(x,\textbf{p}_{\perp}) - \psi ^{+\nu \dagger}_{-0}(x,\textbf{p}_{\perp}) \psi ^{-\nu}_{-0}(x,\textbf{p}_{\perp}) \nonumber\\
	&&+ \psi ^{-\nu \dagger}_{+0}(x,\textbf{p}_{\perp}) \psi ^{+\nu}_{+0}(x,\textbf{p}_{\perp})-\psi^{- \nu \dagger}_{-0}(x,\textbf{p}_{\perp}) \psi ^{+\nu}_{-0}(x,\textbf{p}_{\perp})\Bigg], \label{oeg1tv1} \\
	%
	%
	%g1tv2
	%
	% g_{1T}^{\nu(A)}(x, {\bf p_\perp^2}) 
 &=&  \frac{\iota C_{A}^{2} M}{32 \pi^3 {\textbf{p}_{y}} } \Bigg[
	\psi ^{+\nu \dagger}_{+0}(x,\textbf{p}_{\perp})\psi^{- \nu}_{+0}(x,\textbf{p}_{\perp}) - \psi ^{+\nu \dagger}_{-0}(x,\textbf{p}_{\perp}) \psi ^{-\nu}_{-0}(x,\textbf{p}_{\perp}) \nonumber\\
	&&- \psi ^{-\nu \dagger}_{+0}(x,\textbf{p}_{\perp}) \psi ^{+\nu}_{+0}(x,\textbf{p}_{\perp})+\psi^{- \nu \dagger}_{-0}(x,\textbf{p}_{\perp}) \psi ^{+\nu}_{-0}(x,\textbf{p}_{\perp})\Bigg], \label{oeg1tv2} 
	\\
	%
% \end{eqnarray}
% \begin{eqnarray}
	%h1lpv1
	%
	h_{1L}^{\perp\nu(A)}(x, {\bf p_\perp^2}) &=&\sum_{\lambda^D}  \frac{C_{A}^{2} M}{32 \pi^3 {\textbf{p}_{x}}} \Bigg[\psi ^{+\nu \dagger}_{+\lambda^D}(x,\textbf{p}_{\perp})\psi^{+ \nu}_{-\lambda^D}(x,\textbf{p}_{\perp}) \nonumber \\
	&&+  \psi ^{+\nu \dagger}_{-\lambda^D}(x,\textbf{p}_{\perp})\psi^{+ \nu}_{+\lambda^D}(x,\textbf{p}_{\perp})\Bigg], \label{oeh1lpv1}\\
	%\\
	%
	%h1lpv2
	%
	% h_{1L}^{\perp\nu(A)}(x, {\bf p_\perp^2})
 &=&\sum_{\lambda^D}  \frac{-C_{A}^{2} M}{32 \pi^3 {\textbf{p}_{x}}} \Bigg[\psi ^{-\nu \dagger}_{+\lambda^D}(x,\textbf{p}_{\perp})\psi^{- \nu}_{-\lambda^D}(x,\textbf{p}_{\perp}) \nonumber \\
	&&+ \psi ^{-\nu \dagger}_{-\lambda^D}(x,\textbf{p}_{\perp})\psi^{- \nu}_{+\lambda^D}(x,\textbf{p}_{\perp})\Bigg], \label{oeh1lpv2}\\
	%\\
	%
	%
	%h1lpv3
	%
	% h_{1L}^{\perp\nu(A)}(x, {\bf p_\perp^2}) 
 &=&\sum_{\lambda^D}  \frac{-\iota C_{A}^{2} M}{32 \pi^3 {\textbf{p}_{y}} } \Bigg[
	\psi ^{+\nu \dagger}_{+\lambda^D}(x,\textbf{p}_{\perp})\psi^{+ \nu}_{-\lambda^D}(x,\textbf{p}_{\perp})\nonumber \\
	&&-  \psi ^{+\nu \dagger}_{-\lambda^D}(x,\textbf{p}_{\perp})\psi^{+ \nu}_{+\lambda^D}(x,\textbf{p}_{\perp})\Bigg], \label{oeh1lpv3}\\
	%\\
	%
	%h1lpv4
	%
	% h_{1L}^{\perp\nu(A)}(x, {\bf p_\perp^2}) 
 &=&\sum_{\lambda^D}  \frac{ \iota C_{A}^{2}  M}{32 \pi^3 {\textbf{p}_{y}}} \Bigg[
	\psi ^{-\nu \dagger}_{+\lambda^D}(x,\textbf{p}_{\perp})\psi^{- \nu}_{-\lambda^D}(x,\textbf{p}_{\perp}) \nonumber \\
	&&-  \psi ^{-\nu \dagger}_{-\lambda^D}(x,\textbf{p}_{\perp})\psi^{- \nu}_{+\lambda^D}(x,\textbf{p}_{\perp})\Bigg], \label{oeh1lpv4}
 % \\
	%\\
	%
	%
 \end{eqnarray}
\begin{eqnarray}
	%\ee
	%\be
	%	
	%h1tv1
	%
	h_{1T}^{\nu(A)}(x, {\bf p_\perp^2})+\frac{{\textbf{p}_{x}}^2}{M^2}h_{1T}^{\perp\nu(A)}(x, {\bf p_\perp^2})
	&=&  \frac{C_{A}^{2}}{32 \pi^3} \Bigg[ \psi ^{+\nu \dagger}_{+0}(x,\textbf{p}_{\perp})\psi^{- \nu}_{-0}(x,\textbf{p}_{\perp}) + \psi ^{-\nu \dagger}_{+0}(x,\textbf{p}_{\perp})\psi^{+ \nu}_{-0}(x,\textbf{p}_{\perp}) \nonumber \\
	&&+  \psi ^{+\nu \dagger}_{-0}(x,\textbf{p}_{\perp})\psi^{- \nu}_{+0}(x,\textbf{p}_{\perp}) + \psi ^{-\nu \dagger}_{-0}(x,\textbf{p}_{\perp})\psi^{+ \nu}_{+0}(x,\textbf{p}_{\perp})\Bigg], 
 % \nonumber\\
	\label{oeh1tv1}\\
	%
	%
	%
	%h1tv2
	%
	h_{1T}^{\nu(A)}(x, {\bf p_\perp^2})+\frac{{\textbf{p}_{y}}^2}{M^2}h_{1T}^{\perp\nu(A)}(x, {\bf p_\perp^2})
	&=&  \frac{C_{A}^{2}}{32 \pi^3} \Bigg[\psi ^{+\nu \dagger}_{+0}(x,\textbf{p}_{\perp})\psi^{- \nu}_{-0}(x,\textbf{p}_{\perp}) - \psi ^{-\nu \dagger}_{+0}(x,\textbf{p}_{\perp})\psi^{+ \nu}_{-0}(x,\textbf{p}_{\perp}) \nonumber \\
	&&-\psi ^{+\nu \dagger}_{-0}(x,\textbf{p}_{\perp})\psi^{- \nu}_{+0}(x,\textbf{p}_{\perp}) + \psi ^{-\nu \dagger}_{-0}(x,\textbf{p}_{\perp})\psi^{+ \nu}_{+0}(x,\textbf{p}_{\perp})\Bigg], 
 % \nonumber\\
	\label{oeh1tv2}
	\\
	%
% \end{eqnarray}
% \begin{eqnarray}
	%
	%h1tpv1
	%
	h_{1T}^{\perp\nu(A)}(x, {\bf p_\perp^2}) &=&  \frac{\iota C_{A}^{2} M^2}{32 \pi^3 {\textbf{p}_{x}}{\textbf{p}_{y}}} \Bigg[ \psi ^{+\nu \dagger}_{+0}(x,\textbf{p}_{\perp})\psi^{- \nu}_{-0}(x,\textbf{p}_{\perp}) - \psi ^{-\nu \dagger}_{+0}(x,\textbf{p}_{\perp})\psi^{+ \nu}_{-0}(x,\textbf{p}_{\perp}) \nonumber \\
	&&+  \psi ^{+\nu \dagger}_{-0}(x,\textbf{p}_{\perp})\psi^{- \nu}_{+0}(x,\textbf{p}_{\perp}) - \psi ^{-\nu \dagger}_{-0}(x,\textbf{p}_{\perp})\psi^{+ \nu}_{+0}(x,\textbf{p}_{\perp})\Bigg],
	%	\nonumber\\
	\label{oeh1tpv1}\\
	%
	%
	%h1tpv2
	%
	% h_{1T}^{\perp\nu(A)}(x, {\bf p_\perp^2}) 
 &=&  \frac{-\iota C_{A}^{2} M^2}{32 \pi^3 {\textbf{p}_{x}}{\textbf{p}_{y}}} \Bigg[ \psi ^{+\nu \dagger}_{+0}(x,\textbf{p}_{\perp})\psi^{- \nu}_{-0}(x,\textbf{p}_{\perp}) + \psi ^{-\nu \dagger}_{+0}(x,\textbf{p}_{\perp})\psi^{+ \nu}_{-0}(x,\textbf{p}_{\perp}) \nonumber \\
	&&-  \psi ^{+\nu \dagger}_{-0}(x,\textbf{p}_{\perp})\psi^{- \nu}_{+0}(x,\textbf{p}_{\perp}) - \psi ^{-\nu \dagger}_{-0}(x,\textbf{p}_{\perp})\psi^{+ \nu}_{+0}(x,\textbf{p}_{\perp})\Bigg],
	% \nonumber\\
	\label{oeh1tpv2}\\
	%	\\
	h_{1}^{\nu(A)}(x, {\bf p_\perp^2}) &=&  \frac{C_{A}^{2}}{32 \pi^3} \Bigg[ \psi ^{+\nu \dagger}_{+0}(x,\textbf{p}_{\perp})\psi^{- \nu}_{-0}(x,\textbf{p}_{\perp})+ \psi ^{-\nu \dagger}_{-0}(x,\textbf{p}_{\perp})\psi^{+ \nu}_{+0}(x,\textbf{p}_{\perp})\Bigg],
	 \nonumber\\
	\label{oeh1v1}\\
	%
	%
	%\label{oeh3tv2}
	%\\
	%
% \end{eqnarray}
% %
% %
% \begin{eqnarray}
	{x}~e^{\nu(A)}(x,\textbf{p}_{\perp}^2) &=&\sum_{\lambda^D} \frac{C_{A}^{2}}{16\pi^3}\frac{m}{M}\bigg[|\psi ^{+\nu}_{+\lambda^D}(x,\textbf{p}_{\perp})|^2+|\psi ^{ + \nu}_{-\lambda^D}(x,\textbf{p}_{\perp})|^2\bigg], \label{ev1}\\
	%		
	%
	% {x}~e^{\nu(A)}(x,\textbf{p}_{\perp}^2) 
 &=&\sum_{\lambda^D} \frac{C_{A}^{2}}{16\pi^3}\frac{m}{M}\bigg[|\psi ^{-\nu}_{+\lambda^D}(x,\textbf{p}_{\perp})|^2+|\psi ^{ - \nu}_{-\lambda^D}(x,\textbf{p}_{\perp})|^2\bigg], \label{ev2}\\
	%		
	%
	% f perp
	{x}~f^{\perp\nu(A)}(x,\textbf{p}_{\perp}^2)&=&\sum_{\lambda^D} \frac{C_{A}^{2}}{16\pi^3}\bigg[|\psi ^{+\nu}_{+\lambda^D}(x,\textbf{p}_{\perp})|^2+|\psi ^{ + \nu}_{-\lambda^D}(x,\textbf{p}_{\perp})|^2\bigg],\label{fperpv1} \\
	%		
	%
	% f perp
	% {x}~f^{\perp\nu(A)}(x,\textbf{p}_{\perp}^2)
 &=&\sum_{\lambda^D} \frac{C_{A}^{2}}{16\pi^3}\bigg[|\psi ^{-\nu}_{+\lambda^D}(x,\textbf{p}_{\perp})|^2+|\psi ^{ - \nu}_{-\lambda^D}(x,\textbf{p}_{\perp})|^2\bigg],\label{fperpv2} \\
	%
	%
	% gL perp 1
	{x}~ g_{L}^{\perp\nu(A)}(x, {\bf p_\perp^2}) &=& \sum_{\lambda^D}\frac{C_{A}^{2}}{32\pi^3 {\textbf{p}_{x}}}\bigg({\textbf{p}_{x}}\bigg[|\psi ^{+\nu}_{+\lambda^D}(x,\textbf{p}_{\perp})|^2-|\psi ^{ + \nu}_{-\lambda^D}(x,\textbf{p}_{\perp})|^2\bigg] \nonumber\\
	&& + m\bigg[\psi ^{+\nu \dagger}_{+\lambda^D}(x,\textbf{p}_{\perp})\psi^{+ \nu}_{-\lambda^D}(x,\textbf{p}_{\perp}) + \psi ^{+\nu \dagger}_{-\lambda^D}(x,\textbf{p}_{\perp})\psi^{+ \nu}_{+\lambda^D}(x,\textbf{p}_{\perp})\bigg]\bigg), \nonumber\\
	\label{glperp1v1} 
 % \\
	%		
 \end{eqnarray}
\begin{eqnarray}
	%
	% {x}~ g_{L}^{\perp\nu(A)}(x, {\bf p_\perp^2}) 
 &=& \sum_{\lambda^D}\frac{-C_{A}^{2}}{32\pi^3 {\textbf{p}_{x}}}\bigg({\textbf{p}_{x}}\bigg[|\psi ^{-\nu}_{+\lambda^D}(x,\textbf{p}_{\perp})|^2-|\psi ^{ - \nu}_{-\lambda^D}(x,\textbf{p}_{\perp})|^2\bigg] \nonumber\\
	&& + m\bigg[\psi ^{-\nu \dagger}_{+\lambda^D}(x,\textbf{p}_{\perp})\psi^{- \nu}_{-\lambda^D}(x,\textbf{p}_{\perp}) + \psi ^{-\nu \dagger}_{-\lambda^D}(x,\textbf{p}_{\perp})\psi^{- \nu}_{+\lambda^D}(x,\textbf{p}_{\perp})\bigg]\bigg), \nonumber\\
	\label{glperp1v22} \\
	%
	%
	% gL perp 3
	% {x}~ g_{L}^{\perp\nu(A)}(x, {\bf p_\perp^2}) 
 &=& \sum_{\lambda^D}\frac{C_{A}^{2}}{32\pi^3 {\textbf{p}_{y}}}\bigg({\textbf{p}_{y}}\bigg[|\psi ^{+\nu}_{+\lambda^D}(x,\textbf{p}_{\perp})|^2-|\psi ^{ + \nu}_{-\lambda^D}(x,\textbf{p}_{\perp})|^2\bigg] \nonumber\\
	&& - \iota m\bigg[\psi ^{+\nu \dagger}_{+\lambda^D}(x,\textbf{p}_{\perp})\psi^{+ \nu}_{-\lambda^D}(x,\textbf{p}_{\perp}) - \psi ^{+\nu \dagger}_{-\lambda^D}(x,\textbf{p}_{\perp})\psi^{+ \nu}_{+\lambda^D}(x,\textbf{p}_{\perp})\bigg]\bigg), \nonumber\\
	\label{glperp1v3} \\
	%		
	%
	% {x}~ g_{L}^{\perp\nu(A)}(x, {\bf p_\perp^2}) 
 &=& \sum_{\lambda^D}\frac{-C_{A}^{2}}{32\pi^3 {\textbf{p}_{y}}}\bigg({\textbf{p}_{y}}\bigg[|\psi ^{-\nu}_{+\lambda^D}(x,\textbf{p}_{\perp})|^2-|\psi ^{ - \nu}_{-\lambda^D}(x,\textbf{p}_{\perp})|^2\bigg] \nonumber\\
	&& - \iota m\bigg[\psi ^{-\nu \dagger}_{+\lambda^D}(x,\textbf{p}_{\perp})\psi^{- \nu}_{-\lambda^D}(x,\textbf{p}_{\perp}) - \psi ^{-\nu \dagger}_{-\lambda^D}(x,\textbf{p}_{\perp})\psi^{- \nu}_{+\lambda^D}(x,\textbf{p}_{\perp})\bigg]\bigg), \nonumber\\
	\label{glperp1v4} \\
	%
	%
% \end{eqnarray}
% \begin{eqnarray}
	% gT 1
	{x}\bigg(g_{T}^{'\nu(A)}(x, {\bf p_\perp^2})+\frac{\textbf{p}_{x}^2}{M^2}g_{T}^{\perp (A)}(x, {\bf p_\perp^2})\bigg) &=&\frac{C_{A}^{2}}{32\pi^3 M}\bigg(\textbf{p}_{x}\bigg[\psi ^{+\nu \dagger}_{+0}(x,\textbf{p}_{\perp})\psi^{- \nu}_{+0}(x,\textbf{p}_{\perp}) \nonumber \\
	&&- \psi ^{+\nu \dagger}_{-0}(x,\textbf{p}_{\perp})\psi^{- \nu}_{-0}(x,\textbf{p}_{\perp})+ \psi ^{-\nu \dagger}_{+0}(x,\textbf{p}_{\perp})\psi^{+ \nu}_{+0}(x,\textbf{p}_{\perp})\nonumber \\
	&&- \psi ^{-\nu \dagger}_{-0}(x,\textbf{p}_{\perp})\psi^{+ \nu}_{-0}(x,\textbf{p}_{\perp})\bigg] +m\bigg[\psi ^{+\nu \dagger}_{+0}(x,\textbf{p}_{\perp})\psi^{- \nu}_{-0}(x,\textbf{p}_{\perp}) \nonumber \\
	&&+ \psi ^{+\nu \dagger}_{-0}(x,\textbf{p}_{\perp})\psi^{- \nu}_{+0}(x,\textbf{p}_{\perp})+\psi ^{-\nu \dagger}_{+0}(x,\textbf{p}_{\perp})\psi^{+ \nu}_{-0}(x,\textbf{p}_{\perp}) \nonumber \\
	&&+ \psi ^{-\nu \dagger}_{-0}(x,\textbf{p}_{\perp})\psi^{+ \nu}_{+0}(x,\textbf{p}_{\perp}) \bigg] \bigg), \label{gt1v1}  \\
	%		
	%
	% gT 2
	{x}\bigg(g_{T}^{'\nu(A)}(x, {\bf p_\perp^2})+\frac{\textbf{p}_{y}^2}{M^2}g_{T}^{\perp (A)}(x, {\bf p_\perp^2})\bigg) &=&\frac{\iota C_{A}^{2}}{32\pi^3 M}\bigg(\textbf{p}_{y}\bigg[\psi ^{+\nu \dagger}_{+0}(x,\textbf{p}_{\perp})\psi^{- \nu}_{+0}(x,\textbf{p}_{\perp}) \nonumber \\
	&&- \psi ^{+\nu \dagger}_{-0}(x,\textbf{p}_{\perp})\psi^{- \nu}_{-0}(x,\textbf{p}_{\perp})- \psi ^{-\nu \dagger}_{+0}(x,\textbf{p}_{\perp})\psi^{+ \nu}_{+0}(x,\textbf{p}_{\perp})\nonumber \\
	&&+ \psi ^{-\nu \dagger}_{-0}(x,\textbf{p}_{\perp})\psi^{+ \nu}_{-0}(x,\textbf{p}_{\perp})\bigg] -\iota m\bigg[\psi ^{+\nu \dagger}_{+0}(x,\textbf{p}_{\perp})\psi^{- \nu}_{-0}(x,\textbf{p}_{\perp}) \nonumber \\
	&&- \psi ^{+\nu \dagger}_{-0}(x,\textbf{p}_{\perp})\psi^{- \nu}_{+0}(x,\textbf{p}_{\perp})-\psi ^{-\nu \dagger}_{+0}(x,\textbf{p}_{\perp})\psi^{+ \nu}_{-0}(x,\textbf{p}_{\perp}) \nonumber \\
	&&+ \psi ^{-\nu \dagger}_{-0}(x,\textbf{p}_{\perp})\psi^{+ \nu}_{+0}(x,\textbf{p}_{\perp}) \bigg] \bigg), \label{gt1v2}  
% \\
	%
	%
 \end{eqnarray}
\begin{eqnarray}
	% gT perp 1
	%		
	{x}~g_{T}^{\perp\nu (A)}(x, {\bf p_\perp^2}) &=& \frac{C_{A}^{2} M}{32\pi^3 {\bf p_x} \ {\bf p_y}}\bigg[\textbf{p}_{y}\bigg[\psi ^{+\nu \dagger}_{+0}(x,\textbf{p}_{\perp})\psi^{- \nu}_{+0}(x,\textbf{p}_{\perp}) - \psi ^{+\nu \dagger}_{-0}(x,\textbf{p}_{\perp})\psi^{- \nu}_{-0}(x,\textbf{p}_{\perp})\nonumber \\
	&&+ \psi^{-\nu \dagger}_{+0}(x,\textbf{p}_{\perp})\psi^{+ \nu}_{+0}(x,\textbf{p}_{\perp})- \psi^{-\nu \dagger}_{-0}(x,\textbf{p}_{\perp})\psi^{+ \nu}_{-0}(x,\textbf{p}_{\perp})\bigg]  \nonumber \\
	&& -\iota m\bigg[\psi ^{+\nu \dagger}_{+0}(x,\textbf{p}_{\perp})\psi^{- \nu}_{-0}(x,\textbf{p}_{\perp}) - \psi ^{+\nu \dagger}_{-0}(x,\textbf{p}_{\perp})\psi^{- \nu}_{+0}(x,\textbf{p}_{\perp}) \nonumber \\
	&&+\psi ^{-\nu \dagger}_{+0}(x,\textbf{p}_{\perp})\psi^{+ \nu}_{-0}(x,\textbf{p}_{\perp}) -\psi ^{-\nu \dagger}_{-0}(x,\textbf{p}_{\perp})\psi^{+ \nu}_{+0}(x,\textbf{p}_{\perp}) \bigg] \bigg],  \label{gtperp1v}
	\\
	%		
	%
%  \end{eqnarray}
% \begin{eqnarray}
	% gT perp2
	%				
	% {x}~g_{T}^{\perp\nu (A)}(x, {\bf p_\perp^2})
 &=& \frac{\iota C_{A}^{2} M}{32\pi^3 {\bf p_x} \ {\bf p_y}}\bigg[\textbf{p}_{x}\bigg[\psi ^{+\nu \dagger}_{+0}(x,\textbf{p}_{\perp})\psi^{- \nu}_{+0}(x,\textbf{p}_{\perp}) - \psi ^{+\nu \dagger}_{-0}(x,\textbf{p}_{\perp})\psi^{- \nu}_{-0}(x,\textbf{p}_{\perp})\nonumber \\
	&&- \psi^{-\nu \dagger}_{+0}(x,\textbf{p}_{\perp})\psi^{+ \nu}_{+0}(x,\textbf{p}_{\perp})+ \psi^{-\nu \dagger}_{-0}(x,\textbf{p}_{\perp})\psi^{+ \nu}_{-0}(x,\textbf{p}_{\perp})\bigg]  \nonumber \\
	&&+ m\bigg[\psi ^{+\nu \dagger}_{+0}(x,\textbf{p}_{\perp})\psi^{- \nu}_{-0}(x,\textbf{p}_{\perp}) + \psi ^{+\nu \dagger}_{-0}(x,\textbf{p}_{\perp})\psi^{- \nu}_{+0}(x,\textbf{p}_{\perp}) \nonumber \\
	&&-\psi ^{-\nu \dagger}_{+0}(x,\textbf{p}_{\perp})\psi^{+ \nu}_{-0}(x,\textbf{p}_{\perp})-\psi ^{-\nu \dagger}_{-0}(x,\textbf{p}_{\perp})\psi^{+ \nu}_{+0}(x,\textbf{p}_{\perp}) \bigg] \bigg],  \label{gtperp2v}\\
	%		\\
	%			
	{x}~g_{T}^{\nu(A)}(x, {\bf p_\perp^2})&=&\frac{C_{A}^{2}}{64\pi^3 M}\bigg(\textbf{p}_{x}\bigg[\psi ^{+\nu \dagger}_{+0}(x,\textbf{p}_{\perp})\psi^{- \nu}_{+0}(x,\textbf{p}_{\perp}) - \psi ^{+\nu \dagger}_{-0}(x,\textbf{p}_{\perp})\psi^{- \nu}_{-0}(x,\textbf{p}_{\perp})\nonumber \\
	&&+ \psi ^{-\nu \dagger}_{+0}(x,\textbf{p}_{\perp})\psi^{+ \nu}_{+0}(x,\textbf{p}_{\perp})- \psi ^{-\nu \dagger}_{-0}(x,\textbf{p}_{\perp})\psi^{+ \nu}_{-0}(x,\textbf{p}_{\perp})\bigg]  \nonumber \\
	&&+ \iota\textbf{p}_{y}\bigg[\psi ^{+\nu \dagger}_{+0}(x,\textbf{p}_{\perp})\psi^{- \nu}_{+0}(x,\textbf{p}_{\perp}) - \psi ^{+\nu \dagger}_{-0}(x,\textbf{p}_{\perp})\psi^{- \nu}_{-0}(x,\textbf{p}_{\perp})\nonumber \\
	&&- \psi ^{-\nu \dagger}_{+0}(x,\textbf{p}_{\perp})\psi^{+ \nu}_{+0}(x,\textbf{p}_{\perp})+ \psi ^{-\nu \dagger}_{-0}(x,\textbf{p}_{\perp})\psi^{+ \nu}_{-0}(x,\textbf{p}_{\perp})\bigg]  \nonumber \\
	&& +2m\bigg[\psi ^{+\nu \dagger}_{+0}(x,\textbf{p}_{\perp})\psi^{- \nu}_{-0}(x,\textbf{p}_{\perp}) + \psi ^{-\nu \dagger}_{-0}(x,\textbf{p}_{\perp})\psi^{+ \nu}_{+0}(x,\textbf{p}_{\perp}) \bigg] \bigg), \label{gt1v3}  \\
	%		
	%
	%
% \end{eqnarray}
% \begin{eqnarray}
	%
	% hT perp 1
	{x}~h_{T}^{\perp\nu(A)}(x, {\bf p_\perp^2}) &=&\frac{\iota C_{A}^{2}}{32\pi^3 {\textbf{p}_{y}}} \bigg(({\textbf{p}_{x}}-{\iota}{\textbf{p}_{y}})\bigg[\psi ^{+\nu \dagger}_{+0}(x,\textbf{p}_{\perp})\psi^{- \nu}_{-0}(x,\textbf{p}_{\perp}) + \psi ^{-\nu \dagger}_{+0}(x,\textbf{p}_{\perp})\psi^{+ \nu}_{-0}(x,\textbf{p}_{\perp})\bigg] \nonumber\\
	&&  -({\textbf{p}_{x}}+{\iota}{\textbf{p}_{y}})\bigg[\psi ^{+\nu \dagger}_{-0}(x,\textbf{p}_{\perp})\psi^{- \nu}_{+0}(x,\textbf{p}_{\perp}) + \psi ^{-\nu \dagger}_{-0}(x,\textbf{p}_{\perp})\psi^{+ \nu}_{+0}(x,\textbf{p}_{\perp})\bigg]\bigg),
	% \nonumber\\
	\label{htperp1v1}\\
	%		
	%		
	% hT perp 2
	% {x}~h_{T}^{\perp\nu(A)}(x, {\bf p_\perp^2}) 
 &=&\frac{C_{A}^{2}}{32\pi^3 {\textbf{p}_{x}}} \bigg(({\textbf{p}_{x}}-{\iota} {\textbf{p}_{y}})\bigg[\psi ^{+\nu \dagger}_{+0}(x,\textbf{p}_{\perp})\psi^{- \nu}_{-0}(x,\textbf{p}_{\perp}) - \psi ^{-\nu \dagger}_{+0}(x,\textbf{p}_{\perp})\psi^{+ \nu}_{-0}(x,\textbf{p}_{\perp})\bigg] \nonumber\\
	&&  -({\textbf{p}_{x}}+{\iota}{\textbf{p}_{y}})\bigg[\psi ^{+\nu \dagger}_{-0}(x,\textbf{p}_{\perp})\psi^{- \nu}_{+0}(x,\textbf{p}_{\perp}) - \psi ^{-\nu \dagger}_{-0}(x,\textbf{p}_{\perp})\psi^{+ \nu}_{+0}(x,\textbf{p}_{\perp})\bigg]\bigg),
	% \nonumber\\
	\label{htperp1v2}
 % \\
  \end{eqnarray}
\begin{eqnarray}
	{x}~h_{L}^{\nu(A)}(x, {\bf p_\perp^2}) &=&\sum_{\lambda^D}\frac{C_{A}^{2}}{16\pi^3  M}\bigg(m\bigg[|\psi ^{+\nu}_{+\lambda^D}(x,\textbf{p}_{\perp})|^2-|\psi ^{ + \nu}_{-\lambda^D}(x,\textbf{p}_{\perp})|^2\bigg] \nonumber \\
	&& - ({\textbf{p}_{x}}-\iota {\textbf{p}_{y}})\bigg[\psi ^{+\nu \dagger}_{+\lambda^D}(x,\textbf{p}_{\perp})\psi^{+ \nu}_{-\lambda^D}(x,\textbf{p}_{\perp})\bigg]- ({\textbf{p}_{x}}+\iota {\textbf{p}_{y}})\nonumber\\
	&& \bigg[\psi ^{+\nu \dagger}_{-\lambda^D}(x,\textbf{p}_{\perp})\psi^{+ \nu}_{+\lambda^D}(x,\textbf{p}_{\perp})\bigg]\bigg), \label{hlv1} 
 \\
	%
	%
%  \end{eqnarray}
% \begin{eqnarray}
	%
	% {x}~h_{L}^{\nu(A)}(x, {\bf p_\perp^2}) 
 &=&\sum_{\lambda^D}\frac{-C_{A}^{2}}{16\pi^3  M}\bigg(m\bigg[|\psi ^{-\nu}_{+\lambda^D}(x,\textbf{p}_{\perp})|^2-|\psi ^{ -\nu}_{-\lambda^D}(x,\textbf{p}_{\perp})|^2\bigg] \nonumber \\
	&& - ({\textbf{p}_{x}}-\iota {\textbf{p}_{y}})\bigg[\psi ^{-\nu \dagger}_{+\lambda^D}(x,\textbf{p}_{\perp})\psi^{- \nu}_{-\lambda^D}(x,\textbf{p}_{\perp})\bigg]- ({\textbf{p}_{x}}+\iota {\textbf{p}_{y}})\nonumber\\
	&& \bigg[\psi ^{-\nu \dagger}_{-\lambda^D}(x,\textbf{p}_{\perp})\psi^{- \nu}_{+\lambda^D}(x,\textbf{p}_{\perp})\bigg]\bigg), \label{hlv2} 
 \\
	%
%  \end{eqnarray}
% \begin{eqnarray}
	% hT 1
	{x}~h_{T}^{\nu(A)}(x, {\bf p_\perp^2}) &=&\frac{C_{A}^{2}}{16\pi^3 {\textbf{p}_{x}}}\bigg(m\bigg[\psi ^{+\nu \dagger}_{+0}(x,\textbf{p}_{\perp})\psi^{- \nu}_{+0}(x,\textbf{p}_{\perp}) - \psi ^{+\nu \dagger}_{-0}(x,\textbf{p}_{\perp})\psi^{- \nu}_{-0}(x,\textbf{p}_{\perp})\nonumber \\
	&& +\psi ^{-\nu \dagger}_{+0}(x,\textbf{p}_{\perp})\psi^{+ \nu}_{+0}(x,\textbf{p}_{\perp}) -\psi ^{-\nu \dagger}_{-0}(x,\textbf{p}_{\perp})\psi^{+ \nu}_{-0}(x,\textbf{p}_{\perp}) \bigg] \nonumber\\
	&& -({\textbf{p}_{x}}-\iota {\textbf{p}_{y}})\bigg[\psi ^{+\nu \dagger}_{+0}(x,\textbf{p}_{\perp})\psi^{- \nu}_{-0}(x,\textbf{p}_{\perp}) + \psi ^{-\nu \dagger}_{+0}(x,\textbf{p}_{\perp})\psi^{+ \nu}_{-0}(x,\textbf{p}_{\perp})\bigg] \nonumber \\
	&& -({\textbf{p}_{x}}+\iota {\textbf{p}_{y}})\bigg[\psi ^{+\nu \dagger}_{-0}(x,\textbf{p}_{\perp})\psi^{- \nu}_{+0}(x,\textbf{p}_{\perp}) + \psi ^{-\nu \dagger}_{-0}(x,\textbf{p}_{\perp})\psi^{+ \nu}_{+0}(x,\textbf{p}_{\perp})\bigg]\bigg),
	% \nonumber\\
	\label{ht1v1} \\
	%		\\
	%
	% hT 2
	% {x}~h_{T}^{\nu(A)}(x, {\bf p_\perp^2}) 
 &=&\frac{\iota C_{A}^{2}}{16\pi^3 {\textbf{p}_{y}}}\bigg(m\bigg[\psi ^{+\nu \dagger}_{+0}(x,\textbf{p}_{\perp})\psi^{- \nu}_{+0}(x,\textbf{p}_{\perp}) - \psi ^{+\nu \dagger}_{-0}(x,\textbf{p}_{\perp})\psi^{- \nu}_{-0}(x,\textbf{p}_{\perp})\nonumber \\
	&& -\psi ^{-\nu \dagger}_{+0}(x,\textbf{p}_{\perp})\psi^{+ \nu}_{+0}(x,\textbf{p}_{\perp}) +\psi ^{-\nu \dagger}_{-0}(x,\textbf{p}_{\perp})\psi^{+ \nu}_{-0}(x,\textbf{p}_{\perp}) \bigg] \nonumber\\
	&& -({\textbf{p}_{x}}-\iota {\textbf{p}_{y}})\bigg[\psi ^{+\nu \dagger}_{+0}(x,\textbf{p}_{\perp})\psi^{- \nu}_{-0}(x,\textbf{p}_{\perp}) - \psi ^{-\nu \dagger}_{+0}(x,\textbf{p}_{\perp})\psi^{+ \nu}_{-0}(x,\textbf{p}_{\perp})\bigg] \nonumber \\
	&& -({\textbf{p}_{x}}+\iota {\textbf{p}_{y}})\bigg[\psi ^{+\nu \dagger}_{-0}(x,\textbf{p}_{\perp})\psi^{- \nu}_{+0}(x,\textbf{p}_{\perp}) - \psi ^{-\nu \dagger}_{-0}(x,\textbf{p}_{\perp})\psi^{+ \nu}_{+0}(x,\textbf{p}_{\perp})\bigg]\bigg),
	% \nonumber\\
	\label{ht1v2}\\
	%
% \end{eqnarray}
% %	
% \begin{eqnarray}
	%
	%f3v
	%
	x^2 f_{3}^{\nu(A)}(x, {\bf p_\perp^2}) &=&\sum_{\lambda^D}  \frac{C_{A}^{2}}{16 \pi^3} \bigg(\frac{{\bf p_\perp^2}+m^2}{M^2}\bigg)\Bigg[|\psi ^{+\nu}_{+\lambda^D}(x,\textbf{p}_{\perp})|^2+|\psi ^{ + \nu}_{-\lambda^D}(x,\textbf{p}_{\perp})|^2\Bigg], \label{oef3v1}\\
	%
	%
	%f3v
	%
	% x^2 f_{3}^{\nu(A)}(x, {\bf p_\perp^2}) 
 &=&\sum_{\lambda^D}  \frac{C_{A}^{2}}{16 \pi^3} \bigg(\frac{{\bf p_\perp^2}+m^2}{M^2}\bigg)\Bigg[|\psi ^{-\nu}_{+\lambda^D}(x,\textbf{p}_{\perp})|^2+|\psi ^{- \nu}_{-\lambda^D}(x,\textbf{p}_{\perp})|^2\Bigg], \label{oef3v2}\\
	%
	%g3lv
	%
	x^2 g_{3L}^{\nu(A)}(x, {\bf p_\perp^2}) &=&\sum_{\lambda^D}  \frac{C_{A}^{2}}{32 \pi^3 M^2} \Bigg[\big({{\bf p_\perp^2}-m^2}\big)\bigg[|\psi ^{+\nu}_{+\lambda^D}(x,\textbf{p}_{\perp})|^2-|\psi ^{ + \nu}_{-\lambda^D}(x,\textbf{p}_{\perp})|^2\bigg] \nonumber\\
	&& + 2m ({\textbf{p}_{x}}-\iota {\textbf{p}_{y}})\bigg[\psi ^{+\nu \dagger}_{+\lambda^D}(x,\textbf{p}_{\perp})\psi^{+ \nu}_{-\lambda^D}(x,\textbf{p}_{\perp})\bigg] \nonumber \\
	&&+ 2m ({\textbf{p}_{x}}+\iota {\textbf{p}_{y}}) \bigg[\psi ^{+\nu \dagger}_{-\lambda^D}(x,\textbf{p}_{\perp})\psi^{+ \nu}_{+\lambda^D}(x,\textbf{p}_{\perp})\bigg]\Bigg], \label{oeg3lpv1} 
 % \\
	%
  \end{eqnarray}
\begin{eqnarray}
	%
	% x^2 g_{3L}^{\nu(A)}(x, {\bf p_\perp^2})
 &=&\sum_{\lambda^D}  \frac{-C_{A}^{2}}{32 \pi^3 M^2} \Bigg[\big({{\bf p_\perp^2}-m^2}\big)\bigg[|\psi ^{-\nu}_{+\lambda^D}(x,\textbf{p}_{\perp})|^2-|\psi ^{- \nu}_{-\lambda^D}(x,\textbf{p}_{\perp})|^2\bigg] \nonumber\\
	&& + 2m ({\textbf{p}_{x}}-\iota {\textbf{p}_{y}})\bigg[\psi ^{-\nu \dagger}_{+\lambda^D}(x,\textbf{p}_{\perp})\psi^{- \nu}_{-\lambda^D}(x,\textbf{p}_{\perp})\bigg] \nonumber \\
	&&+ 2m ({\textbf{p}_{x}}+\iota {\textbf{p}_{y}}) \bigg[\psi ^{-\nu \dagger}_{-\lambda^D}(x,\textbf{p}_{\perp})\psi^{- \nu}_{+\lambda^D}(x,\textbf{p}_{\perp})\bigg]\Bigg], \label{oeg3lpv2} \\
 % \\
	%
%  \end{eqnarray}
% \begin{eqnarray}
	%g3tv1
	%
	x^2 g_{3T}^{\nu(A)}(x, {\bf p_\perp^2}) &=&  \frac{C_{A}^{2}}{32 \pi^3 M  {\textbf{p}_{x}}} \Bigg[\big({{\bf p_\perp^2}-m^2}\big)\bigg[
	\psi ^{+\nu \dagger}_{+0}(x,\textbf{p}_{\perp})\psi^{- \nu}_{+0}(x,\textbf{p}_{\perp}) - \psi ^{+\nu \dagger}_{-0}(x,\textbf{p}_{\perp}) \psi ^{-\nu}_{-0}(x,\textbf{p}_{\perp}) \nonumber\\
	&&+ \psi ^{-\nu \dagger}_{+0}(x,\textbf{p}_{\perp}) \psi ^{+\nu}_{+0}(x,\textbf{p}_{\perp})-\psi^{- \nu \dagger}_{-0}(x,\textbf{p}_{\perp}) \psi ^{+\nu}_{-0}(x,\textbf{p}_{\perp})\bigg] \nonumber\\
	&& + 2m ({\textbf{p}_{x}}-\iota {\textbf{p}_{y}})\bigg[\psi ^{+\nu \dagger}_{+0}(x,\textbf{p}_{\perp})\psi^{- \nu}_{-0}(x,\textbf{p}_{\perp}) + \psi ^{-\nu \dagger}_{+0}(x,\textbf{p}_{\perp})\psi^{+ \nu}_{-0}(x,\textbf{p}_{\perp})\bigg] \nonumber \\
	&&+ 2m ({\textbf{p}_{x}}+\iota {\textbf{p}_{y}}) \bigg[\psi ^{+\nu \dagger}_{-0}(x,\textbf{p}_{\perp})\psi^{- \nu}_{+0}(x,\textbf{p}_{\perp}) + \psi ^{-\nu \dagger}_{-0}(x,\textbf{p}_{\perp})\psi^{+ \nu}_{+0}(x,\textbf{p}_{\perp})\bigg]\Bigg], \label{oeg3tv1}
 % \nonumber 
 \\ 
 % \\
	%
%  \end{eqnarray}
% \begin{eqnarray}
	%
	%g3tv2
	%
	% x^2 g_{3T}^{\nu(A)}(x, {\bf p_\perp^2}) 
 &=&  \frac{\iota C_{A}^{2}}{32 \pi^3 M {\textbf{p}_{y}} } \Bigg[\big({{\bf p_\perp^2}-m^2}\big)\bigg[
	\psi ^{+\nu \dagger}_{+0}(x,\textbf{p}_{\perp})\psi^{- \nu}_{+0}(x,\textbf{p}_{\perp}) - \psi ^{+\nu \dagger}_{-0}(x,\textbf{p}_{\perp}) \psi ^{-\nu}_{-0}(x,\textbf{p}_{\perp}) \nonumber\\
	&&- \psi ^{-\nu \dagger}_{+0}(x,\textbf{p}_{\perp}) \psi ^{+\nu}_{+0}(x,\textbf{p}_{\perp})+\psi^{- \nu \dagger}_{-0}(x,\textbf{p}_{\perp}) \psi ^{+\nu}_{-0}(x,\textbf{p}_{\perp})\bigg] \nonumber\\
	&& + 2m ({\textbf{p}_{x}}-\iota {\textbf{p}_{y}})\bigg[\psi ^{+\nu \dagger}_{+0}(x,\textbf{p}_{\perp})\psi^{- \nu}_{-0}(x,\textbf{p}_{\perp}) - \psi ^{-\nu \dagger}_{+0}(x,\textbf{p}_{\perp})\psi^{+ \nu}_{-0}(x,\textbf{p}_{\perp})\bigg] \nonumber \\
	&&+ 2m ({\textbf{p}_{x}}+\iota {\textbf{p}_{y}}) \bigg[\psi ^{+\nu \dagger}_{-0}(x,\textbf{p}_{\perp})\psi^{- \nu}_{+0}(x,\textbf{p}_{\perp}) - \psi ^{-\nu \dagger}_{-0}(x,\textbf{p}_{\perp})\psi^{+ \nu}_{+0}(x,\textbf{p}_{\perp})\bigg]\Bigg], \label{oeg3tv2} 
 % \nonumber 
 \\
	% \\
	%
% \end{eqnarray}
% \begin{eqnarray}
	%h3lpv1
	%
	x^2~h_{3L}^{\perp\nu(A)}(x, {\bf p_\perp^2}) &=&\sum_{\lambda^D}  \frac{C_{A}^{2}}{32 \pi^3 {\textbf{p}_{x}} M} \Bigg[2m{\textbf{p}_{x}} \bigg[|\psi ^{+\nu}_{+\lambda^D}(x,\textbf{p}_{\perp})|^2-|\psi ^{ + \nu}_{-\lambda^D}(x,\textbf{p}_{\perp})|^2\bigg] \nonumber\\
	&& + \bigg( m^2 -({\textbf{p}_{x}}-\iota {\textbf{p}_{y}})^2 \bigg)
	\bigg[\psi ^{+\nu \dagger}_{+\lambda^D}(x,\textbf{p}_{\perp})\psi^{+ \nu}_{-\lambda^D}(x,\textbf{p}_{\perp})\bigg] \nonumber \\
	&&+ \bigg( m^2 -({\textbf{p}_{x}}+\iota {\textbf{p}_{y}})^2 \bigg) \bigg[\psi ^{+\nu \dagger}_{-\lambda^D}(x,\textbf{p}_{\perp})\psi^{+ \nu}_{+\lambda^D}(x,\textbf{p}_{\perp})\bigg]\Bigg], \label{oeh3lpv1}\\
	%\\
	%
	%h3lpv2
	%
	% x^2~h_{3L}^{\perp\nu(A)}(x, {\bf p_\perp^2}) 
 &=&\sum_{\lambda^D}  \frac{-C_{A}^{2}}{32 \pi^3 {\textbf{p}_{x}} M} \Bigg[2m{\textbf{p}_{x}} \bigg[|\psi ^{-\nu}_{+\lambda^D}(x,\textbf{p}_{\perp})|^2-|\psi ^{ - \nu}_{-\lambda^D}(x,\textbf{p}_{\perp})|^2\bigg] \nonumber\\
	&& + \bigg( m^2 -({\textbf{p}_{x}}-\iota {\textbf{p}_{y}})^2 \bigg)
	\bigg[\psi ^{-\nu \dagger}_{+\lambda^D}(x,\textbf{p}_{\perp})\psi^{- \nu}_{-\lambda^D}(x,\textbf{p}_{\perp})\bigg] \nonumber \\
	&&+ \bigg( m^2 -({\textbf{p}_{x}}+\iota {\textbf{p}_{y}})^2 \bigg) \bigg[\psi ^{-\nu \dagger}_{-\lambda^D}(x,\textbf{p}_{\perp})\psi^{- \nu}_{+\lambda^D}(x,\textbf{p}_{\perp})\bigg]\Bigg], \label{oeh3lpv2}
 % \\
	%\\
	%
	%
  \end{eqnarray}
\begin{eqnarray}
	%h3lpv3
	%
	% x^2~h_{3L}^{\perp\nu(A)}(x, {\bf p_\perp^2}) 
 &=&\sum_{\lambda^D}  \frac{C_{A}^{2}}{32 \pi^3 {\textbf{p}_{y}} M} \Bigg[2m{\textbf{p}_{y}} \bigg[|\psi ^{+\nu}_{+\lambda^D}(x,\textbf{p}_{\perp})|^2-|\psi ^{ + \nu}_{-\lambda^D}(x,\textbf{p}_{\perp})|^2\bigg] \nonumber\\
	&&-\iota \bigg( m^2 +({\textbf{p}_{x}}-\iota {\textbf{p}_{y}})^2 \bigg)
	\bigg[\psi ^{+\nu \dagger}_{+\lambda^D}(x,\textbf{p}_{\perp})\psi^{+ \nu}_{-\lambda^D}(x,\textbf{p}_{\perp})\bigg] \nonumber \\
	&&+\iota \bigg( m^2 +({\textbf{p}_{x}}+\iota {\textbf{p}_{y}})^2 \bigg) \bigg[\psi ^{+\nu \dagger}_{-\lambda^D}(x,\textbf{p}_{\perp})\psi^{+ \nu}_{+\lambda^D}(x,\textbf{p}_{\perp})\bigg]\Bigg], \label{oeh3lpv3}\\
	%\\
	%
	%h3lpv4
	%
	% x^2~h_{3L}^{\perp\nu(A)}(x, {\bf p_\perp^2}) 
 &=&\sum_{\lambda^D}  \frac{-C_{A}^{2}}{32 \pi^3 {\textbf{p}_{y}} M} \Bigg[2m{\textbf{p}_{y}} \bigg[|\psi ^{-\nu}_{+\lambda^D}(x,\textbf{p}_{\perp})|^2-|\psi ^{ - \nu}_{-\lambda^D}(x,\textbf{p}_{\perp})|^2\bigg] \nonumber\\
	&& -\iota \bigg( m^2 +({\textbf{p}_{x}}-\iota {\textbf{p}_{y}})^2 \bigg)
	\bigg[\psi ^{-\nu \dagger}_{+\lambda^D}(x,\textbf{p}_{\perp})\psi^{- \nu}_{-\lambda^D}(x,\textbf{p}_{\perp})\bigg] \nonumber \\
	&&+\iota \bigg( m^2 +({\textbf{p}_{x}}+\iota {\textbf{p}_{y}})^2 \bigg) \bigg[\psi ^{-\nu \dagger}_{-\lambda^D}(x,\textbf{p}_{\perp})\psi^{- \nu}_{+\lambda^D}(x,\textbf{p}_{\perp})\bigg]\Bigg], \label{oeh3lpv4}	\\
	%\\
	%
% \end{eqnarray}
% \begin{eqnarray}
	%
	%\ee
	%\be
	%h3tv1
	%
	x^2 \bigg[h_{3T}^{\nu(A)}(x, {\bf p_\perp^2})+\frac{{\textbf{p}_{x}}^2}{M^2}h_{3T}^{\perp\nu(A)}(x, {\bf p_\perp^2})\bigg]
	&=&  \frac{C_{A}^{2}}{32 \pi^3 M^2} \Bigg[2m{\textbf{p}_{x}} \bigg[\psi ^{+\nu \dagger}_{+0}(x,\textbf{p}_{\perp})\psi^{- \nu}_{+0}(x,\textbf{p}_{\perp}) \nonumber\\
	&&- \psi ^{+\nu \dagger}_{-0}(x,\textbf{p}_{\perp}) \psi ^{-\nu}_{-0}(x,\textbf{p}_{\perp}) + \psi ^{-\nu \dagger}_{+0}(x,\textbf{p}_{\perp}) \psi ^{+\nu}_{+0}(x,\textbf{p}_{\perp})\nonumber\\
	&&-\psi^{- \nu \dagger}_{-0}(x,\textbf{p}_{\perp}) \psi ^{+\nu}_{-0}(x,\textbf{p}_{\perp})\bigg] + \bigg( m^2 -({\textbf{p}_{x}}-\iota {\textbf{p}_{y}})^2 \bigg)\nonumber\\
	&&
	\bigg[\psi ^{+\nu \dagger}_{+0}(x,\textbf{p}_{\perp})\psi^{- \nu}_{-0}(x,\textbf{p}_{\perp}) + \psi ^{-\nu \dagger}_{+0}(x,\textbf{p}_{\perp})\psi^{+ \nu}_{-0}(x,\textbf{p}_{\perp})\bigg] \nonumber \\
	&&+ \bigg( m^2 -({\textbf{p}_{x}}+\iota {\textbf{p}_{y}})^2 \bigg) \bigg[\psi ^{+\nu \dagger}_{-0}(x,\textbf{p}_{\perp})\psi^{- \nu}_{+0}(x,\textbf{p}_{\perp}) \nonumber\\
	&&+ \psi ^{-\nu \dagger}_{-0}(x,\textbf{p}_{\perp})\psi^{+ \nu}_{+0}(x,\textbf{p}_{\perp})\bigg]\Bigg], 
 % \nonumber\\
	\label{oeh3tv11}\\
	%
	%
	%
	%h3tv2
	%
	x^2 \bigg[h_{3T}^{\nu(A)}(x, {\bf p_\perp^2})+\frac{{\textbf{p}_{y}}^2}{M^2}h_{3T}^{\perp\nu(A)}(x, {\bf p_\perp^2})\bigg]
	&=&  \frac{\iota C_{A}^{2}}{32 \pi^3 M^2} \Bigg[2m{\textbf{p}_{y}} \bigg[\psi ^{+\nu \dagger}_{+0}(x,\textbf{p}_{\perp})\psi^{- \nu}_{+0}(x,\textbf{p}_{\perp}) \nonumber\\
	&&- \psi ^{+\nu \dagger}_{-0}(x,\textbf{p}_{\perp}) \psi ^{-\nu}_{-0}(x,\textbf{p}_{\perp}) - \psi ^{-\nu \dagger}_{+0}(x,\textbf{p}_{\perp}) \psi ^{+\nu}_{+0}(x,\textbf{p}_{\perp})\nonumber\\
	&&+\psi^{- \nu \dagger}_{-0}(x,\textbf{p}_{\perp}) \psi ^{+\nu}_{-0}(x,\textbf{p}_{\perp})\bigg] -\iota \bigg( m^2 +({\textbf{p}_{x}}-\iota {\textbf{p}_{y}})^2 \bigg)\nonumber\\
	&& \bigg[\psi ^{+\nu \dagger}_{+0}(x,\textbf{p}_{\perp})\psi^{- \nu}_{-0}(x,\textbf{p}_{\perp}) - \psi ^{-\nu \dagger}_{+0}(x,\textbf{p}_{\perp})\psi^{+ \nu}_{-0}(x,\textbf{p}_{\perp})\bigg] \nonumber \\
	&&+\iota \bigg( m^2 +({\textbf{p}_{x}}+\iota {\textbf{p}_{y}})^2 \bigg) \bigg[\psi ^{+\nu \dagger}_{-0}(x,\textbf{p}_{\perp})\psi^{- \nu}_{+0}(x,\textbf{p}_{\perp}) \nonumber\\
	&&- \psi ^{-\nu \dagger}_{-0}(x,\textbf{p}_{\perp})\psi^{+ \nu}_{+0}(x,\textbf{p}_{\perp})\bigg]\Bigg], 
 % \nonumber\\
	\label{oeh3tv2}
 % \\
	%
 \end{eqnarray}
\begin{eqnarray}
	%
	%h3tpv1
	%
	x^2~h_{3T}^{\perp\nu(A)}(x, {\bf p_\perp^2}) &=&  \frac{C_{A}^{2}}{32 \pi^3 {\textbf{p}_{x}}{\textbf{p}_{y}}} \Bigg[2m{\textbf{p}_{y}} \bigg[\psi ^{+\nu \dagger}_{+0}(x,\textbf{p}_{\perp})\psi^{- \nu}_{+0}(x,\textbf{p}_{\perp}) - \psi ^{+\nu \dagger}_{-0}(x,\textbf{p}_{\perp}) \psi ^{-\nu}_{-0}(x,\textbf{p}_{\perp}) \nonumber\\
	&&+ \psi ^{-\nu \dagger}_{+0}(x,\textbf{p}_{\perp}) \psi ^{+\nu}_{+0}(x,\textbf{p}_{\perp})-\psi^{- \nu \dagger}_{-0}(x,\textbf{p}_{\perp}) \psi ^{+\nu}_{-0}(x,\textbf{p}_{\perp})\bigg] -\iota \bigg( m^2 \nonumber\\
	&&+({\textbf{p}_{x}}-\iota {\textbf{p}_{y}})^2 \bigg)
	\bigg[\psi ^{+\nu \dagger}_{+0}(x,\textbf{p}_{\perp})\psi^{- \nu}_{-0}(x,\textbf{p}_{\perp}) + \psi ^{-\nu \dagger}_{+0}(x,\textbf{p}_{\perp})\psi^{+ \nu}_{-0}(x,\textbf{p}_{\perp})\bigg] +\iota \bigg( m^2 \nonumber\\
	&&+({\textbf{p}_{x}}+\iota {\textbf{p}_{y}})^2 \bigg) \bigg[\psi ^{+\nu \dagger}_{-0}(x,\textbf{p}_{\perp})\psi^{- \nu}_{+0}(x,\textbf{p}_{\perp}) + \psi ^{-\nu \dagger}_{-0}(x,\textbf{p}_{\perp})\psi^{+ \nu}_{+0}(x,\textbf{p}_{\perp})\bigg]\Bigg],
	\label{oeh3tpv1}
 % \nonumber \\ 
 \\
	%
	%
	%h3tpv2
	%
	% x^2~h_{3T}^{\perp\nu(A)}(x, {\bf p_\perp^2}) 
 &=&  \frac{\iota C_{A}^{2}}{32 \pi^3 {\textbf{p}_{x}}{\textbf{p}_{y}}} \Bigg[2m{\textbf{p}_{x}} \bigg[\psi ^{+\nu \dagger}_{+0}(x,\textbf{p}_{\perp})\psi^{- \nu}_{+0}(x,\textbf{p}_{\perp}) - \psi ^{+\nu \dagger}_{-0}(x,\textbf{p}_{\perp}) \psi ^{-\nu}_{-0}(x,\textbf{p}_{\perp}) \nonumber\\
	&&- \psi ^{-\nu \dagger}_{+0}(x,\textbf{p}_{\perp}) \psi ^{+\nu}_{+0}(x,\textbf{p}_{\perp})+\psi^{- \nu \dagger}_{-0}(x,\textbf{p}_{\perp}) \psi ^{+\nu}_{-0}(x,\textbf{p}_{\perp})\bigg] + \bigg( m^2 \nonumber\\
	&&-({\textbf{p}_{x}}-\iota {\textbf{p}_{y}})^2 \bigg)
	\bigg[\psi ^{+\nu \dagger}_{+0}(x,\textbf{p}_{\perp})\psi^{- \nu}_{-0}(x,\textbf{p}_{\perp}) - \psi ^{-\nu \dagger}_{+0}(x,\textbf{p}_{\perp})\psi^{+ \nu}_{-0}(x,\textbf{p}_{\perp})\bigg] + \bigg( m^2 \nonumber\\
	&&-({\textbf{p}_{x}}+\iota {\textbf{p}_{y}})^2 \bigg) \bigg[\psi ^{+\nu \dagger}_{-0}(x,\textbf{p}_{\perp})\psi^{- \nu}_{+0}(x,\textbf{p}_{\perp}) - \psi ^{-\nu \dagger}_{-0}(x,\textbf{p}_{\perp})\psi^{+ \nu}_{+0}(x,\textbf{p}_{\perp})\bigg]\Bigg],
	\label{oeh3tpv2}
 % \nonumber \\ 
 \\
	%
	%
%  \end{eqnarray}
% \begin{eqnarray}
	%h3v
	%
	x^2 h_{3}^{\nu(A)}(x, {\bf p_\perp^2})
	&=&  \frac{C_{A}^{2}}{64 \pi^3 M^2} \Bigg[2m\bigg[{\textbf{p}_{x}} \Big[\psi ^{+\nu \dagger}_{+0}(x,\textbf{p}_{\perp})\psi^{- \nu}_{+0}(x,\textbf{p}_{\perp}) - \psi ^{+\nu \dagger}_{-0}(x,\textbf{p}_{\perp}) \psi ^{-\nu}_{-0}(x,\textbf{p}_{\perp}) \nonumber\\
	&&+ \psi ^{-\nu \dagger}_{+0}(x,\textbf{p}_{\perp}) \psi ^{+\nu}_{+0}(x,\textbf{p}_{\perp})-\psi^{- \nu \dagger}_{-0}(x,\textbf{p}_{\perp}) \psi ^{+\nu}_{-0}(x,\textbf{p}_{\perp})\Big] \nonumber\\
	&&+\iota{\textbf{p}_{y}} \Big[\psi ^{+\nu \dagger}_{+0}(x,\textbf{p}_{\perp})\psi^{- \nu}_{+0}(x,\textbf{p}_{\perp}) - \psi ^{+\nu \dagger}_{-0}(x,\textbf{p}_{\perp}) \psi ^{-\nu}_{-0}(x,\textbf{p}_{\perp}) \nonumber\\
	&&- \psi ^{-\nu \dagger}_{+0}(x,\textbf{p}_{\perp}) \psi ^{+\nu}_{+0}(x,\textbf{p}_{\perp})+\psi^{- \nu \dagger}_{-0}(x,\textbf{p}_{\perp}) \psi ^{+\nu}_{-0}(x,\textbf{p}_{\perp})\Big]\bigg] \nonumber\\
	&& + \bigg( m^2 -({\textbf{p}_{x}}-\iota {\textbf{p}_{y}})^2 \bigg)
	\bigg[\psi ^{+\nu \dagger}_{+0}(x,\textbf{p}_{\perp})\psi^{- \nu}_{-0}(x,\textbf{p}_{\perp}) + \psi ^{-\nu \dagger}_{+0}(x,\textbf{p}_{\perp})\psi^{+ \nu}_{-0}(x,\textbf{p}_{\perp})\bigg] \nonumber \\
	&&+ \bigg( m^2 -({\textbf{p}_{x}}+\iota {\textbf{p}_{y}})^2 \bigg) \bigg[\psi ^{+\nu \dagger}_{-0}(x,\textbf{p}_{\perp})\psi^{- \nu}_{+0}(x,\textbf{p}_{\perp}) + \psi ^{-\nu \dagger}_{-0}(x,\textbf{p}_{\perp})\psi^{+ \nu}_{+0}(x,\textbf{p}_{\perp})\bigg], \nonumber\\
	&& + \bigg( m^2 +({\textbf{p}_{x}}-\iota {\textbf{p}_{y}})^2 \bigg)
	\bigg[\psi ^{+\nu \dagger}_{+0}(x,\textbf{p}_{\perp})\psi^{- \nu}_{-0}(x,\textbf{p}_{\perp}) - \psi ^{-\nu \dagger}_{+0}(x,\textbf{p}_{\perp})\psi^{+ \nu}_{-0}(x,\textbf{p}_{\perp})\bigg] \nonumber \\
	&&- \bigg( m^2 +({\textbf{p}_{x}}+\iota {\textbf{p}_{y}})^2 \bigg) \bigg[\psi ^{+\nu \dagger}_{-0}(x,\textbf{p}_{\perp})\psi^{- \nu}_{+0}(x,\textbf{p}_{\perp}) - \psi ^{-\nu \dagger}_{-0}(x,\textbf{p}_{\perp})\psi^{+ \nu}_{+0}(x,\textbf{p}_{\perp})\bigg]\Bigg],\nonumber\\
	\label{oeh3v1}
\end{eqnarray}
where $\lambda^D$ runs over $0,\pm1$, which is nothing but the summation over vector diquark's helicity.
\section{Explicit Expressions of TMDs}
\label{sec_exf}
By substituting the expressions of LFWFs from Table \ref{tab_LFWF} into the scalar diquark overlap Eqs. {\eqref{oef1s1}-\eqref{oeh3v}}, we have obtained the explicit expressions of T-even TMDs for scalar diquark  as
\begin{eqnarray}
	%f1s done ss
	%
	f_{1}^{\nu(S)}(x, {\bf p_\perp^2}) &=&  \frac{C_{S}^{2} N_s^2}{16 \pi^3} \Bigg[ \mathcal{T}_{11}^{\nu} + \frac{\bf p_\perp^2}{x^2 M^2} \mathcal{T}_{22}^{\nu}\Bigg], \label{eef1s} \\
	%
	%g1ls done ss
	%
	g_{1L}^{\nu(S)}(x, {\bf p_\perp^2}) &=&  \frac{C_{S}^{2} N_s^2}{32 \pi^3} \Bigg[ \mathcal{T}_{11}^{\nu} - \frac{\bf p_\perp^2}{x^2 M^2} \mathcal{T}_{22}^{\nu}\Bigg],  \label{eeg1ls} \\
	%
	%g1ts
	%
	g_{1T}^{\nu(S)}(x, {\bf p_\perp^2}) &=&  \frac{C_{S}^{2} N_s^2}{8 \pi^3} \Bigg[\frac{1}{x} \mathcal{T}_{12}^{\nu} \Bigg], \label{eeg3ts} \\
	%hlps
	%
	h_{1L}^{\perp\nu(S)}(x, {\bf p_\perp^2}) &=&  -\frac{C_{S}^{2} N_s^2}{16 \pi^3} \Bigg[\frac{1}{x} \mathcal{T}_{12}^{\nu} \Bigg],
	\label{eeh3lps}\\ 
	%
	%\ee
	%\be
	% \nonumber % Remove numbering (before each equation)
	%
	%
	% h1ts done ss
	%
	h_{1T}^{\nu(S)}(x, {\bf p_\perp^2})&=&  \frac{C_{S}^{2} N_s^2}{16 \pi^3} \Bigg[ \mathcal{T}_{11}^{\nu} + \frac{\bf p_\perp^2}{x^2 M^2} \mathcal{T}_{22}^{\nu}\Bigg], \label{eehts} \\
	%htps
	%
	h_{1T}^{\perp \nu(S)}(x, {\bf p_\perp^2}) &=&  -\frac{C_{S}^{2} N_s^2}{8 \pi^3}\Bigg[ \frac{1}{x^2} \mathcal{T}_{22}^{\nu}\Bigg], \label{eeh1tps}\\
	h_{1}^{ \nu(S)}(x, {\bf p_\perp^2}) &=&  \frac{C_{S}^{2}N_s^2}{16 \pi^3} \Bigg[\mathcal{T}_{11}^{\nu}\Bigg],  \label{eeh1s}
 \\
	%
% 	\end{eqnarray}
% \begin{eqnarray}
	% e
	x e^{\nu (S)}(x, {\bf p_\perp^2}) &=& \frac{{C_{S}^{2} N_s^2}}{16\pi^3} \frac{m}{M} \bigg[ \mathcal{T}_{11}^{\nu} + \frac{\bf p_\perp^2}{x^2 M^2} \mathcal{T}_{22}^{\nu}\bigg], \label{ef1} \\
	% fperp
	x f^{\perp\nu (S)}(x, {\bf p_\perp^2}) &=&  \frac{{C_{S}^{2} N_s^2}}{16 \pi^3}  \bigg[ \mathcal{T}_{11}^{\nu} + \frac{\bf p_\perp^2}{x^2 M^2} \mathcal{T}_{22}^{\nu}\bigg], \label{fperpf}\\
	% gLperp
	x g_{L}^{\perp\nu (S)}(x, {\bf p_\perp^2}) &=&  \frac{C_{S}^{2} N_s^2}{32 \pi^3} \bigg[ \mathcal{T}_{11}^{\nu} - \frac{\bf p_\perp^2}{x^2 M^2} \mathcal{T}_{22}^{\nu}-\frac{2 m}{x M} \mathcal{T}_{12}^{\nu} \bigg], \label{glperps}\\
	% gT
	x g_{T}^{'\nu (S)}(x, {\bf p_\perp^2}) &=&  \frac{{C_{S}^{2} N_s^2}}{16 \pi^3} \frac{m}{M} \bigg[ \mathcal{T}_{11}^{\nu} + \frac{\bf p_\perp^2}{x^2 M^2} \mathcal{T}_{22}^{\nu}\bigg], \label{gts}\\
	% gTperp
	x g_{T}^{\perp\nu (S)}(x, {\bf p_\perp^2}) &=&  \frac{{C_{S}^{2} N_s^2}}{8 \pi^3} {M} \bigg[ \frac{1}{xM} \mathcal{T}_{12}^{\nu}- \frac{m}{x^2 M^2} \mathcal{T}_{22}^{\nu} \bigg],\label{gtperps}\\
	x g_{T}^{\nu (S)}(x, {\bf p_\perp^2}) &=& \frac{{C_{S}^{2} N_s^2}}{16 \pi^3}   \frac{1}{M} \bigg[m~\mathcal{T}_{11}^{\nu} +\frac{\bf p_\perp^2}{x M} \mathcal{T}_{12}^{\nu} \bigg], \label{gtsf} 
  \\
	%	\end{eqnarray}
%   \end{eqnarray}
% \begin{eqnarray}
% hTperp
x h_{T}^{\perp\nu (S)}(x, {\bf p_\perp^2}) &=&  \frac{{C_{S}^{2} N_s^2}}{16 \pi^3}
\bigg[ \mathcal{T}_{11}^{\nu} + \frac{\bf p_\perp^2}{x^2 M^2} \mathcal{T}_{22}^{\nu}\bigg], \label{htperpf}\\
% hL
x h_{L}^{\nu (S)}(x, {\bf p_\perp^2}) &=& \frac{{C_{S}^{2} N_s^2}}{16 \pi^3}   \frac{1}{M} \bigg[m\bigg( \mathcal{T}_{11}^{\nu} - \frac{\bf p_\perp^2}{x^2 M^2} \mathcal{T}_{22}^{\nu}\bigg)
% \nonumber \\
%&&
+\frac{2 p_\perp^2}{x M} \mathcal{T}_{12}^{\nu} \bigg], \label{hlf}
% \\
 \end{eqnarray}
\begin{eqnarray}
% hT
x h_{T}^{\nu (S)}(x, {\bf p_\perp^2}) &=&  -\frac{{C_{S}^{2} N_s^2}}{8 \pi^3}   \bigg[ \mathcal{T}_{11}^{\nu} - \frac{\bf p_\perp^2}{x^2 M^2} \mathcal{T}_{22}^{\nu}-\frac{2 m}{x M} \mathcal{T}_{12}^{\nu} \bigg],  \label{hts}\\
% \end{eqnarray}
% % Twist-4 Scalar
% \begin{eqnarray}
%f3s
%
x^2 f_{3}^{\nu(S)}(x, {\bf p_\perp^2}) &=&  \frac{C_{S}^{2} N_s^2}{16 \pi^3} \bigg(\frac{{\bf p_\perp^2}+m^2}{M^2}\bigg) \Bigg[ \mathcal{T}_{11}^{\nu} + \frac{\bf p_\perp^2}{x^2 M^2} \mathcal{T}_{22}^{\nu}\Bigg], \label{eef3s} \\
%
%g3ls
%
x^2 g_{3L}^{\nu(S)}(x, {\bf p_\perp^2}) &=&  \frac{C_{S}^{2} N_s^2}{32 \pi^3 M^2} \Bigg[\big({{\bf p_\perp^2}-m^2}\big)\bigg[ \mathcal{T}_{11}^{\nu} - \frac{\bf p_\perp^2}{x^2 M^2} \mathcal{T}_{22}^{\nu}\bigg]-\frac{4m {\bf p_\perp^2}}{x M} \mathcal{T}_{12}^{\nu} \Bigg],  \label{eeg3ls} \\
%
%g3ts
%
x^2 g_{3T}^{\nu(S)}(x, {\bf p_\perp^2}) &=&  \frac{C_{S}^{2} N_s^2}{8 \pi^3 M} \Bigg[m\bigg[ \mathcal{T}_{11}^{\nu} - \frac{\bf p_\perp^2}{x^2 M^2} \mathcal{T}_{22}^{\nu}\bigg]+\frac{\big({{\bf p_\perp^2}-m^2}\big)}{x M} \mathcal{T}_{12}^{\nu} \Bigg], \label{eeg3ts1} \\
%
%h3lps
%
x^2 h_{3L}^{\perp\nu(S)}(x, {\bf p_\perp^2}) &=&  \frac{C_{S}^{2} N_s^2}{16 \pi^3 M} \Bigg[m\bigg[ \mathcal{T}_{11}^{\nu} - \frac{\bf p_\perp^2}{x^2 M^2} \mathcal{T}_{22}^{\nu}\bigg]+\frac{\big({\bf p_\perp^2}-m^2\big)}{x M} \mathcal{T}_{12}^{\nu} \Bigg], \label{eeh3lps1}\\ 
%
%\ee
%\be
% \nonumber % Remove numbering (before each equation)
%
%h3ts
%
x^2 h_{3T}^{\nu(S)}(x, {\bf p_\perp^2})&=&  \frac{C_{S}^{2} N_s^2}{16 \pi^3} \bigg(\frac{{\bf p_\perp^2}+m^2}{M^2}\bigg) \Bigg[ \mathcal{T}_{11}^{\nu} + \frac{\bf p_\perp^2}{x^2 M^2} \mathcal{T}_{22}^{\nu}\Bigg], \label{eeh3ts} \\
%h3tps
%
x^2 h_{3T}^{\perp \nu(S)}(x, {\bf p_\perp^2}) &=&  -\frac{C_{S}^{2} N_s^2}{8 \pi^3} \Bigg[ \mathcal{T}_{11}^{\nu} + \frac{m^2}{x^2 M^2} \mathcal{T}_{22}^{\nu}-\frac{2m}{x M} \mathcal{T}_{12}^{\nu} \Bigg], \label{eeh3tps}\\
x^2 h_{3}^{ \nu(S)}(x, {\bf p_\perp^2}) &=&  \frac{C_{S}^{2}N_s^2}{16 \pi^3 M^2} \Bigg[m^2 \mathcal{T}_{11}^{\nu} + \frac{{\bf p_\perp^4}}{x^2 M^2} \mathcal{T}_{22}^{\nu}+\frac{2m{\bf p_\perp^2}}{x M} \mathcal{T}_{12}^{\nu} \Bigg]. \label{eeh3ps1}
\end{eqnarray}
$\mathcal{T}_{ij}^{\nu}$ in above equations is defined for convenience as
\begin{eqnarray}
\mathcal{T}_{ij}^{(\nu)}(x,\bfp)&=&\varphi_i^{(\nu) \dagger}(x,\bfp) \varphi_j^{(\nu)}(x,\bfp)
\label{Tij1},
\end{eqnarray}
where, $i,j=1,2$. As a direct consequence of Eq. \eqref{LFWF_phi} and Eq. \eqref{Tij1}, we get
\begin{eqnarray}
\mathcal{T}_{ij}^{(\nu)}(x,\bfp)&=&\mathcal{T}_{ji}^{(\nu)}(x,\bfp)\label{Tij2},\\
\varphi_i^{(\nu)\dagger}(x,\bfp)&=&\varphi_i^{(\nu)}(x,\bfp)\label{Tij3}.
\end{eqnarray}
Similarly, by substituting the expressions of LFWFs from Table \ref{tab_LFWF} into the vector diquark overlap Eqs. {\eqref{oef1v1}-\eqref{oeh3v1}, we have obtained the explicit expressions of T-even TMDs for vector diquark  as
\begin{eqnarray}
	% \nonumber % Remove numbering (before each equation)
	%
	%f1v
	%
	f_{1}^{\nu(A)}(x, {\bf p_\perp^2}) &=& \frac{C_{A}^{2}}{16 \pi^3}  \bigg(\frac{1}{3} |N_0^\nu|^2+\frac{2}{3}|N_1^\nu|^2 \bigg) \Bigg[ \mathcal{T}_{11}^{\nu} + \frac{\bf p_\perp^2}{x^2 M^2} \mathcal{T}_{22}^{\nu}\Bigg], \label{eef1v}\\
	%
	%g1lv
	%
	g_{1L}^{\nu(A)}(x, {\bf p_\perp^2}) &=& \frac{C_{A}^{2}}{32 \pi^3}  \bigg(\frac{1}{3} |N_0^\nu|^2-\frac{2}{3}|N_1^\nu|^2 \bigg) \Bigg[ \mathcal{T}_{11}^{\nu} - \frac{\bf p_\perp^2}{x^2 M^2} \mathcal{T}_{22}^{\nu} \Bigg],  \label{eeg1lv} \\
	%
	%g1tv
	%
	g_{1T}^{\nu(A)}(x, {\bf p_\perp^2}) &=&  -\frac{C_{A}^{2}}{8 \pi^3}\bigg(\frac{1}{3} |N_0^\nu|^2\bigg) \Bigg[\frac{1}{x} \mathcal{T}_{12}^{\nu} \Bigg],  \label{eeg3tv}
 % \\
	%
   \end{eqnarray}
\begin{eqnarray}
	%h1lpv
	%
	h_{1L}^{\perp\nu(A)}(x, {\bf p_\perp^2}) &=& \frac{C_{A}^{2}}{16 \pi^3}  \bigg(\frac{1}{3} |N_0^\nu|^2-\frac{2}{3}|N_1^\nu|^2 \bigg) \Bigg[-\frac{M}{x M} \mathcal{T}_{12}^{\nu} \Bigg], \label{eeh3lpv}\\
	%
	% \nonumber % Remove numbering (before each equation)
	%
	%h1tv
	%
	h_{1T}^{\nu(A)}(x, {\bf p_\perp^2})&=&  -\frac{C_{A}^{2}}{16 \pi^3}\bigg(\frac{1}{3} |N_0^\nu|^2\bigg) \Bigg[ \mathcal{T}_{11}^{\nu} + \frac{\bf p_\perp^2}{x^2 M^2} \mathcal{T}_{22}^{\nu}\Bigg], \label{eeh1tv} \\
	%\\
	%\ee
	%\be
	%h1tpv
	%
	h_{1T}^{\perp \nu(A)}(x, {\bf p_\perp^2}) &=&  \frac{C_{A}^{2}}{8 \pi^3}\bigg(\frac{1}{3} |N_0^\nu|^2\bigg) \Bigg[ \frac{1}{x^2} \mathcal{T}_{22}^{\nu}\Bigg], \label{eeh1tpv}\\
	h_{1}^{ \nu(A)}(x, {\bf p_\perp^2}) &=&  -\frac{C_{A}^{2}}{16 \pi^3}\bigg(\frac{1}{3} |N_0^\nu|^2\bigg) \Bigg[ \mathcal{T}_{11}^{\nu}  \Bigg],  \label{eeh1v}\\
	%
	%
% \end{eqnarray}
% % 
% \begin{eqnarray}
	% e
	x e^{\nu (A)}(x, {\bf p_\perp^2}) &=& \frac{{C_{A}^{2}}}{16\pi^3}\bigg(\frac{2}{3}|N_1^\nu|^2 + \frac{1}{3} |N_0^\nu|^2\bigg) \frac{m}{M} \bigg[ \mathcal{T}_{11}^{\nu} + \frac{\bf p_\perp^2}{x^2 M^2} \mathcal{T}_{22}^{\nu}\bigg], \label{ef} \\
	% fperp
	x f^{\perp\nu (A)}(x, {\bf p_\perp^2}) &=&  \frac{{C_{A}^{2}}}{16 \pi^3} \bigg(\frac{2}{3}|N_1^\nu|^2 + \frac{1}{3} |N_0^\nu|^2\bigg) \bigg[ \mathcal{T}_{11}^{\nu} + \frac{\bf p_\perp^2}{x^2 M^2} \mathcal{T}_{22}^{\nu}\bigg], \label{fperpf1}\\
	% gLperp
	x g_{L}^{\perp\nu (A)}(x, {\bf p_\perp^2}) &=&  \frac{{C_{A}^{2}}}{32 \pi^3}\bigg(-\frac{2}{3}|N_1^\nu|^2 + \frac{1}{3} |N_0^\nu|^2\bigg) \bigg[ \mathcal{T}_{11}^{\nu} - \frac{\bf p_\perp^2}{x^2 M^2} \mathcal{T}_{22}^{\nu}-\frac{2 m}{x M} \mathcal{T}_{12}^{\nu} \bigg], \label{glperpv}\\
	% gT
	x g_{T}^{'\nu (A)}(x, {\bf p_\perp^2}) &=&  \frac{C_{A}^{2}}{16 \pi^3}\bigg(  -\frac{1}{3} {{} |N_0^\nu|^2}\bigg)\frac{m}{M} \bigg[ \mathcal{T}_{11}^{\nu} + \frac{\bf p_\perp^2}{x^2 M^2} \mathcal{T}_{22}^{\nu}\bigg], \label{gtv}\\
	% gTperp
	x g_{T}^{\perp\nu (A)}(x, {\bf p_\perp^2}) &=&  \frac{C_{A}^{2}}{8 \pi^3} \bigg( -\frac{1}{3} {{} |N_0^\nu|^2})\bigg){M} \bigg[ \frac{1}{xM} \mathcal{T}_{12}^{\nu}- \frac{m}{x^2 M^2} \mathcal{T}_{22}^{\nu} \bigg],\label{gtperpv}\\
	x g_{T}^{\nu (A)}(x, {\bf p_\perp^2}) &=& \frac{C_{A}^{2}}{16 \pi^3} \bigg( -\frac{1}{3} {{} |N_0^\nu|^2})\bigg)  \frac{1}{M} \bigg[m~\mathcal{T}_{11}^{\nu} +\frac{\bf p_\perp^2}{x M} \mathcal{T}_{12}^{\nu} \bigg], \label{gtvf} \\
	%	\end{eqnarray}
%	\begin{eqnarray}
	% hTperp
	x h_{T}^{\perp\nu (A)}(x, {\bf p_\perp^2}) &=&  \frac{{C_{A}^{2}}}{16 \pi^3}
	\bigg(  -\frac{1}{3} { |N_0^\nu|^2}\bigg) \bigg[ \mathcal{T}_{11}^{\nu} + \frac{\bf p_\perp^2}{x^2 M^2} \mathcal{T}_{22}^{\nu}\bigg], \label{htperpf1}\\
	% hL
	x h_{L}^{\nu (A)}(x, {\bf p_\perp^2}) &=& \frac{C_{A}^{2}}{16 \pi^3} \bigg(-\frac{2}{3}|N_1^\nu|^2 + \frac{1}{3} |N_0^\nu|^2\bigg) \frac{1}{M} \bigg[m\bigg( \mathcal{T}_{11}^{\nu} - \frac{\bf p_\perp^2}{x^2 M^2} \mathcal{T}_{22}^{\nu}\bigg) \nonumber \\
	&&+\frac{2 p_\perp^2}{x M} \mathcal{T}_{12}^{\nu} \bigg], \label{hlf2} \\
	% hT
	x h_{T}^{\nu (A)}(x, {\bf p_\perp^2}) &=&  \frac{{C_{A}^{2}}}{8 \pi^3} \bigg( +\frac{1}{3} { |N_0^\nu|^2} \bigg) \bigg[ \mathcal{T}_{11}^{\nu} - \frac{\bf p_\perp^2}{x^2 M^2} \mathcal{T}_{22}^{\nu}-\frac{2 m}{x M} \mathcal{T}_{12}^{\nu} \bigg],  \label{htv}\\
	% \nonumber % Remove numbering (before each equation)
	%
	%f3v
	%
	x^2 f_{3}^{\nu(A)}(x, {\bf p_\perp^2}) &=& \frac{C_{A}^{2}}{16 \pi^3}  \bigg(\frac{1}{3} |N_0^\nu|^2+\frac{2}{3}|N_1^\nu|^2 \bigg)\bigg(\frac{{\bf p_\perp^2}+m^2}{M^2}\bigg) \Bigg[ \mathcal{T}_{11}^{\nu} + \frac{\bf p_\perp^2}{x^2 M^2} \mathcal{T}_{22}^{\nu}\Bigg], \label{eef3v1}
 \\
	%
%   \end{eqnarray}
% % 
% \begin{eqnarray}
	%g3lv
	%
	x^2 g_{3L}^{\nu(A)}(x, {\bf p_\perp^2}) &=& \frac{C_{A}^{2}}{32 \pi^3 M^2}  \bigg(\frac{1}{3} |N_0^\nu|^2-\frac{2}{3}|N_1^\nu|^2 \bigg) \Bigg[\big({{\bf p_\perp^2}-m^2}\big)\bigg[ \mathcal{T}_{11}^{\nu} - \frac{\bf p_\perp^2}{x^2 M^2} \mathcal{T}_{22}^{\nu}\bigg]\nonumber \\
	&&-\frac{4m {\bf p_\perp^2}}{x M} \mathcal{T}_{12}^{\nu} \Bigg],  \label{eeg3lv}
 % \\
	%
   \end{eqnarray}
\begin{eqnarray}
	%g3tv
	%
	x^2 g_{3T}^{\nu(A)}(x, {\bf p_\perp^2}) &=&  -\frac{C_{A}^{2}}{8 \pi^3 M}\bigg(\frac{1}{3} |N_0^\nu|^2\bigg) \Bigg[m\bigg[ \mathcal{T}_{11}^{\nu} - \frac{\bf p_\perp^2}{x^2 M^2} \mathcal{T}_{22}^{\nu}\bigg]+\frac{\big({{\bf p_\perp^2}-m^2}\big)}{x M} \mathcal{T}_{12}^{\nu} \Bigg],  \label{eeg3tv1} \\
	%
	%h3lpv
	%
	x^2 h_{3L}^{\perp\nu(A)}(x, {\bf p_\perp^2}) &=& \frac{C_{A}^{2}}{16 \pi^3 M}  \bigg(\frac{1}{3} |N_0^\nu|^2-\frac{2}{3}|N_1^\nu|^2 \bigg) \Bigg[m\bigg[ \mathcal{T}_{11}^{\nu} - \frac{\bf p_\perp^2}{x^2 M^2} \mathcal{T}_{22}^{\nu}\bigg]\nonumber\\ &&+\frac{\big({{\bf p_\perp^2}-m^2}\big)}{x M} \mathcal{T}_{12}^{\nu} \Bigg], \label{eeh3lpv1}\\
	%
	% \nonumber % Remove numbering (before each equation)
	%
	%h3tv
	%
	x^2 h_{3T}^{\nu(A)}(x, {\bf p_\perp^2})&=&  -\frac{C_{A}^{2}}{16 \pi^3}\bigg(\frac{1}{3} |N_0^\nu|^2\bigg) \bigg(\frac{{\bf p_\perp^2}+m^2}{M^2}\bigg) \Bigg[ \mathcal{T}_{11}^{\nu} + \frac{\bf p_\perp^2}{x^2 M^2} \mathcal{T}_{22}^{\nu}\Bigg], \label{eeh3tv1} \\
	%\\
	%\ee
	%\be
	%h3tpv
	%
	x^2 h_{3T}^{\perp \nu(A)}(x, {\bf p_\perp^2}) &=&  \frac{C_{A}^{2}}{8 \pi^3}\bigg(\frac{1}{3} |N_0^\nu|^2\bigg) \Bigg[ \mathcal{T}_{11}^{\nu} + \frac{m^2}{x^2 M^2} \mathcal{T}_{22}^{\nu}-\frac{2m}{x M} \mathcal{T}_{12}^{\nu} \Bigg], \label{eeh3tpv}\\
	%
	%
	%\ee
	%\be
	%h3tpv
	%
	x^2 h_{3}^{ \nu(A)}(x, {\bf p_\perp^2}) &=&  \frac{-C_{A}^{2}}{16 \pi^3 M^2}\bigg(\frac{1}{3} |N_0^\nu|^2\bigg) \Bigg[m^2 \mathcal{T}_{11}^{\nu} + \frac{{\bf p_\perp^4}}{x^2 M^2} \mathcal{T}_{22}^{\nu}+\frac{2m{\bf p_\perp^2}}{x M} \mathcal{T}_{12}^{\nu} \Bigg].  \label{eeh3v}
\end{eqnarray}
%
	% $ \mathcal{F}\enspace  \mathbcal{F}$
	% $ \mathcal{F}\enspace $
	% %
	% %
	% $ \mathcal{T}$
	% $ \mathcal{T}\enspace$
	% $  \mathbcal{T}$
	% $ \mathcal{T}\enspace $
	%
	% $ \ltimes \varpi  \varsigma \varrho$ \\
	%  $\merge$
		\section{Relations among TMDs}
\label{sec_rel}
To check the consistency of model results, equation of motion results are important \cite{Sharma:2023wha, Avakian:2010br}. Other than that, model relations have also been explored in the literature
\cite{Maji:2015vsa,Sharma:2023wha,Sharma:2023fbb,Sharma:2022ylk,Wei:2016far,Liu:2021ype}. We have taken this opportunity to explore the model-dependent relations at all twist levels. For ease of calculation, analysis and comparison purposes we have given the relations of scalar and vector diquark separately.  
\subsection{Scalar Diquark Relations}
\label{sec_rels}

\subsubsection{Linear Relations of Scalar Diquark}
\label{sec_lsrels}
From scalar diquark explicit Eqs. \eqref{eef1s}-\eqref{eeh3ps1}, the linear relations for the scalar diquark can be established as
% $\mho$
% $\Game$
% $\vartheta$
\begin{eqnarray}
	% 
	% alpha scalar
	% 
	\mho_{S} (say)&=& f_1 = h_{1T} =(\frac{M}{m})~x~e=x~f^\perp=(\frac{M}{m})~x~g_T'\nonumber\\
	&=&x~h_{T}^\perp
	=\big(\frac{M^2}{{\bf p_\perp^2}+m^2}\big)x^2~f_3=(\frac{M^2}{{\bf p_\perp^2}+m^2}\big)~x^2~h_{3T},\\
	%
	% 
	%	% alpha vector 1
	%	% 
	%	\mho_{V_{1}} (say)&=& f_1 =(\frac{M}{m})~x~e=x~f^\perp=\big(\frac{M^2}{{\bf p_\perp^2}+m^2}\big)x^2~f_3 ,\\
	%	%
	%			% 
	%	% alpha vector 2
	%	% 
	%	\mho_{V_{2}} (say)&=&  h_{1T} =(\frac{M}{m})~x~g_T'=x~h_{T}^\perp
	%	=(\frac{M^2}{{\bf p_\perp^2}+m^2}\big)~x^2~h_{3T},\\
	%	
	%	Beta Scalar
	%	
	\varrho_{S} (say)&=& g_{1T} =-2 h_{1L}^\perp,\\
	%	
	%%	Beta Vector
	%%	
	%	\varrho_{V} (say)&=& g_{1T} =2 \bigg(\frac{|N_0^\nu|^2}{|N_0^\nu|^2-{2}|N_1^\nu|^2}\bigg) h_{1L}^\perp,\\
	%	
	%	
	%	Gamma Scalar
	%	
	\varsigma_{S} (say)&=& 4~x~g_{L}^\perp =-{x~h_{T}},\\
	%	
	%	
	%	%   Gamma Vector
	%	%	
	%	\varsigma_{V} (say)&=& 4\bigg(\frac{|N_0^\nu|^2}{|N_0^\nu|^2-{2}|N_1^\nu|^2}\bigg)x~g_{L}^\perp ={x~h_{T}},\\
	%	
	%	
	%	Circle Scalar
	%			
	\varpi_{S} (say)&=& x^2~g_{3T} =2~x^2~ h_{3L}^\perp,\\
	%
	%	
	%	
	%	%	Circle Vector
	%	%			
	%	\varpi_{V} (say)&=& x^2~g_{3T} =-2\bigg(\frac{|N_0^\nu|^2}{|N_0^\nu|^2-{2}|N_1^\nu|^2}\bigg)x^2~ h_{3L}^\perp,\\
	%
	%	Simba Scalar
	%	
	\mho_{S}&=& h_1 - \frac{\bf p_\perp^2}{2M^2}~h_{1T}^\perp,
 % \\
	%
	%%	Simba Vector
	%%	
	%		\mho_{V}&=& h_1 - \frac{\bf p_\perp^2}{2M^2}~h_{1T}^\perp,\\
	%%
  \end{eqnarray}
\begin{eqnarray}
	%	Pocco Scalar
	%	
	\mho_{S}&=& 2 h_{1}-2 g_{1L},\\
	%
	%%	Pocco Vector
	%%	
	%	\mho_{V}&=& 2 h_{1}-2 g_{1L},\\
	%	
	% Kayo Scalar
	%
	2 g_{1L} &=& h_1 + \frac{\bf p_\perp^2}{2M^2}~h_{1T}^\perp,\\
	%	
	%% Kayo Vector
	%%
	%	-2 g_{1L}\bigg(\frac{|N_0^\nu|^2}{|N_0^\nu|^2-{2}|N_1^\nu|^2}\bigg) &=& h_1 + \frac{\bf p_\perp^2}{2M^2}~h_{1T}^\perp,\\
	%	
	% Nana Scalar only
	%
	\mho_{S}&=& 2 g_{1L} - \frac{\bf p_\perp^2}{M^2}~h_{1T}^\perp,\\
	%
	% S
	%
	\varsigma_{S}&=& 4 g_{1L}-\frac{2m}{M} \varrho_{S},\\
	%
	%% 	V
	%%
	%\varsigma_{V}&=& 4\bigg(\frac{|N_0^\nu|^2}{|N_0^\nu|^2-{2}|N_1^\nu|^2}\bigg) g_{1L}+\frac{2m}{M} \varrho_{V},\\
	%
	% 	S
	%
	x~g_{T}^\perp &=& \varrho_{S}+ \frac{m}{M} h_{1T}^\perp,\\
	%
	%
	%%	V
	%%
	%x~g_{T}^\perp &=& \varrho_{V}+ \frac{m}{M} h_{1T}^\perp,\\
	%	
	%		S
	%	
	x~g_{T}&=& \frac{m}{M}~h_{1}+\frac{\bf p_\perp^2}{2M^2} \varrho_{S},	\\
	%	
	%%		V	
	%%
	%x~g_{T}&=& \frac{m}{M}~h_{1}+\frac{\bf p_\perp^2}{2M^2} \varrho_{V},
	%	%	\\
	%	
	%		\end{eqnarray}	
%	\begin{eqnarray}
	%		
	%	S
	%
	x~h_{L}&=& \frac{2m}{M}~g_{1L}+\frac{\bf p_\perp^2}{M^2} \varrho_{S},\\
	%
	%	%	V	
	%	%
	%x~h_{L}&=& \frac{2m}{M}~g_{1L}-\frac{\bf p_\perp^2}{M^2}\bigg(\frac{|N_0^\nu|^2}{|N_0^\nu|^2-{2}|N_1^\nu|^2}\bigg) \varrho_{V},\\
	%	
	%	S
	%
	x^2~g_{3L}&=& \big(\frac{{\bf p_\perp^2}-m^2}{M^2}\big) g_{1L}  -\big(\frac{m p_\perp^2}{M^3}\big) \varrho_{S},\\
	%
	%	
	%	%	V
	%	%
	%x^2~g_{3L}&=& \big(\frac{{\bf p_\perp^2}-m^2}{M^2}\big) g_{1L}  -\big(\frac{m p_\perp^2}{M^3}\big)\bigg(\frac{|N_0^\nu|^2}{|N_0^\nu|^2-{2}|N_1^\nu|^2}\bigg) \varrho_{V},\\
	%%
	%		S
	%
	%
	\varpi_{S} &=& \big(\frac{{\bf p_\perp^2}-m^2}{M^2}\big) \varrho_{S} +\frac{4 m}{M}~g_{1L},\\
	%
	%%		V
	%%
	%%	
	%\varpi_{V} &=& \big(\frac{{\bf p_\perp^2}-m^2}{M^2}\big) \varrho_{V} -\frac{4 m}{M}\bigg(\frac{|N_0^\nu|^2}{|N_0^\nu|^2-{2}|N_1^\nu|^2}\bigg)g_{1L},\\
	%
	%	S
	%
	x^2 h_{3T}^{\perp}&=& -2 h_1 + \frac{m^2}{M^2} h_{1T}^\perp +\frac{2m}{M} \varrho_{S},\\
	%
	%%	V
	%%	
	%%
	%x^2 h_{3T}^{\perp}&=& -2 h_1 + \frac{m^2}{M^2} h_{1T}^\perp +\frac{2m}{M} \varrho_{V},\\
	%
	%
	x^2~h_{3}&=& \big(\frac{m^2}{M^2}\big) h_{1}-\big(\frac{\bf p_\perp^4}{2 M^4}\big) h_{1T}^\perp  +\big(\frac{m p_\perp^2}{M^3}\big) \varrho_{S}.
	%	\\
	%	
	%
	%%
	%x^2~h_{3}&=& \big(\frac{m^2}{M^2}\big) h_{1}-\big(\frac{\bf p_\perp^4}{2 M^4}\big) h_{1T}^\perp  +\big(\frac{m p_\perp^2}{M^3}\big) \varrho_{V}.
\end{eqnarray}
It is worthwhile to mention that it is obvious that the above relations correspond to the scalar diquark only, so we have omitted the superscript $\nu(S)$ for simplification. Similarly, in the proceeding work, we have chosen to do so wherever required.
\subsubsection{Quadratic Relations of Scalar Diquark}
\label{sec_lsre2s}
Similarly, the scalar diquark's quadratic relations can be established using the scalar diquark explicit Eqs. \eqref{eef1s}-\eqref{eeh3ps1} as
\begin{eqnarray}
	%	
	%		S
	%
	\mho_{S}^2 &=& h_1^2 +\frac{\bf p_\perp^4}{4M^4}  h_{1T}^{\perp~2}+\frac{\bf p_\perp^2}{2M^2}\varrho_{S}^2,\\
	%	
	%%		V
	%%
	%\mho_{V}^2 &=& h_1^2 +\frac{\bf p_\perp^4}{4M^4}  h_{1T}^{\perp~2}+\frac{\bf p_\perp^2}{2M^2}\varrho_{V}^2,\\
	%%
	%		S
	%
	\varrho_{S}^2 &=& -2 h_1 h_{1T}^{\perp},\\
	%	
	%
	%%		V
	%%
	%\varrho_{V}^2 &=& -2 h_1 h_{1T}^{\perp}\\
	%	
	%		S
	%
	4 g_{1L}^{2} &=& h_1^2 +\frac{\bf p_\perp^4}{4M^4}  h_{1T}^{\perp~2}-\frac{\bf p_\perp^2}{2M^2}\varrho_{S}^2,\\
	%		
	%%		V
	%%
	%4 g_{1L}^{2} &=& \bigg(\frac{|N_0^\nu|^2-{2}|N_1^\nu|^2}{|N_0^\nu|^2}\bigg)^2 \bigg(h_1^2 +\frac{\bf p_\perp^4}{4M^4}  h_{1T}^{\perp~2}-\frac{\bf p_\perp^2}{2M^2}\varrho_{V}^2 \bigg),\\
	%	
	%		S
	%
	\varsigma_{S}^2 &=& 4 \bigg(h_1^2 +\frac{\bf p_\perp^4}{4M^4}  h_{1T}^{\perp~2} 
	+(\frac{2m^2-p_\perp^2}{2M^2})\varrho_{S}^2
	-(\frac{m p_\perp^2}{M^3} h_{1T}^{\perp}
	+\frac{2m}{M} h_{1} )\varrho_{S}\bigg),
 % \\
	%	
	%%		V
	%%
	%\varsigma_{V}^2 &=& 4 \bigg(h_1^2 +\frac{\bf p_\perp^4}{4M^4}  h_{1T}^{\perp~2} 
	%	+(\frac{2m^2-p_\perp^2}{2M^2})\varrho_{V}^2
	%-(\frac{m p_\perp^2}{M^3} h_{1T}^{\perp}
	%+\frac{2m}{M} h_{1} )\varrho_{V}\bigg),\\
	%
   \end{eqnarray}
\begin{eqnarray}
	%		S
	%	
	(xh_{L})^2&=& \frac{m^2}{M^2}(h_1^2 +\frac{\bf p_\perp^4}{4M^4}  h_{1T}^{\perp~2}-\frac{\bf p_\perp^2}{2M^2}\varrho_{S}^2)+\frac{\bf p_\perp^4}{M^4} \varrho_{S}^2
	+\frac{2 m p_\perp^2 \varrho_{S}}{M^3}(
	h_1 + \frac{\bf p_\perp^2}{2M^2}~h_{1T}^\perp)
	,\nonumber\\ 
 \\
	%
	%		V
	%%	
	%(xh_{L})^2&=& \bigg(\frac{|N_0^\nu|^2-{2}|N_1^\nu|^2}{|N_0^\nu|^2}\bigg)^2\bigg(\frac{m^2}{M^2}(h_1^2 +\frac{\bf p_\perp^4}{4M^4}  h_{1T}^{\perp~2}-\frac{\bf p_\perp^2}{2M^2}\varrho_{V}^2)+\frac{\bf p_\perp^4}{M^4} \varrho_{V}^2
	%\nonumber\\
	%&&+\frac{2 m p_\perp^2 \varrho_{V}}{M^3}(
	%h_1 + \frac{\bf p_\perp^2}{2M^2}~h_{1T}^\perp)\bigg)
	%, \\
	%
	%		S
	%
	(x^2h_{3})^2&=& (\frac{m^2}{M^2}h_1)^2 +(\frac{\bf p_\perp^4}{2M^4} h_{1T}^{\perp})^2+\frac{3}{2}(\frac{m p_\perp^2}{M^3}\varrho_{S})^2 -(\frac{m p_\perp^6 }{M^7})\varrho_{S} h_{1T}^\perp  \nonumber\\&&+ (\frac{2 m^3 p_\perp^2}{M^5}) \varrho_{S} h_{1}
	, 
  \\
	%	
%    \end{eqnarray}
% % 
% \begin{eqnarray}
	%%		V
	%%
	%(x^2h_{3})^2&=& (\frac{m^2}{M^2}h_1)^2 +(\frac{\bf p_\perp^4}{2M^4} h_{1T}^{\perp})^2+\frac{3}{2}(\frac{m p_\perp^2}{M^3}\varrho_{V})^2
	%-(\frac{m p_\perp^6 }{M^7})\varrho_{V} h_{1T}^\perp  \nonumber\\&&+ (\frac{2 m^3 p_\perp^2}{M^5}) \varrho_{V} h_{1}
	%, \\
	%
	%		S
	%
	(x^2 h_{3T}^{\perp})^2&=& 4 \bigg( h_1^2 +(\frac{m^2}{2M^2} h_{1T}^{\perp})^2+\frac{3}{2}(\frac{m}{M}\varrho_{S})^2+(\frac{m }{M})^3 \varrho_{S} h_{1T}^{\perp}
	- (\frac{2 m}{M}) \varrho_{S} h_{1}\bigg)
	, \\
	%
	%%		V
	%%
	%(x^2 h_{3T}^{\perp})^2&=& 4 \bigg( h_1^2 +(\frac{m^2}{2M^2} h_{1T}^{\perp})^2+\frac{3}{2}(\frac{m}{M}\varrho_{V})^2+(\frac{m }{M})^3 \varrho_{V} h_{1T}^{\perp}
	%- (\frac{2 m}{M}) \varrho_{V} h_{1}\bigg)
	%, 
	%%\\
	%%
	%\end{eqnarray}
	%\begin{eqnarray}
	%		S
	%
	({x^2 g_{3L}})^2&=&  \bigg((\frac{{\bf p_\perp^2}-m^2}{2M^2}) \bigg(h_1^2 +\frac{\bf p_\perp^4}{4M^4}  h_{1T}^{\perp~2} 
	-\frac{\bf p_\perp^2}{2M^2}\varrho_{S}^2\bigg)
	+(\frac{m p_\perp^2}{M^3}\varrho_{S})^2
	\nonumber\\
	&&-\frac{m p_\perp^2}{M^5}({{\bf p_\perp^2}-m^2}) \varrho_{S}(
	h_1 + \frac{\bf p_\perp^2}{2M^2}~h_{1T}^\perp)
	\bigg)
	, \\
	%
	%
	%%		V
	%%
	%%
	%({x^2 g_{3L}})^2&=&  \bigg(\frac{|N_0^\nu|^2-{2}|N_1^\nu|^2}{|N_0^\nu|^2}\bigg)^2\bigg((\frac{{\bf p_\perp^2}-m^2}{2M^2}) \bigg(h_1^2 +\frac{\bf p_\perp^4}{4M^4}  h_{1T}^{\perp~2} 
	%-\frac{\bf p_\perp^2}{2M^2}\varrho_{V}^2\bigg)
	%+(\frac{m p_\perp^2}{M^3}\varrho_{V})^2
	%\nonumber\\
	%&&-\frac{m p_\perp^2}{M^5}({{\bf p_\perp^2}-m^2}) \varrho_{V}(
	%h_1 + \frac{\bf p_\perp^2}{2M^2}~h_{1T}^\perp)
	%\bigg)
	%, \\
	%%
	%	
	%		S
	%
	\varpi_{S}^2 &=& \frac{4m^2}{M^2} \bigg(h_1^2 +\frac{\bf p_\perp^4}{4M^4}  h_{1T}^{\perp~2} 
	-\frac{\bf p_\perp^2}{2M^2}\varrho_{S}^2\bigg)+(\frac{{\bf p_\perp^2}-m^2}{M^2})^2 \varrho_{S}^2
	\nonumber\\
	&&+\frac{4m}{M^3}({{\bf p_\perp^2}-m^2}) \varrho_{S}(
	h_1 + \frac{\bf p_\perp^2}{2M^2}~h_{1T}^\perp).
	%\\
	%	
	%%		V
	%%
	%\varpi_{V}^2 &=& \frac{4m^2}{M^2} \bigg(h_1^2 +\frac{\bf p_\perp^4}{4M^4}  h_{1T}^{\perp~2} 
	%-\frac{\bf p_\perp^2}{2M^2}\varrho_{V}^2\bigg)+(\frac{{\bf p_\perp^2}-m^2}{M^2})^2 \varrho_{V}^2
	%\nonumber\\
	%&&+\frac{4m}{M^3}({{\bf p_\perp^2}-m^2}) \varrho_{V}(
	%h_1 + \frac{\bf p_\perp^2}{2M^2}~h_{1T}^\perp),
	%%\\
	%
\end{eqnarray}

\subsubsection{Inequality Relations for Scalar Diquark}
\label{sec_lsre3s}
It is important to check the inta-twist and inter-twist inequalities existing between TMDs \cite{Maji:2015vsa}. The inequality relations obtained from scalar diquark explicit Eqs. \eqref{eef1s}-\eqref{eeh3ps1} can be written as
\begin{eqnarray}
	% 
	% alpha scalar
	% 
	\mho_{S} \ge 0 \label{kai1}\\
	\varrho_{S} \ge 0,\\
	\varpi_{S} \ge 0,\\
	h_{1T}^\perp \le 0,\\
	h_1 \ge 0 ,\\
	x~g_{T}^\perp \ge 0,\\
	x~g_{T} \ge 0,	\\
	x~h_{L} \ge 0, \\
	%
%    \end{eqnarray}
% \begin{eqnarray}
	%
	%
	x^2 h_{3T}^{\perp} \le 0,\\
	x^2~h_{3} \ge 0 ,
 % \\
	%	\\
	%	
   \end{eqnarray}
\begin{eqnarray}
 | \mho_{S} | \ge 2 |g_{1L}|, \label{kai2}\\
 | \mho_{S} | \ge \frac{\bfp^2}{2M^2} | {h_{1T}^\perp} |, \\
 | \mho_{S} | \ge |h_{1}|, \label{kai3} \\
 |h_{1}| \ge 2 |g_{1L}|, \\
 |\varsigma_{S}| \le 4 |g_{1L}|.
	%%
	%x^2~h_{3}&=& \big(\frac{m^2}{M^2}\big) h_{1}-\big(\frac{\bf p_\perp^4}{2 M^4}\big) h_{1T}^\perp  +\big(\frac{m p_\perp^2}{M^3}\big) \varrho_{V}.
\end{eqnarray}
Eqs. \eqref{kai1}, \eqref{kai2}, and \eqref{kai3} are the generalized form of relations that are found to be valid for all models and QCD \cite{Avakian:2010br,Bacchetta:1999kz}. However, these relations have been established for $f_1$ TMD only instead of $\mho_{S}$ in literature.

\subsection{Vector Diquark Relations}
\label{sec_relv}

\subsubsection{Linear Relations of Vector Diquark}
\label{sec_lsrelv}
The linear relations for the vector diquark can be established using its explicit Eqs. \eqref{eef1v}-\eqref{eeh3tpv}. We have
\begin{eqnarray}
	%	% 
	%	% alpha scalar
	%	% 
	%	\mho_{S} (say)&=& f_1 = h_{1T} =(\frac{M}{m})~x~e=x~f^\perp=(\frac{M}{m})~x~g_T'\nonumber\\
	%	&=&x~h_{T}^\perp
	%	=x^2~\big(\frac{M^2}{{\bf p_\perp^2}+m^2}\big)x^2~f_3=(\frac{M^2}{{\bf p_\perp^2}+m^2}\big)~x^2~h_{3T},\\
	%	%
	% 
	% alpha vector 1
	% 
	\mho_{V_{1}} (say)&=& f_1 =(\frac{M}{m})~x~e=x~f^\perp=\big(\frac{M^2}{{\bf p_\perp^2}+m^2}\big)x^2~f_3 ,\\
	%
	% 
	% alpha vector 2
	% 
	\mho_{V_{2}} (say)&=&  h_{1T} =(\frac{M}{m})~x~g_T'=x~h_{T}^\perp
	=(\frac{M^2}{{\bf p_\perp^2}+m^2}\big)~x^2~h_{3T},\\
	%	
	%	%	Beta Scalar
	%	%	
	%	\varrho_{S} (say)&=& g_{1T} =-2 h_{1L}^\perp,\\
	%	
	%	Beta Vector
	%	
	\varrho_{V} (say)&=& g_{1T} =2 \bigg(\frac{|N_0^\nu|^2}{|N_0^\nu|^2-{2}|N_1^\nu|^2}\bigg) h_{1L}^\perp,\\
	%	
	%	
	%	%	Gamma Scalar
	%	%	
	%	\varsigma_{S} (say)&=& 4~x~g_{L}^\perp =-{x~h_{T}},\\
	%	
	%	
	%   Gamma Vector
	%	
	\varsigma_{V} (say)&=& 4\bigg(\frac{|N_0^\nu|^2}{|N_0^\nu|^2-{2}|N_1^\nu|^2}\bigg)x~g_{L}^\perp ={x~h_{T}}, \\
	%	
	%	
	%	%	Circle Scalar
	%	%			
	%	\varpi_{S} (say)&=& x^2~g_{3T} =2~x^2~ h_{3L}^\perp,\\
	%	%
	%
%    \end{eqnarray}
% % 
% \begin{eqnarray}
	%	
	%	Circle Vector
	%			
	\varpi_{V} (say)&=& x^2~g_{3T} =-2\bigg(\frac{|N_0^\nu|^2}{|N_0^\nu|^2-{2}|N_1^\nu|^2}\bigg)x^2~ h_{3L}^\perp,\\
	%
	%	%	Simba Scalar
	%	%	
	%	\mho_{S}&=& h_1 - \frac{\bf p_\perp^2}{2M^2}~h_{1T}^\perp,\\
	%	%
	%	Simba Vector
	%	
	\mho_{V_2}&=& h_1 - \frac{\bf p_\perp^2}{2M^2}~h_{1T}^\perp,\\
	%
	%	%	Pocco Scalar
	%	%	
	%	\mho_{S}&=& 2 h_{1}-2 g_{1L},\\
	%	%
	%	Pocco Vector
	%	
	\mho_{V_1}&=& 2 h_{1}-2 g_{1L},\\
	%	
	%	% Kayo Scalar
	%	%
	%	2 g_{1L} &=& h_1 + \frac{\bf p_\perp^2}{2M^2}~h_{1T}^\perp,\\
	%	
	% Kayo Vector
	%
	-2 g_{1L}\bigg(\frac{|N_0^\nu|^2}{|N_0^\nu|^2-{2}|N_1^\nu|^2}\bigg) &=& h_1 + \frac{\bf p_\perp^2}{2M^2}~h_{1T}^\perp,
 % \\
	%	
	%	% Nana Scalar only
	%	%
	%	\mho_{S}&=& 2 g_{1L} - \frac{\bf p_\perp^2}{M^2}~h_{1T}^\perp,\\
	%
	%	% S
	%	%
	%	\varsigma_{S}&=& 4 g_{1L}-\frac{2m}{M} \varrho_{S},\\
	%
	% 	V
	%
   \end{eqnarray}
\begin{eqnarray}
	\varsigma_{V}&=& 4\bigg(\frac{|N_0^\nu|^2}{|N_0^\nu|^2-{2}|N_1^\nu|^2}\bigg) g_{1L}+\frac{2m}{M} \varrho_{V},\\
	%
	%	% 	S
	%	%
	%	x~g_{T}^\perp &=& \varrho_{S}+ \frac{m}{M} h_{1T}^\perp,\\
	%	%
	%
	%	V
	%
	x~g_{T}^\perp &=& \varrho_{V}+ \frac{m}{M} h_{1T}^\perp,\\
	%	
	%	%		S
	%	%	
	%	x~g_{T}&=& \frac{m}{M}~h_{1}+\frac{\bf p_\perp^2}{2M^2} \varrho_{S},
	%	\\
	%	
	%		V	
	%
	x~g_{T}&=& \frac{m}{M}~h_{1}+\frac{\bf p_\perp^2}{2M^2} \varrho_{V},	\\
	%	
	%\end{eqnarray}	
	%\begin{eqnarray}
	%	%		
	%	%	S
	%	%
	%	x~h_{L}&=& \frac{2m}{M}~g_{1L}+\frac{\bf p_\perp^2}{M^2} \varrho_{S},\\
	%
	%	V	
	%
	x~h_{L}&=& \frac{2m}{M}~g_{1L}-\frac{\bf p_\perp^2}{M^2}\bigg(\frac{|N_0^\nu|^2-{2}|N_1^\nu|^2}{|N_0^\nu|^2}\bigg) \varrho_{V},\\
	%	
	%	%	S
	%	%
	%	x^2~g_{3L}&=& \big(\frac{{\bf p_\perp^2}-m^2}{M^2}\big) g_{1L}  -\big(\frac{m p_\perp^2}{M^3}\big) \varrho_{S},\\
	%
	%	
	%	V
	%
	x^2~g_{3L}&=& \big(\frac{{\bf p_\perp^2}-m^2}{M^2}\big) g_{1L}  + \big(\frac{m p_\perp^2}{M^3}\big)\bigg(\frac{|N_0^\nu|^2-{2}|N_1^\nu|^2}{|N_0^\nu|^2}\bigg) \varrho_{V},\\
	%
	%	%		S
	%	%
	%	%
	%	\varpi_{S} &=& \big(\frac{{\bf p_\perp^2}-m^2}{M^2}\big) \varrho_{S} +\frac{4 m}{M}~g_{1L},\\
	%
	%		V
	%
	%	
	\varpi_{V} &=& \big(\frac{{\bf p_\perp^2}-m^2}{M^2}\big) \varrho_{V} -\frac{4 m}{M}\bigg(\frac{|N_0^\nu|^2}{|N_0^\nu|^2-{2}|N_1^\nu|^2}\bigg)g_{1L},\\
	%
	%	%	S
	%	%
	%	x^2 h_{3T}^{\perp}&=& -2 h_1 + \frac{m^2}{M^2} h_{1T}^\perp +\frac{2m}{M} \varrho_{S},\\
	%	%
	%	V
	%	
	%
	x^2 h_{3T}^{\perp}&=& -2 h_1 + \frac{m^2}{M^2} h_{1T}^\perp +\frac{2m}{M} \varrho_{V},\\
	%
	%
	%	x^2~h_{3}&=& \big(\frac{m^2}{M^2}\big) h_{1}-\big(\frac{\bf p_\perp^4}{2 M^4}\big) h_{1T}^\perp  +\big(\frac{m p_\perp^2}{M^3}\big) \varrho_{S},\\
	%	%	
	%
	%
	x^2~h_{3}&=& \big(\frac{m^2}{M^2}\big) h_{1}-\big(\frac{\bf p_\perp^4}{2 M^4}\big) h_{1T}^\perp  +\big(\frac{m p_\perp^2}{M^3}\big) \varrho_{V}.
\end{eqnarray}

\subsubsection{Quadratic Relations of Vector Diquark}
\label{sec_lsre2v}
From vector diquark's explicit Eqs. \eqref{eef1v}-\eqref{eeh3tpv}, their quadratic relations can also be established and we have
\begin{eqnarray}
	%	
	%	%		S
	%	%
	%	\mho_{S}^2 &=& h_1^2 +\frac{\bf p_\perp^4}{4M^4}  h_{1T}^{\perp~2}+\frac{\bf p_\perp^2}{2M^2}\varrho_{S}^2,\\
	%	
	%		V
	%
	\mho_{V_2}^2 &=& h_1^2 +\frac{\bf p_\perp^4}{4M^4}  h_{1T}^{\perp~2}+\frac{\bf p_\perp^2}{2M^2}\varrho_{V}^2,\\
	%
	%	%		S
	%	%
	%	\varrho_{S}^2 &=& -2 h_1 h_{1T}^{\perp}\\
	%	%	
	%
	%		V
	%
	\varrho_{V}^2 &=& -2 h_1 h_{1T}^{\perp},\\
	%	
	%	%		S
	%	%
	%	4 g_{1L}^{2} &=& h_1^2 +\frac{\bf p_\perp^4}{4M^4}  h_{1T}^{\perp~2}-\frac{\bf p_\perp^2}{2M^2}\varrho_{S}^2,\\
	%		
	%		V
	%
	4 g_{1L}^{2} &=& \bigg(\frac{|N_0^\nu|^2-{2}|N_1^\nu|^2}{|N_0^\nu|^2}\bigg)^2 \bigg(h_1^2 +\frac{\bf p_\perp^4}{4M^4}  h_{1T}^{\perp~2}-\frac{\bf p_\perp^2}{2M^2}\varrho_{V}^2 \bigg),\\
	%	
	%	%		S
	%	%
	%	\varsigma_{S}^2 &=& 4 \bigg(h_1^2 +\frac{\bf p_\perp^4}{4M^4}  h_{1T}^{\perp~2} 
	%	+(\frac{2m^2-p_\perp^2}{2M^2})\varrho_{S}^2
	%	-(\frac{m p_\perp^2}{M^3} h_{1T}^{\perp}
	%	+\frac{2m}{M} h_{1} )\varrho_{S}\bigg),\\
	%	%	
	%		V
	%
	\varsigma_{V}^2 &=& 4 \bigg(h_1^2 +\frac{\bf p_\perp^4}{4M^4}  h_{1T}^{\perp~2} 
	+(\frac{2m^2-p_\perp^2}{2M^2})\varrho_{V}^2
	-(\frac{m p_\perp^2}{M^3} h_{1T}^{\perp}
	+\frac{2m}{M} h_{1} )\varrho_{V}\bigg),\\
	%
	%	%		S
	%	%	
	%	(xh_{L})^2&=& \frac{m^2}{M^2}(h_1^2 +\frac{\bf p_\perp^4}{4M^4}  h_{1T}^{\perp~2}-\frac{\bf p_\perp^2}{2M^2}\varrho_{S}^2)+\frac{\bf p_\perp^4}{M^4} \varrho_{S}^2
	%	+\frac{2 m p_\perp^2 \varrho_{S}}{M^3}(
	%	h_1 + \frac{\bf p_\perp^2}{2M^2}~h_{1T}^\perp)
	%	,\nonumber\\ \\
	%
	%		V
	%	
	(xh_{L})^2&=& \bigg(\frac{|N_0^\nu|^2-{2}|N_1^\nu|^2}{|N_0^\nu|^2}\bigg)^2\bigg(\frac{m^2}{M^2}(h_1^2 +\frac{\bf p_\perp^4}{4M^4}  h_{1T}^{\perp~2}-\frac{\bf p_\perp^2}{2M^2}\varrho_{V}^2)+\frac{\bf p_\perp^4}{M^4} \varrho_{V}^2
	\nonumber\\
	&&+\frac{2 m p_\perp^2 \varrho_{V}}{M^3}(
	h_1 + \frac{\bf p_\perp^2}{2M^2}~h_{1T}^\perp)\bigg)
	, 
 % \\
	%	
   \end{eqnarray}
\begin{eqnarray}
	%	%		S
	%	%
	%	(x^2h_{3})^2&=& (\frac{m^2}{M^2}h_1)^2 +(\frac{\bf p_\perp^4}{2M^4} h_{1T}^{\perp})^2+\frac{3}{2}(\frac{m p_\perp^2}{M^3}\varrho_{S})^2 -(\frac{m p_\perp^6 }{M^7})\varrho_{S} h_{1T}^\perp  \nonumber\\&&+ (\frac{2 m^3 p_\perp^2}{M^5}) \varrho_{S} h_{1}
	%	, \\
	%	
	%		V
	%
	(x^2h_{3})^2&=& (\frac{m^2}{M^2}h_1)^2 +(\frac{\bf p_\perp^4}{2M^4} h_{1T}^{\perp})^2+\frac{3}{2}(\frac{m p_\perp^2}{M^3}\varrho_{V})^2
	-(\frac{m p_\perp^6 }{M^7})\varrho_{V} h_{1T}^\perp  \nonumber\\&&+ (\frac{2 m^3 p_\perp^2}{M^5}) \varrho_{V} h_{1}
	, \\
	%	%
	%	%		S
	%	%
	%	(x^2 h_{3T}^{\perp})^2&=& 4 \bigg( h_1^2 +(\frac{m^2}{2M^2} h_{1T}^{\perp})^2+\frac{3}{2}(\frac{m}{M}\varrho_{S})^2+(\frac{m }{M})^3 \varrho_{S} h_{1T}^{\perp}
	%	- (\frac{2 m}{M}) \varrho_{S} h_{1}\bigg)
	%	, \\
	%
	%		V
	%
	(x^2 h_{3T}^{\perp})^2&=& 4 \bigg( h_1^2 +(\frac{m^2}{2M^2} h_{1T}^{\perp})^2+\frac{3}{2}(\frac{m}{M}\varrho_{V})^2+(\frac{m }{M})^3 \varrho_{V} h_{1T}^{\perp}
	- (\frac{2 m}{M}) \varrho_{V} h_{1}\bigg)
	, 
	%\\
	%
\end{eqnarray}
\begin{eqnarray}
	%	%		S
	%	%
	%	({x^2 g_{3L}})^2&=&  \bigg(\frac{|N_0^\nu|^2-{2}|N_1^\nu|^2}{|N_0^\nu|^2}\bigg)^2\bigg((\frac{{\bf p_\perp^2}-m^2}{2M^2}) \bigg(h_1^2 +\frac{\bf p_\perp^4}{4M^4}  h_{1T}^{\perp~2} 
	%	-\frac{\bf p_\perp^2}{2M^2}\varrho_{S}^2\bigg)
	%	+(\frac{m p_\perp^2}{M^3}\varrho_{S})^2
	%	\nonumber\\
	%	&&-\frac{m p_\perp^2}{M^5}({{\bf p_\perp^2}-m^2}) \varrho_{S}(
	%	h_1 + \frac{\bf p_\perp^2}{2M^2}~h_{1T}^\perp)
	%	\bigg)
	%	, \\
	%
	%
	%		V
	%
	%
	({x^2 g_{3L}})^2&=&  \bigg(\frac{|N_0^\nu|^2-{2}|N_1^\nu|^2}{|N_0^\nu|^2}\bigg)^2\bigg((\frac{{\bf p_\perp^2}-m^2}{2M^2}) \bigg(h_1^2 +\frac{\bf p_\perp^4}{4M^4}  h_{1T}^{\perp~2} 
	-\frac{\bf p_\perp^2}{2M^2}\varrho_{V}^2\bigg)
	+(\frac{m p_\perp^2}{M^3}\varrho_{V})^2
	\nonumber\\
	&&-\frac{m p_\perp^2}{M^5}({{\bf p_\perp^2}-m^2}) \varrho_{V}(
	h_1 + \frac{\bf p_\perp^2}{2M^2}~h_{1T}^\perp)
	\bigg)
	, \\
	%
	%	
	%	%		S
	%	%
	%	\varpi_{S}^2 &=& \frac{4m^2}{M^2} \bigg(h_1^2 +\frac{\bf p_\perp^4}{4M^4}  h_{1T}^{\perp~2} 
	%	-\frac{\bf p_\perp^2}{2M^2}\varrho_{S}^2\bigg)+(\frac{{\bf p_\perp^2}-m^2}{M^2})^2 \varrho_{S}^2
	%	\nonumber\\
	%	&&+\frac{4m}{M^3}({{\bf p_\perp^2}-m^2}) \varrho_{S}(
	%	h_1 + \frac{\bf p_\perp^2}{2M^2}~h_{1T}^\perp),\\
	%	%	
	%		V
	%
	\varpi_{V}^2 &=& \frac{4m^2}{M^2} \bigg(h_1^2 +\frac{\bf p_\perp^4}{4M^4}  h_{1T}^{\perp~2} 
	-\frac{\bf p_\perp^2}{2M^2}\varrho_{V}^2\bigg)+(\frac{{\bf p_\perp^2}-m^2}{M^2})^2 \varrho_{V}^2
	\nonumber\\
	&&+\frac{4m}{M^3}({{\bf p_\perp^2}-m^2}) \varrho_{V}(
	h_1 + \frac{\bf p_\perp^2}{2M^2}~h_{1T}^\perp).
	%\\
	%
\end{eqnarray}

%
%
%$f_1 = h_{1T} =(\frac{M}{m})~x~e=x~f^\perp=(\frac{M}{m})~x~g_T'=x~h_{T}^\perp=x^2~\big(\frac{M^2}{{\bf p_\perp^2}+m^2}\big)x^2~f_3=(\frac{M^2}{{\bf p_\perp^2}+m^2}\big)~x^2~h_{3T}=\ltimes$
%
%

%
%
%%%%%%%%%%%%%%%%%%%%%%%%%%%%%%%%
%
% $ \mathcal{F}\enspace  \mathbcal{F}$
%$ \mathcal{F}\enspace $
%%
%%
%$ \mathcal{T}$
%$ \mathcal{T}\enspace$
%$  \mathbcal{T}$
%$ \mathcal{T}\enspace $
%%
%$ \ltimes \varpi  \varsigma \varrho$ \\
%%  $\merge$
%
%$f_1 = h_{1T} =(\frac{M}{m})~x~e=x~f^\perp=(\frac{M}{m})~x~g_T'=x~h_{T}^\perp=x^2~\big(\frac{M^2}{{\bf p_\perp^2}+m^2}\big)x^2~f_3=(\frac{M^2}{{\bf p_\perp^2}+m^2}\big)~x^2~h_{3T}=\ltimes$
%
%

\subsubsection{Inequality Relations for Vector Diquark}
\label{sec_lsre3v}
Using explicit Eqs. \eqref{eef1v}-\eqref{eeh3tpv}, the inequality relations existing among vector diquark TMDs can be written as
\begin{eqnarray}
	\mho_{V_{1}} \ge 0, \\
	\mho_{V_{2}} \le 0, \\
	\varrho_{V} \le 0,\\
	\varpi_{V} \le 0,\\
	h_{1T}^\perp \ge 0,\\
	h_1 \le 0 ,\\
	x~g_{T}^\perp \le 0,\\
	x~g_{T} \le 0,	\\
	x~h_{L} \ge 0,\\
	x^2 h_{3T}^{\perp} \ge 0,\\
	x^2~h_{3} \le 0,\\
| \mho_{V_{1}} | \ge \frac{\bfp^2}{2M^2} | {h_{1T}^\perp} |, \\
| \mho_{V_{2}} | \ge |h_{1}|, \\
 |\varsigma_{V}  \bigg(\frac{|N_0^\nu|^2-{2}|N_1^\nu|^2}{|N_0^\nu|^2}\bigg) | \le 4|g_{1L}|.
	%	\\
	%	
	%
\end{eqnarray}

\subsection{TMD Amplitude Matrix}
\label{sec_rel2}
In this subsection, we have explored the amplitude matrix for various diquark possibilities within the context of TMDs. The generalized form of amplitude matrix for diquark spectator $\lambda^{Sp}$ can be expressed as
%$\lambda^{Sp}$
%
\[
A^{\lambda^{Sp}}
= 
\begin{bmatrix}
	A_{++,++}^{\lambda^{Sp}}  &  A_{++,+-}^{\lambda^{Sp}} & A_{++,-+}^{\lambda^{Sp}} & A_{++,--}^{\lambda^{Sp}}     \\
	A_{+-,++}^{\lambda^{Sp}}  &  A_{+-,+-}^{\lambda^{Sp}} & A_{+-,-+}^{\lambda^{Sp}} & A_{+-,--}^{\lambda^{Sp}}     \\
	A_{-+,++}^{\lambda^{Sp}}  &  A_{-+,+-}^{\lambda^{Sp}} & A_{-+,-+}^{\lambda^{Sp}} & A_{-+,--}^{\lambda^{Sp}}     \\
	A_{--,++}^{\lambda^{Sp}}  &  A_{--,+-}^{\lambda^{Sp}} & A_{--,-+}^{\lambda^{Sp}} & A_{--,--}^{\lambda^{Sp}}    
\end{bmatrix}.
\]
The generalized element of the matrix is corresponding to the wave function product as
\begin{eqnarray}
	A_{{\Lambda^{N_f}}{\lambda^{q_f}},~{\Lambda^{N_i}}{\lambda^{q_i}}}^{\lambda^{Sp}}= \psi^{\Lambda^{N_f}\dagger}_{\lambda^{q_f} \lambda^{Sp}}(x,\bfp)\psi^{\Lambda^{N_i}}_{\lambda^{q_i}\lambda^{Sp}}(x,\bfp).
\end{eqnarray}
We have four such matrices because the helicity of the spectator for the case of scalar diquark $\lambda^{Sp}=\lambda_{S}=0$ (singlet) and that for vector diquark is $\lambda^{Sp}=\lambda^{D}=\pm 1,0$ (triplet). The TMD amplitude matrix for scalar diquark is given by
\[
A^{\lambda_{S}}
= \frac{1}{\frac{C_{S}^{2}}{16 \pi^3}}
\begin{bmatrix}
	h_{1}  & -\frac{({\textbf{p}_{x}}+\iota {\textbf{p}_{y}})}{2M} 	\varrho_{S} & \frac{({\textbf{p}_{x}}-\iota {\textbf{p}_{y}})}{2M} 	\varrho_{S} & h_{1}     \\
	-\frac{({\textbf{p}_{x}}-\iota {\textbf{p}_{y}})}{2M}	\varrho_{S}  &  -\frac{\bf p_\perp^2}{2M^2}	h_{1T}^{\perp } & \frac{({\textbf{p}_{x}}-\iota {\textbf{p}_{y}})^2}{2M^2}	h_{1T}^{\perp } & -\frac{({\textbf{p}_{x}}-\iota {\textbf{p}_{y}})}{2M} \varrho_{S}    \\
	\frac{({\textbf{p}_{x}}+\iota {\textbf{p}_{y}})}{2M}	\varrho_{S}  &  \frac{({\textbf{p}_{x}}+\iota {\textbf{p}_{y}})^2}{2M^2}	h_{1T}^{\perp } & 	-\frac{\bf p_\perp^2}{2M^2} h_{1T}^{\perp } & \frac{({\textbf{p}_{x}}+\iota {\textbf{p}_{y}})}{2M} \varrho_{S}    \\
	h_{1}  & -\frac{({\textbf{p}_{x}}+\iota {\textbf{p}_{y}})}{2M} \varrho_{S} & \frac{({\textbf{p}_{x}}-\iota {\textbf{p}_{y}})}{2M} \varrho_{S} & h_{1}     
\end{bmatrix}.
\]
%
%\[
%\begin{bmatrix}
%	A_{++,++}^{S}  &  A_{++,+-}^{S} & A_{++,-+}^{S} & A_{++,--}^{S}     \\
%	A_{+-,++}^{S}  &  A_{+-,+-}^{S} & A_{+-,-+}^{S} & A_{+-,--}^{S}     \\
%	A_{-+,++}^{S}  &  A_{-+,+-}^{S} & A_{-+,-+}^{S} & A_{-+,--}^{S}     \\
%	A_{--,++}^{S}  &  A_{--,+-}^{S} & A_{--,-+}^{S} & A_{--,--}^{S}    
%\end{bmatrix}
%= \frac{1}{\frac{C_{S}^{2}}{16 \pi^3}}
%\begin{bmatrix}
%	a  &  b & c & d     \\
%    a  &  b & c & d     \\
%	a  &  b & c & d     \\
%	a  &  b & c & d     
%\end{bmatrix} .
%\]
%
%
Now, for the case of vector diquark, there are three possible amplitude matrices corresponding to the helicity possibilities as
\[
A^{0}
= \frac{1}{\frac{C_{A}^{2}}{16 \pi^3}}
\begin{bmatrix}
	-h_{1}  & \frac{({\textbf{p}_{x}}+\iota {\textbf{p}_{y}})}{2M} g_{1T} & \frac{({\textbf{p}_{x}}-\iota {\textbf{p}_{y}})}{2M} g_{1T} & h_{1}     \\
	\frac{({\textbf{p}_{x}}-\iota {\textbf{p}_{y}})}{2M}	g_{1T}  &  \frac{\bf p_\perp^2}{2M^2}	h_{1T}^{\perp } & \frac{({\textbf{p}_{x}}-\iota {\textbf{p}_{y}})^2}{2M^2}	h_{1T}^{\perp } & -\frac{({\textbf{p}_{x}}-\iota {\textbf{p}_{y}})}{2M} g_{1T}    \\
	\frac{({\textbf{p}_{x}}+\iota {\textbf{p}_{y}})}{2M}	g_{1T}  &  \frac{({\textbf{p}_{x}}+\iota {\textbf{p}_{y}})^2}{2M^2}	h_{1T}^{\perp } & 	\frac{\bf p_\perp^2}{2M^2} h_{1T}^{\perp } & -\frac{({\textbf{p}_{x}}+\iota {\textbf{p}_{y}})}{2M} g_{1T}    \\
	h_{1}  & -\frac{({\textbf{p}_{x}}+\iota {\textbf{p}_{y}})}{2M} g_{1T} & -\frac{({\textbf{p}_{x}}-\iota {\textbf{p}_{y}})}{2M} g_{1T} & -h_{1}     
\end{bmatrix},
\]
%
%
%\frac{({\textbf{p}_{x}}+\iota {\textbf{p}_{y}})}{2M}
%t
%
%
%   ({\textbf{p}_{x}}+\iota {\textbf{p}_{y}})^2
%
%\frac{({\textbf{p}_{x}}+\iota {\textbf{p}_{y}})^2}{2M^2}
%
%\frac{\bf p_\perp^2}{M^2}
%
\[
A^{+}
= \frac{1}{\frac{C_{A}^{2}}{16 \pi^3}}
\begin{bmatrix}
	\frac{1}{2} \big[	f_{1} + 2	g_{1L} + 2 	h_{1} \big]  &  	\frac{({\textbf{p}_{x}}+\iota {\textbf{p}_{y}})}{2M} \big[2h_{1L}^{\perp} - g_{1T}\big] & 0 & 0     \\
	\frac{({\textbf{p}_{x}}-\iota {\textbf{p}_{y}})}{2M} \big[2h_{1L}^{\perp} - g_{1T}\big]  &  \frac{1}{2} \big[	f_{1} - 2	g_{1L} - \frac{1}{M^2} \frac{\bf p_\perp^2}{x^2 M^2} 	h_{1T}^{\perp} \big] & 0 & 0     \\
	0  &  0 & 0 & 0     \\
	0  &  0 & 0 & 0     
\end{bmatrix}, 
\]
\[
A^{-} 
= \frac{1}{\frac{C_{A}^{2}}{16 \pi^3}}
\begin{bmatrix}
	0  &  0 & 0 & 0     \\
	0  &  0 & 0 & 0     \\
	0  &  0 & \frac{1}{2} \big[	f_{1} - 2	g_{1L} - \frac{1}{M^2} \frac{\bf p_\perp^2}{x^2 M^2} 	h_{1T}^{\perp} \big] & 	\frac{-({\textbf{p}_{x}}+\iota {\textbf{p}_{y}})}{2M} \big[2h_{1L}^{\perp} - g_{1T}\big]     \\
	0  &  0 & 	\frac{-({\textbf{p}_{x}}-\iota {\textbf{p}_{y}})}{2M} \big[2h_{1L}^{\perp} - g_{1T}\big] & \frac{1}{2} \big[	f_{1} + 2	g_{1L} + 2 	h_{1} \big]     
\end{bmatrix}. 
\]
These matrices, along with the use of TMD relations discussed before, can easily be used to map TMD expressions \cite{Maji:2017bcz,Sharma:2023llg,Sharma:2023wha,Sharma:2023fbb,Sharma:2022caa,Sharma:2022ylk} within LFQDM \cite{Maji:2016yqo}. 
				\section{Summary and Conclusions}
				\label{sec_conclusion}
In this work, we have investigated the T-even TMDs for proton at all levels of twist using systematic computations within the context of the LFQDM. From the parameterization equations of TMDs, we have found that there are multiple ways through which a TMD can be expressed in terms of the helicities of the proton in the initial and the final state. Acknowledging the choices of different sets of independent TMDs arising from the boundness of a few transversely polarized TMDs, we chose to provide results for all of them. For the first time, we have provided a parameterization table that can be used for the helicity recognition and derivation of proton TMDs. This can be used for the quark TMDs determination of every spin $1/2$ particle. Using this, we have provided multiple overlap forms for each TMD by exploiting the unintegrated SIDIS quark-quark correlator. This can be used for the study of the probability amplitudes. In the case of twist-$3$ and twist-$4$, the amplitude of TMDs is divided by a factor of $x$ and $x^2$, which arises from the spinor product of the correlator at higher twist where $x$ is the longitudinal momentum fraction. The higher twist TMDs contribute less compared to the leading twist but play an important role in measuring the scattering cross-section. For completion, we have presented all the T-even TMDs in the explicit expressions for both cases of diquark, being a scalar or a vector. Due to the fact that some TMD relations established in LFQDM and other models are being followed in LFQDM  for scalar diquark possibility only, for discussing such scenarios, we have provided relations of scalar and vector diquark possibilities separately. The unpolarized quark TMDs in Wandzura-Wilczek approximation are related to each other and shows a positive distributions for all order twist, while some TMDs shows negative distributions. Within the same model, we have examined the linear and quadratic relationships of TMDs at both intra-twist and inter-twist levels. Equation of motion relations are found to be naturally followed by LFQDM. We have also explored the inequality relations among TMDs. 
To provide aid for the derivation and analysis of TMDs in LFQDM, we have presented the amplitude matrix in the form of TMDs. These are $4$ in number, $1$ for scalar diquark's singlet state, and $3$ for vector diquark's triplet state. However, including quark-gluon interactions in LFQDM can improve the results from a phenomenology point of view. We believe our work will help future model-dependent and independent TMD calculations. The upcoming electron-Ion collider (EIC) and Nuclotron-based Ion Collider fAcility (NICA) are going to provide more insight about the transverse structure of proton. We plan to further add gluon contributions to the higher order TMDs along with sea
quark contributions in our future work. We are currently focusing on the pure twist-$3$ calculations using some other model, where we can introduce gluon contributions.

\section*{Acknowledgment}

N.K. and H.D. would like to thank the Science and Engineering Research Board, Department of Science and Technology, Government of India through the grant (Ref No. TAR/2021/000157).} S.S. and H.D. are grateful to O. V. Teryaev for useful discussions and acknowledge the hospitality at Joint Institute for Nuclear Research (JINR) during ``India-JINR workshop on elementary particle and nuclear physics and condensed matter research''.

% can use a bibliography generated by BibTeX as a .bbl file
% BibTeX documentation can be easily obtained at:
% http://www.ctan.org/tex-archive/biblio/bibtex/contrib/doc/
	 \bibliography{sample}{}
				\bibliographystyle{unsrt}

\end{document}